\documentclass[11pt]{article}
\pdfoutput=1
\usepackage{myoptions}
\usepackage{jheppub}
\usepackage{float, extarrows, tikz-cd}
\usepackage{graphicx}
\usepackage{slashed}
\usepackage{tabularx,ragged2e}
\usepackage{amssymb}
\usepackage{amsmath,amssymb}
\usepackage{slashed}
\usepackage{hyperref}
\usepackage{caption}
\usepackage{xcolor}
\usepackage{dsfont}
\usepackage{verbatim}
\usepackage{subfig}
\usepackage{mathtools, xcolor,ytableau, amsfonts,tikz}
\usepackage{graphicx}
\usepackage{physics}
\usepackage{placeins}
\newcolumntype{C}{>{\Centering\arraybackslash}X}

\ytableausetup{centertableaux,boxsize=.5em}
\title{Pole skipping away from maximal chaos}

											 \author[a,b,c]{Changha Choi,}   
                                          \author[a]{M\'ark Mezei,}                                       
                                           \author[d]{G\'abor S\'arosi}
                                           \affiliation[a]{Simons Center for Geometry and Physics, SUNY, Stony Brook, NY 11794, USA}                                                                    
                                            \affiliation[b]{C.N. Yang Institute for Theoretical Physics, SUNY, Stony Brook, NY 11794, USA}
                                           \affiliation[c]{Kavli Institute for Theoretical Physics, University of California, Santa Barbara, CA 93106, USA}
										 \affiliation[d]{CERN, Theoretical Physics Department, 1211 Geneva 23, Switzerland}	\emailAdd{changhachoi@gmail.com}  
                                           \emailAdd{mmezei@scgp.stonybrook.edu}                                           
                                           \emailAdd{gabor.sarosi@cern.ch}

\abstract{
Pole skipping is a recently discovered subtle effect in the thermal energy density retarded two point function at a special point in the complex $(\omega,p)$ planes. We propose that pole skipping is determined by the stress tensor contribution to many-body chaos, and the special point is at
$(\omega,p)_\text{p.s.}= i \lambda^{(T)}(1,1/u_B^{(T)})$, where $\lambda^{(T)}=2\pi/\beta$ and $u_B^{(T)}$ are the stress tensor contributions to the Lyapunov exponent and the butterfly velocity respectively. While this proposal is consistent with previous studies conducted for maximally chaotic theories, where the stress tensor dominates chaos, it clarifies that one cannot use pole skipping to extract the Lyapunov exponent of a theory, which obeys $\lambda\leq \lambda^{(T)}$. On the other hand, in a large class of strongly coupled but non-maximally chaotic theories $u_B^{(T)}$ is the true butterfly velocity and we conjecture that $u_B\leq u_B^{(T)}$ is a universal bound. While it remains a challenge to explain pole skipping in a general framework, we provide a stringent test of our proposal in the large-$q$ limit of the SYK chain, where we determine $\lam,\, u_B,$ and the energy density two point function in closed form for all values of the coupling, interpolating between the free and maximally chaotic limits. Since such an explicit expression for a thermal correlator is one of a kind, we take the opportunity to analyze many of its properties: the coupling dependence of the diffusion constant, the dispersion relations of poles, and the convergence properties of all order hydrodynamics. 
}
\begin{document}
\begin{flushright}
\hfill{\tt CERN-TH-2020-171}
\end{flushright}

\maketitle
\renewcommand{\arraystretch}{1.5}
\newcommand{\cev}[1]{\reflectbox{\ensuremath{\vec{\reflectbox{\ensuremath{#1}}}}}}
\newcommand{\lag}{\mathcal{L}}
\newcommand{\beq}{\begin{equation}\begin{aligned}}
\newcommand{\eeq}{\end{aligned}\end{equation}}
\newcommand{\ti}{\frac{2i}{3}}
\newcommand{\tp}{\frac{1}{2\pi}}
\newcommand{\tb}{\tilde{b}}
\newcommand{\fp}{\frac{1}{4\pi}}
\newcommand{\dc}{\Delta c}
\newcommand{\lz}{\mathcal{L}_0[A]}
\newcommand{\lh}{\hat{\mathcal{L}}}
\newcommand{\gam}{\gamma}
\newcommand{\Gam}{\Gamma}
\newcommand{\til}{\tilde}
\newcommand{\pp}[2]{\frac{#1}{\sqrt{1+(#2)^2}}}
\newcommand{\bpm}{\begin{pmatrix}}
\newcommand{\epm}{\end{pmatrix}}
\newcommand{\texr}{\textcolor{red}}
\newcommand{\lr}{\left\langle}
\newcommand{\rr}{\right\rangle}

\section{Introduction and summary of results}

There is a rich variety of phenomena and signatures associated to quantum chaos. Sharpening our understanding of one of these, the growth of operator complexity and its probe, the out-of-time order correlation function (OTOC)  has lead to great advances in the understanding of many-body chaos and its relation to gravity through AdS/CFT in recent years. 

Recently, a novel signature of chaos was proposed, the pole skipping phenomenon \cite{Grozdanov:2017ajz,Blake:2017ris}. While this effect was very convincingly demonstrated in \cite{Grozdanov:2017ajz,Blake:2017ris,Blake:2018leo,Haehl:2018izb,Grozdanov:2018kkt,Haehl:2019eae} for theories saturating the chaos bound of \cite{Maldacena:2015waa}, it was not understood what the fate of this effect was away from maximal chaos. In this paper, we propose a generalization of pole skipping away from maximal chaos. We provide a stringent test of the proposal by exact computations in a solvable chaotic theory, the large-$q$ SYK chain \cite{Gu:2016oyy,Maldacena:2016hyu}.

We spend the rest of the introduction reviewing the behavior of OTOCs in large-$N$ theories, introducing the pole skipping phenomenon, formulating it away from maximal chaos, and providing evidence for the proposal. Since our investigation led us to a precious simple closed form thermal correlator in an interacting chaotic theory, we close the introduction with this formula and a brief discussion of its properties.

\subsection{Pole skipping at maximal chaos}

In maximally chaotic theories in the regime $\beta\ll t,\abs{x}\ll t_\text{scr}=O(\log N)$  the OTOC behaves as
\es{OTOCMax}{
\text{OTOC}(t,x)=1-{\#\over N}\, \exp\left[{2\pi \ov \beta} \left(t-{\abs{x}/u_B}\right)\right]\,, \qquad \color{red}{\boxed{\text{maximal chaos}}}
} 
where the $\lambda_L={2\pi / \beta}$ is the maximal Lyapunov exponent and $u_B$ is the butterfly velocity that determines the boundary of the butterfly cone, the region in which the OTOC grows.

Recently the pole skipping phenomenon in the thermal energy density retarded two point function $G^R_{\varepsilon\varepsilon}(\om,p)$ was discovered \cite{Grozdanov:2017ajz,Blake:2017ris} and tested in maximally chaotic theories \cite{Grozdanov:2017ajz,Blake:2017ris,Blake:2018leo,Haehl:2018izb,Grozdanov:2018kkt,Haehl:2019eae}. The  energy density retarded two point function has a family of hydrodynamic poles on the complex $\omega$ plane as we vary $p$:
\es{HydroPole}{
\omega_\text{pole}(p)=\begin{cases}
\pm c_s p+\dots\quad &\text{(sound)}\\
-iD p^2+\dots\quad &\text{(energy diffusion)}
\end{cases}
}
where the dispersion relation depends on whether the system conserves momentum or not. The discovery of \cite{Grozdanov:2017ajz} was that this family of poles goes through a special point determined by $\lambda_L$ and $u_B$:
\es{PoleSkip}{
(\omega,p)_\text{p.s.}=i\lambda_L\left(1,{1\over u_B}\right)\,, \qquad \color{red}{\boxed{\text{maximal chaos}}}
}
 and at this precise location its residue vanishes. The proposed explanation of this pole skipping is that the growth of the OTOC comes from the same hydrodynamic mode that is responsible for energy transport \cite{Blake:2017ris}. In the OTOC, this pole gives rise to the exponential growth, while in the energy-energy correlator, the growth must be absent, and therefore the pole must be skipped. This prediction of pole skipping in the thermal energy density two point function for maximally chaotic system was verified in a wide class of holographic systems dual to Einstein gravity with a general matter content \cite{Blake:2018leo}.
 
 Hence it is natural to expect pole skipping to take place at the point \eqref{PoleSkip} also away from maximal chaos. This would make pole skipping extremely exciting, as it would suggest that one can extract some data about chaos from a two point function, which is a much simpler observable than the OTO four point function. However, the naive conjecture \eqref{PoleSkip}  cannot be true. The most elementary way to see this, is to realize that while not all 2d CFTs are maximally chaotic, their  retarded energy density two point function $G^R_{\varepsilon\varepsilon}$ is universal \cite{Haehl:2018izb},\footnote{However, pole skipping for a 2d CFT on a compact spatial manifold is not universal, instead it depends on the stress tensor one point function \cite{Ramirez:2020qer}.}
 \es{2dCFT}{
 G^R_{\varepsilon\varepsilon}(\om,p)=-{\pi c\ov 6}\,\om\le(\om^2+(2\pi/\beta)^2\ri)\le[{1\ov \om-p+i\epsilon}+{1\ov \om+p+i\epsilon}\ri]\,,
 }
 where $\epsilon>0$ is infinitesimal. This expression exhibits pole skipping at the point $(\omega,p)_\text{p.s.}=i\,(2\pi/\beta)(1,1)$, which is unrelated to $\lam_L$ in general. Note that $u_B=1$ in any 2d CFT \cite{Mezei:2019dfv}. We explain below how to modify \eqref{PoleSkip} so that it holds away from maximal chaos.

\subsection{Pole skipping away from maximal chaos}

Away from maximal chaos \eqref{OTOCMax} is replaced by 
\es{OTOCNonMax}{
\text{OTOC}(t,x)=1-{\#\over N}\, \exp\left[\lambda\left(\abs{x}\over t\right)\,t\right]\,,
} 
where we introduced the velocity dependent Lyapunov exponent (VDLE) $\lambda\left(u\right)$ \cite{Xu:2018xfz,Khemani:2018sdn,Mezei:2019dfv}. The ordinary Lyapunov exponent $\lambda_L$ is obtained by setting $u=0$: $\lambda_L=\lambda\left(0\right)$, and $u_B$ is defined by the edge of the growing region, $\lambda\left(u_B\right)=0$. Note that while in the maximally chaotic case chaos is characterized by two numbers, $\lambda_L$ and $u_B$, in the non-maximal case there is a whole function $\lambda\left(u\right)$ worth of freedom. In terms of this function, the chaos bound translates to the pointwise bound $\lambda(u)\leq 2\pi \beta^{-1}(1-|u|/u_B)$ \cite{Mezei:2019dfv}.

Importantly, in all examples we know  $\lambda\left(u\right)$ is a result of a competition between two terms, one coming from the ``leading Regge trajectory'' and another describing the contribution of the stress tensor.\footnote{The stress tensor itself is part of the leading Regge trajectory, but it plays a distinguished role.}
For small $\abs{u}$ the Regge contribution dominates. For larger $\abs{u}\leq u_B$ there are two possible scenarios depending on the details of $\lambda\left(u\right)$ \cite{Mezei:2019dfv}: In the first case the Regge contribution dominates for all $0\leq \abs{u}\leq u_B$. In the second case, above some critical velocity $u_*\leq \abs{u}\leq u_B$ the stress tensor dominates, and we have
\es{lamvT}{
\lam(u)=\lam_L^{(T)}\le(1-{\abs{u}\ov u_B^{(T)}}\ri)\qquad \le(u_*\leq \abs{u}\leq u_B^{(T)}\ri)\,.
}
The stress tensor contribution to chaos is determined by two numbers $\lam_L^{(T)}={2\pi/ \beta}$ and $ u_B^{(T)}$. In the second scenario described in \eqref{lamvT}, we have $u_B= u_B^{(T)}$, while in the first case we have $u_B< u_B^{(T)}$.\footnote{This inequality follows from the concavity of the contribution of the leading Regge trajectory to the VDLE. In all the examples we know, this is true, and we suspect it holds generally.} In maximally chaotic theories one has  $u_*=0$, i.e. the stress tensor dominates for all $u$, and we recover \eqref{OTOCMax}.

We are now ready to formulate our proposal. We conjecture that the pole skipping point in the retarded energy density two point function is at\footnote{This formula is valid in the presence of (possibly discrete) translational symmetry.} 
\es{PoleSkip2}{
\boxed{(\omega,p)_\text{p.s.}=i\lambda_L^{(T)}\left(1,{1\over u_B^{(T)}}\right)\,,}
}
i.e. it encodes the stress tensor contribution to chaos. While the stress tensor determines chaotic dynamics at maximal chaos, its contribution can be decreased in non-maximally chaotic theories or completely cancelled in integrable systems; an example of the latter is provided in \cite{Perlmutter:2016pkf}. 
The proposal is consistent with the literature on pole skipping in maximally chaotic theories, since in that case $\lambda_L=\lambda_L^{(T)}$ and $u_B=u_B^{(T)}$. It is also consistent with the 2d CFT result discussed around \eqref{2dCFT}. 

Let us discuss the existing evidence for \eqref{PoleSkip2} in the literature. The main evidence comes from conformal field theories on Rindler space. Rindler space is a patch of flat spacetime but it is conformal to $S^1\times \mathbb{H}_{d-1}$ and therefore it is an example of thermal physics that can be studied using vacuum correlators in a conformal field theory. For example, the thermal energy density two point function can be obtained from the vacuum one via a Weyl transformation and therefore is universal to all CFTs. Pole skipping in this context was recently studied in \cite{Haehl:2019eae} and it was shown that it happens at the values corresponding to maximal chaos.\footnote{Since the spatial manifold $\mathbb{H}_{d-1}$ is curved, \eqref{PoleSkip2} must be interpreted so that it applies to the appropriate notion of Fourier transform. In flat space the Fourier mode at the poles skipping point takes the form $\exp\le[i p_\text{p.s.} x\ri]=\exp\le[- \frac{\lambda_L^{(T)}}{u^{(T)}_B} x\ri]$. In analogy with this result, the right interpretation in hyperbolic space  is that for large geodesic separations, the Fourier mode at the pole skipping point must behave as $\exp[{-\frac{\lambda_L^{(T)}}{u^{(T)}_B}\rho}]$, where $\rho$ is the spatial geodesic distance.} Note that most CFTs on Rindler space \textit{do not} display maximal chaos, instead the Lyapunov exponent is determined by the resummation of operators on the leading conformal Regge trajectory and is related to the analytic continuation of the spin of this trajectory to non-physical operator dimensions \cite{Maldacena:2015waa,Mezei:2019dfv}. However, for a large class of theories the stress tensor contribution dominates near the butterfly front and $u^{(T)}_B=(d-1)^{-1}$ gives the true butterfly speed; for example this happens in planar $\mathcal{N}=4$ SYM theory whenever the 't Hooft coupling is greater than 37.7384 \cite{Mezei:2019dfv}, while the theory is only maximally chaotic at infinite coupling.

   Another piece of non-trivial evidence comes from holographic theories with higher derivative corrections performed using shockwaves \cite{Roberts:2014isa,Maldacena:2015waa,Mezei:2016wfz,Alishahiha:2016cjk}.\footnote{Shockwave computations only capture $t$-channel graviton exchange, but correctly compute the VDLE $\lam(v)$ in Einstein gravity. It was found in \cite{Chowdhury:2019kaq} that there are contributions to the four graviton S-matrix in higher derivative gravity theories that are not captured by  the  shockwave computations and that change the value of the Lyapunov exponent. It is an open problem to determine how these terms change the VDLE  and $v_B$. We thank Shiraz Minwalla and Douglas Stanford for a discussion on this point.} Such corrections do not affect the Lyapunov exponent but they change the butterfly speed. It was shown in \cite{Grozdanov:2018kkt} that both for Gauss-Bonnet coupling and the leading $\alpha'^3 R^4$ correction the change in the butterfly speed and the energy-energy correlator are such that pole skipping remains valid. This is consistent with \eqref{PoleSkip2} since the Lyapunov exponent stays maximal. In fact, when we consider finite coupling corrections to the thermal OTOC in $\mathcal{N}=4$ SYM\footnote{Note that here we put the theory on $S^1\times \mathbb{R}_{d-1}$ as opposed to the Rindler discussion of the previous paragraph.} we need to include worldsheet effects that correct the Lyapunov exponent \cite{Shenker:2014cwa} in addition to the leading higher derivative term $\alpha'^3 R^4$ that corrects the butterfly speed.\footnote{The first correction to the Lyapunov exponent comes at order $1/\sqrt{\lambda}$ while to the butterfly speed at order $1/\lambda^{3/2}$, where $\lambda$ is the 't Hooft coupling.} The fact that in this case pole skipping still happens at maximal Lyapunov exponent is additional evidence for \eqref{PoleSkip2} (and contradicts the naive propsal \eqref{PoleSkip}).

\begin{figure}[!h]
\begin{center}
\includegraphics[width=0.5\textwidth]{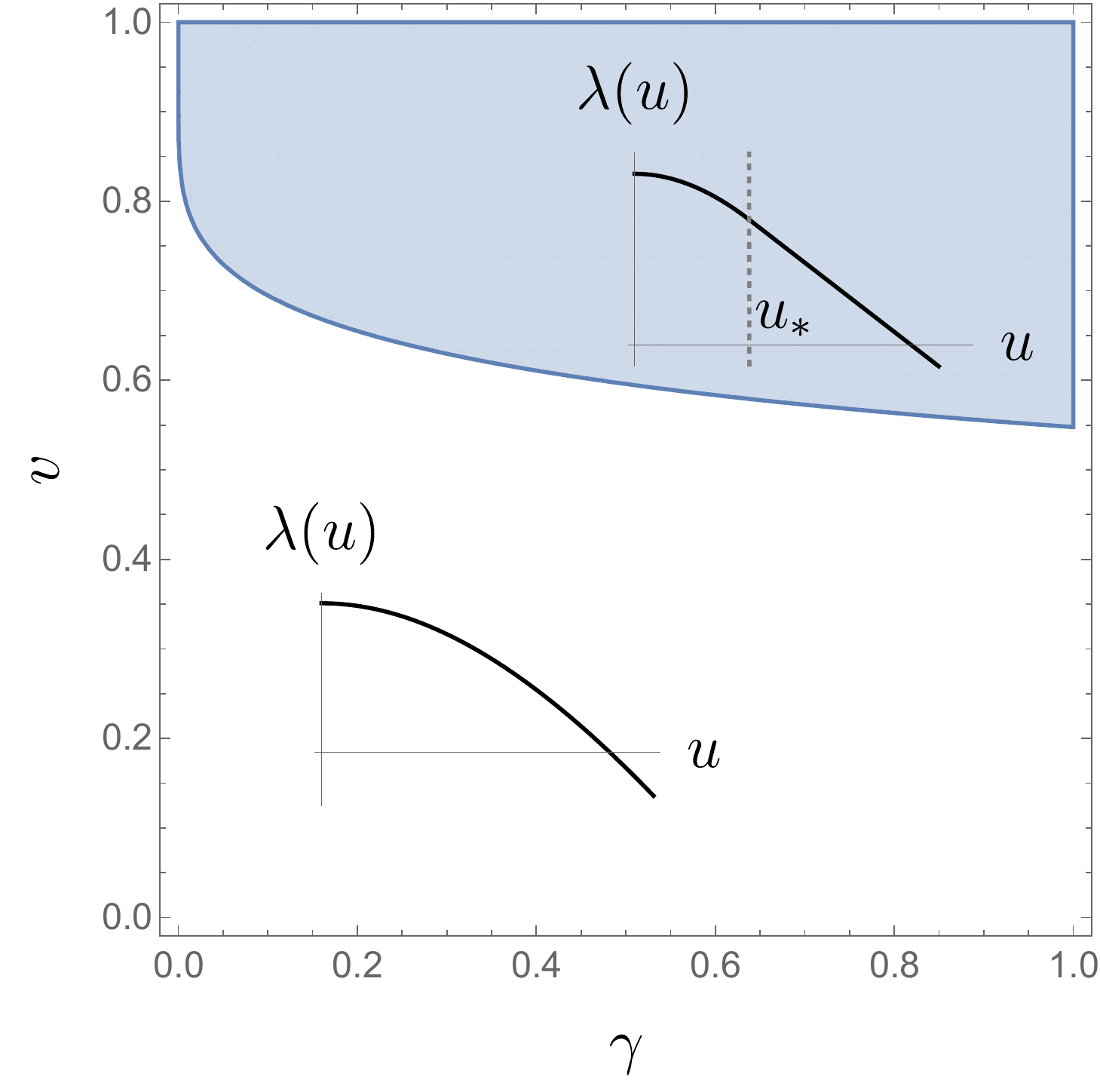}
\caption{The two dimensional coupling space of the model. There is a critical line $v_*(\ga)$ that separates the blue region, where above a critical velocity $u_*$ the VDLE is dominated by the pole and is maximal, from the white region, where the saddle always dominates, and the VDLE is nowhere maximal. }
\label{fig:vdle}
\end{center}
\end{figure}

The main technical result of this paper is the confirmation of \eqref{PoleSkip2} in the large-$q$ limit of the SYK chain introduced in \cite{Gu:2016oyy}, where we obtain both $\lam(u)$ and $ G^R_{\varepsilon\varepsilon}$ in closed form, and can prove \eqref{PoleSkip2} analytically. The model has two  dimensionless coupling constants $0< v <1$ and $0<\gamma\leq1$. The latter controls the strength of inter-site coupling, while the former controls the interaction strength; as a function of $v$ the model interpolates between a free and a maximally chaotic theory. There exists a critical line in coupling space $v_*(\gamma)$, above which (in the regime $v_*(\gamma)\leq v <1$) chaos is non-maximal, but the butterfly speed is maximal $u_B=u_B^{(T)}$, while for $0< v <v_*(\gamma)$ we get that both $\lam_L<\lam_L^{(T)}$ and $u_B<u_B^{(T)}$, see Fig.~\ref{fig:vdle}. Contrary to the 2d CFT case, here both $u_B$ and $u_B^{(T)}$ are nontrivial functions of the couplings. 

\pagebreak

We end this section with some comments about \eqref{PoleSkip2}.
\begin{itemize}

\item The proposal is somewhat anticlimactic, as it shows that just from determining the pole skipping point, we cannot read off the true $\lam_L$ and $u_B$ characterizing the OTOC. On the other hand, while $\lambda^{(T)}_L\equiv 2\pi/\beta$ does not carry any non-trivial information, $u_B^{(T)}$ still depends on the theory. Moreover, for a large set of strongly coupled but not maximally chaotic theories, one has $u_B=u_B^{(T)}$. In these theories, the OTOC butterfly speed can be read from the location of the pole skipping point, however, we caution that there is no way to tell whether  $u_B^{(T)}$ is the true butterfly speed just by having access to the energy density two point function.  

\item 
The proposal \eqref{PoleSkip2} refers to the stress tensor contribution to the growth of the OTOC. This can be defined in models where we can analytically compute the OTOC, but it is not entirely clear how to define it in complete generality. However, when there is a critical velocity $u_*<u_B$, we can read this data directly from the behavior of the VDLE near $u=u_B$.

\item There is strong evidence that $u_B\leq u_B^{(T)}$ is true generally \cite{Mezei:2019dfv}, in which case we can read a bound on the butterfly effect from the pole skipping point using the VDLE bound of  \cite{Mezei:2019dfv} (given in the text after \eqref{OTOCNonMax}).

\item It would be very interesting to prove  \eqref{PoleSkip2} in full generality. This would likely require the development of an effective field theory framework for the four point functions of non-maximally chaotic theories generalizing the works \cite{Blake:2017ris,Kitaev:2017awl} that apply at maximal chaos. The theory would have to apply for large Lorentzian times for OTOC configurations, and would also have to govern energy transport at finite frequency and momentum.

\end{itemize}

\subsection{The energy density two point function in the large-$q$ SYK chain}

A byproduct of our investigation is a closed form expression for the energy density two point function in the large-$q$ SYK chain, see Sec.~\ref{sec:SYKreview} for the definition of the model. It is given by the simple formula:
\es{2ptCorr}{
\boxed{G_{\varepsilon \varepsilon }^R(\omega)=-{Nv\ov2q^2}\le(\p_\theta \log  \psi_n(\theta_v) +\tan{\pi v\ov 2}{h(h-1)\ov2}\ri)\Big\vert_{\, n\rightarrow -i\omega+\epsilon}\,,}
}
where $v$ is the SYK coupling constant, we chose the inverse temperature $\beta=2\pi$ (without loss of generality), and the function $\psi_n(\theta)$  is a sum of two hypergeometric functions given explicitly by:
\es{masterfuncDef}{
\psi_n(\theta)&=c_o \psi_n^o(\theta)+c_e \psi_n^e(\theta)\,,\\
c_o&={\Gamma\left(1-{h\ov 2}-{n\ov 2v}\right) \sin\left({\pi h \ov 2}+{\pi n \ov 2v} \right) \sin\left({n\pi\ov 2}\right)\ov  \Gamma\left({1\ov 2}-{h\ov 2}+{n\ov 2v}\right)}\,,\\
  c_e&= {\Gamma\left({1\ov 2}-{h\ov 2}-{n\ov 2v}\right)\cos \left({\pi h \ov 2}+{\pi n \ov 2v} \right) \cos\left({n\pi\ov 2}\right)\ov 2\Gamma\left(1-{h\ov 2}+{n\ov 2v}\right)}\,,\\
  \psi^e_{n}&=\sin( \theta)^h  {_2 F_1}\left({h-n/v\ov 2},{h+n/v \ov 2},{1 \ov 2},\cos^2\theta\right)\,,\\
\psi^o_{n}&= \cos( \theta)\sin( \theta)^h {_2F_1}\left({1+h-n/v\ov 2},{1+h+n/v\ov 2},{3 \ov 2},\cos^2\theta\right)\,,\\
h&=\frac12\le(1+\sqrt{9+4\gamma (\cos(p)-1)}\ri)\,,\\
\theta_v &=\frac{\pi}{2}(1-v)\,,
}
where $\gamma$ is the inter site coupling constant and $p$ is the momentum.
We believe this result is precious: this is the only closed form thermal correlator in a chaotic system, which changes nontrivially as we go from weak to strong coupling.
The only analogous known expressions is \eqref{2dCFT} in  2d CFT (and similar results for higher dimensional CFTs in hyperbolic space \cite{Haehl:2019eae}), which however is universal for any CFT integrable or chaotic, and is completely determined by symmetry. Beyond this, there exist results either for small coupling \cite{Hartnoll:2005ju,Romatschke:2015gic,Kurkela:2017xis,Moore:2018mma,Grozdanov:2018atb}, or in the holographic setting corresponding to large coupling \cite{Son:2002sd,Starinets:2002br,Kovtun:2005ev,Hartnoll:2005ju,Alday:2020eua}.

We extract \eqref{2ptCorr} from a complicated formula for the four point function by taking an OPE limit. 
The main technical result of the paper is that we find the unique analytic continuation of this Euclidean Green function defined at discrete Matsubara frequencies to the entire complex plane, which allows us to  prove \eqref{PoleSkip2} and to study its analytic structure. 

In more detail, is easy to verify that \eqref{2ptCorr} exhibits pole skipping at
\es{omhPS}{
(\omega,h(p))_\text{p.s.}=\le(i,1+{1\ov v}\ri)\,.
}
With the explicit  simple formulas  for $\lam(u),\, u_B,\, \lam_L^{(T)},\, u_B^{(T)}$ determined in Sec.~\ref{sec:Lyapunov}, we can verify that this confirms the proposal \eqref{PoleSkip2} for all values of the couplings $v$ and $\gamma$.

The only singularities of the correlator \eqref{2ptCorr} are poles. For small $p$, the closest pole to the real $\om$ axis is a diffusion pole with dispersion relation $\om(p)=-i D p^2+\dots$. As we increase $p$ this pole can collide with a non-hydrodynamic pole on the imaginary $\om$ axis, and subsequently the poles can depart the imaginary $\om$ axis to the complex plane. While this scenario is reminiscent of what happens in theories with weakly broken momentum conservation  \cite{Grozdanov:2018fic}, in our model these poles never become propagating quasiparticles (sound modes), as they always have comparable real and imaginary frequencies. We also show that besides a Drude peak (present for small momentum) the spectral function is featureless: it does not show the presence of quasiparticles. 

The collision of poles delineates the range of momenta for which all order hydrodynamics converges \cite{Grozdanov:2019kge,Grozdanov:2019uhi}. We determine the radius of convergence by finding pole collisions and the reconnection of pole trajectories for complex momenta. Unlike in more familiar examples, here the radius of convergence of hydrodynamics decreases with increasing coupling $v$.

\subsection{Outline}

The outline of the paper is as follows. In Sec.~\ref{sec:SYKreview} we review the SYK chain and its large-$q$ limit, where the quantum many-body problem can be reduced to solving differential equations. In Sec.~\ref{sec:Lyapunov} we use the retarded kernel approach for computing $\lam(u)$. This is considerably simpler than computing the full four point function that we undertake in Sec.~\ref{sec:4point}. In Sec.~\ref{sec:2point} we extract a simple formula for the Euclidean energy density two point function from the complicated formula for the four point function by taking an OPE limit. In Sec.~\ref{sec:poleskip} we find all the pole skipping points in the energy density correlator, and verify the proposal \eqref{PoleSkip2}. In Sec.~\ref{sec:hydro}, we perform a detailed study of the analytic structure of this correlator, and extract information about the (all order) hydrodynamics of the SYK chain. In Appendix~\ref{app:sykdot}, we give a new derivation of the coordinate space four point function of the large-$q$ SYK dot first derived in \cite{Streicher:2019wek,Choi:2019bmd} using two different methods.

\pagebreak

\section{Large-$q$ SYK chain}
\label{sec:SYKreview}

We will be studying the SYK chain introduced in \cite{Gu:2016oyy}, which is a 1+1 dimensional generalization of the SYK dot \cite{sachdev1993gapless,kitaev2014hidden,Polchinski:2016xgd,Maldacena:2016hyu}. It is a disordered chain of Majorana fermions with a large $N$ number of on-site degrees of freedom that is chaotic, yet solvable using methods resembling Dynamical Mean Field Theory.
The fermions interact in groups of $q$, and by taking this parameter to be large, we can analytically solve the model for all values of the two remaining dimensionless coupling constants of the theory: $v$ that controls the strength of interactions (and can be thought of as a proxy for temperature measured in units of the dimensionful coupling constant ${\cal J}$) and $\gamma$ that controls the relative strength between on-site and inter-site interactions and that enters in all results in a simple kinematical way.

The Hamiltonian of the system is given by
\beq
\label{eq:Hamiltonian}
H&=i^{q/2} \sum_{x=0}^{M-1} \Big( \sum_{1\leq i_1<...<i_q\leq N}J_{i_1...i_q,x}\chi_{i_1,x}...\chi_{i_q,x}\\ &+ \sum_{\substack{1\leq i_1<...<i_{q/2} \leq N\\1\leq  j_1<...<j_{q/2}\leq N}}J'_{i_1...i_{q/2}j_1...j_{q/2},x}\chi_{i_1,x}...\chi_{i_{q/2},x}\chi_{j_1,x+1}...\chi_{j_{q/2},x+1}\Big)\,.
\eeq
Here, $\chi_{i,x}$, $i=1,...,N$ are Majorana fermions obeying the commutation relation $\lbrace \chi_{i,x},\chi_{j,x}\rbrace=\delta_{xy}\delta_{ij}$ and periodic boundary condition, $\chi_{i,0}=\chi_{i,M}$.  $q$ is an even integer, and the first line of \eqref{eq:Hamiltonian} is an on-site term, while the second line is an interaction between two groups of $q/2$ fermions on neighboring sites. $J_{i_1...i_q,x}$, $J'_{i_1...i_{q/2}j_1...j_{q/2},x}$ are independent Gaussian random variables with zero mean. Their variances are given by
\beq
\overline{{J_{i_1...i_q,x}}^2} = \frac{(q-1)!}{N^{q-1}}J_0^2\,, \quad \quad \overline{{J'_{i_1...i_{q/2}j_1...j_{q/2},x}}^2} = \frac{[(q/2)!]^2}{q N^{q-1}}J_1^2\,.
\eeq

The SYK model is known to be self-averaging \cite{sachdev1993gapless, kitaev2014hidden, Michel:2016kwn}. This means that while we are interested in the theory with quenched disorder, we can solve the technically much simpler problem of the theory with annealed disorder.  That is, we average the generator of disconnected correlation functions instead of correlation functions themselves.  After integrating out the disorder and the fermions, one obtains an effective action for bilocal fields
\beq
S_{\rm eff}[G,\Sigma]=&N\sum_{x=0}^{M-1} \Big[-\log \text{Pf}(\pa_\tau-\Sigma_x)
\\&+{1\ov 2} \int_0^\beta d\tau_1d\tau_2\left( \Sig_x G_x -{J_0^2\ov q} G_x ^q -{J_1^2\ov q} G_x^{q/2} G_{x+1}^{q/2}\right) \Big]\,,
\eeq
where $G_x(\tau_1,\tau_2)$ is the fermion two point function, while $\Sigma_x(\tau_1,\tau_2)$ is the self energy.\footnote{The fermion two point function vanishes between fermions on different sites and is only nonzero for fermions on the same site $x$ in the disorder averaged theory, making the SYK chain locally critical \cite{Gu:2016oyy}.}

This action is further simplified and becomes local (in the two time dimensions) in the large $q$ limit, when $N\gg q^2 \gg 1$.\footnote{It would also be interesting to analyse the regime $q^2\sim N$ using the methods of \cite{Berkooz:2018qkz}.} The non-trivial limit is obtained by fixing $\mathcal J_{0,1}^2\equiv q J_{0,1}^2/2^{q-1}$. In this limit, the Schwinger-Dyson equations for $G_x,\Sigma_x$ suggest that the system is almost free, with interactions entering at order $1/q$. Hence following \cite{Cotler:2016fpe}, we proceed by changing variables according to ($\sigma_x,g_x\sim O(1)$)
\beq
\Sigma_x(\tau_1,\tau_2)=\frac{1}{q}\sigma_x(\tau_1,\tau_2), \quad G_x(\tau_1,\tau_2)=\frac{\text{sgn}(\tau_{12})}{2}\left( 1+\frac{g_x(\tau_1,\tau_2)}{q}\right)\,,
\eeq
expanding the action in $1/q$ and integrating out $\sigma_x$. The resulting action at leading order in $1/q$ is a periodic lattice of coupled Liouville theories
\beq
S_{\rm eff}={N \ov 4q^2}\sum_{x=0}^{M-1} \int d\tau_1 d\tau_2 \left[  \frac{1}{4}\pa_{\tau_1} g_x\pa_{\tau_2}g_x -  \mathcal J_0^2 e^{g_x} -  \mathcal J_1^2 e^{{1 \ov 2}(g_x+g_{x+1})} \right]\,.
\eeq
The field $g_x(\tau_1,\tau_2)$ is symmetric under the exchange $\tau_1 \leftrightarrow \tau_2$ and satisfies the boundary condition $g_x(\tau_1,\tau_1)=0$.

\subsection{Linearized action}

We are interested in fluctuations around the large $N$ saddle point. Following \cite{Gu:2016oyy}, we look for a  translation invariant saddle, $g_x(\tau_1,\tau_2)=g^s(\tau_1-\tau_2)$ that is independent of $x$. The saddle point equation then is just Liouville's equation
\beq
\pa_\tau ^2  g^s= 2\mathcal{J}^2 e^{ g^s}\,, \quad \mathcal{J}^2\equiv \mathcal J_0^2 + \mathcal J_1^2 \,.
\eeq
We look for a finite temperature solution satisfying $g(0)=g(\beta)=0$. We will set $\beta=2\pi$ for simplicity, this can be easily restored by dimensional analysis. The solution satisfying the boundary conditions is \cite{Maldacena:2016hyu}
\es{Saddle}{
e^{g^s(\tau)}=\left[{\cos(\pi v /2 ) \ov \cos (\pi v ({1\ov 2}-{\vert \tau \vert \ov 2\pi}))} \right]^2\,,
\qquad 2\pi \mathcal J={\pi v  \ov \cos{\pi v \ov 2}}\,.
}
We now expand around the saddle $g_x=g^s+\delta g_x$ so that the second order action becomes
\beq
\delta S_{\rm eff}&={N \ov 4q^2} \int d\tau_1 d\tau_2\sum_{x=0}^{M-1} \le[\delta g_x \left( -\frac{1}{4}\pa_{\tau_1}\pa_{\tau_2} - {\mathcal J_0^2 \ov 2}e^{g^s}  \right)\delta g_x-{\mathcal J_1^2 \ov 2}e^{g^s}\le({\delta g_x+\delta g_{x+1}\ov 2}\ri)^2\ri]
\\&=-{N \ov 4q^2} \int d\tau_1 d\tau_2\sum_{x,y} \delta g_x \left( \frac{1}{4}\pa_{\tau_1}\pa_{\tau_2}\delta_{x,y}+\le({\mathcal J_0^2 \ov 2}+{\mathcal J_1^2 \ov 4}\ri) e^{g^s} \delta_{x,y}+{\mathcal J_1^2 \ov 8}e^{g^s} \delta_{x,y\pm1}  \right)\delta g_y\,.
\eeq
The inverse of the kernel in this bilinear action is related to the Euclidean time-ordered connected four point function $\mathcal F_{xy}$:
\beq
\label{eq:4ptfunc}
 {1\ov N}\mathcal{F}_{xy}(\tau_1,\tau_2;\tau_3,\tau_4)&\equiv \frac{1}{N^2}\sum_{ij}\langle \mathcal T \chi_{i,x}(\tau_1)\chi_{i,x}(\tau_2)\chi_{j,y}(\tau_3)\chi_{j,y}(\tau_4)\rangle-\langle G_x(\tau_1,\tau_2)\rangle \langle G_y(\tau_3,\tau_4)\rangle
 \\&=\frac{\text{sgn}(\tau_{12})\text{sgn}(\tau_{34})}{4q^2} \langle \delta g_x(\tau_1,\tau_2)\delta g_y (\tau_3,\tau_4) \rangle +O(1/N)\,.
\eeq
We proceed by Fourier transforming $\delta g_x=\frac{1}{M} \sum_{ k=0}^{M-1} \delta g_p\, e^{i\frac{2\pi k x}{M}}$ (from now on we set the lattice spacing to be 1) and taking the limit of an infinite lattice $M\rightarrow \infty$ so that the momentum variable $p=2\pi k/M$ is continuous. We obtain
\beq
\label{eq:Liouvilleaction}
\delta S_{\rm eff}&=-{N \ov 4q^2} \int d\tau_1 d\tau_2\int _{-\pi}^{\pi} {dp\ov 2\pi}~\delta g_p^* \left({1\ov 4} \pa_{\tau_1} \pa_{\tau_2}+\left[{\mathcal J_0^2 \ov 2}+{\mathcal J_1^2 \ov 4}\right] e^{g^s}+{\mathcal J_1^2 \ov 4}e^{g^s} \cos(p) \right)\delta g_{p}
\\&=-{N \ov 4q^2} \int d\tau_1 d\tau_2\int _{-\pi}^{\pi} {dp\ov 2\pi}~\delta g_p^* \left({1\ov 4} \pa_{\tau_1} \pa_{\tau_2}+\frac{{\mathcal J_0^2 \ov 2}+{\mathcal J_1^2 \ov 4}\big[1+\cos (p)\big]}{\cos^2\le(\pi v \le({1\ov 2}-{\vert \tau _{12} \vert \ov 2\pi}\ri)\ri)}{v^2 \ov 4 \mathcal J^2} \right)\delta g_{p}\,.
\eeq
Note that each momentum sector is decoupled at leading order and thus we can evaluate the fermion four point function in the momentum sector $p$ as
\beq
\mathcal{G}_p(\tau_1,\tau_2;\tau_3,\tau_4)\equiv \langle \delta g_p^*(\tau_1,\tau_2) \delta g_p(\tau_3,\tau_4)\rangle= {4q^2 \ov N}{\text{sgn}(\tau_{12})\text{sgn}(\tau_{34})}\,\mathcal F_p(\tau_1,\tau_2;\tau_3,\tau_4)\,,
\eeq
where we used translational invariance to define
\beq
 \langle \delta g_x(\tau_1,\tau_2)\delta g_y (\tau_3,\tau_4) \rangle =\int _{-\pi}^\pi{dp\ov 2\pi}e^{ip(x-y)}\langle \delta g_p^*(\tau_1,\tau_2) \delta g_p(\tau_3,\tau_4)\rangle\,, \quad\mathcal{F}_{xy}=\int _{-\pi}^\pi{dp\ov 2\pi}e^{ip(x-y)}\mathcal F_p\,.
\eeq
The boundary condition and symmetries of $\delta g$ imply
\beq
~&\mathcal F_p(\tau_1,\tau_2;\tau_3,\tau_4)=\mathcal F_p(2\pi-\tau_1,\tau_2+\pi;\tau_3,\tau_4)= \mathcal F_p(\tau_1,\tau_2;2\pi-\tau_3,\tau_4+\pi)\,,
\\& \mathcal F_p(\tau_1,\tau_2;\tau_3,\tau_4)=- \mathcal F_p(\tau_2,\tau_1;\tau_3,\tau_4)=-\mathcal F_p(\tau_1,\tau_2;\tau_4,\tau_3)\,.
\eeq 
just as in the case of the single SYK dot. 

\subsection{Ladder kernel}

It will be useful to Fourier transform in the center of mass coordinate according to $\delta g_p(\tau_1 , \tau_2)=\sum_{n\in 
\mathbb{Z}} e^{-in Y} g_{p,n}( X)$, where $Y=\frac{\tau_1+\tau_2}{2}$ and $X={\tau_2-\tau_1}$. 
In terms of these, the boundary condition $g_x(\tau_1,\tau_1)=0$ and symmetries reduce to $g_{p,n}(0)=0,~g_{p,n}(X)=(-1)^n g_{p,n}(2\pi-X)$.  The kernel of the quadratic action \eqref{eq:Liouvilleaction} reads as 
\beq \label{eq:kernel}
K\equiv{N\ov q^2}\left(\pa_X^2+{n^2\ov 4}-{h(h-1) v^2  \ov 4\cos^2\le({v(\pi-X)\ov 2}\ri)}\right),
\eeq
 where 
\es{eq:defh}{
\frac{h(h-1)}{2}&=1+\frac{\gamma }{2}[\cos(p)-1]\,,\\
\gamma &\equiv \frac{\mathcal{J}_1^2}{\mathcal{J}^2}\leq 1\,.
}
 For real momentum $p$, we have $1\leq h\leq 2$.\footnote{For generic $\ga$ the range of $h$ is smaller, and we only fill the range for $\ga=1$. }

To further simplify the kernel, we introduce $2\theta \equiv vX+(1-v)\pi=\pi-v(\pi-X)$. Then the eigenvalue problem for the $g_{p,n}(X)$ reduces to the following equation (we separate the overall factor of $N/q^2$ for simplicity)
\beq
\label{eq:redef}
v^2 \left(\pa_{\theta}^2+{\tilde n}^2-{h(h-1)  \ov \sin^2(\theta )}\right)g_{p,n}=\mu_{p,n}\,g_{p,n}\,, \quad \tilde n={n\ov v}\,.
\eeq
The above kernel admits the following zero modes:
\beq
\label{eq:eigenfunctions}
~&\psi^e_{n}=\sin( \theta)^h  {_2 F_1}\left({h-n/v\ov 2},{h+n/v \ov 2},{1 \ov 2},\cos^2\theta\right)\,,
\\&\psi^o_{n}= \cos( \theta)\sin( \theta)^h {_2F_1}\left({1+h-n/v\ov 2},{1+h+n/v\ov 2},{3 \ov 2},\cos^2\theta\right).
\eeq
These eigenfunctions are even/odd under $\theta \to \pi-\theta$. For integer $n$, these modes do not satisfy the boundary condition that the fluctuations must vanish at $X=0,2\pi$, or $\theta=\frac{\pi}{2}(1\pm v)$. We can obtain eigenfunctions instead by replacing $n$ by a non-integral $m$ that is defined by requiring these boundary conditions to hold. In this case, the eigenvalue $\mu_{p,n}$ in \eqref{eq:redef} is given by $(n^2-m^2)/4$ (this is the route taken in \cite{Choi:2019bmd} for the SYK dot). This will turn out to be impractical below, instead, we will invert the kernel using the zero modes \eqref{eq:eigenfunctions} with integer $n$.

\section{Velocity dependent Lyapunov exponent}
\label{sec:Lyapunov}
 
We would like to extract the velocity dependent Lyapunov exponent from the ladder kernel \eqref{eq:kernel}.  The simplest way to read the leading growing contribution of a mode with momentum $p$ to the four point function is to follow the retarded kernel method of \cite{kitaev2014hidden,Maldacena:2016hyu}. The rule is to replace the sides of the ladder with retarded two point functions (which are now just Heaviside theta functions) and the rungs with left-right propagators (corresponding to the Lorentzian continuation $X\rightarrow i\chi+\pi$ in the two point function, or $\theta \to i\xi+\pi/2$ in \eqref{eq:redef}).  This gives rise to a Schr\"{o}dinger problem.  The maximal growth exponent in Lorentzian $Y$ admitted by the kernel corresponds to the  ground state energy $-\kappa^2$ of
\beq
\label{eq:retardedkernel}
\left[-v^2\pa_{\xi}^2-{h(h-1) v^2 \ov \cosh^2(\xi)}\right]\psi(\xi)=-{\kappa}^2 \psi(\xi)\,.
\eeq
The ground state wave function of this problem is $\psi(\xi) \sim \frac{1}{(\cosh \xi )^{h-1}}$, and the corresponding ground state energy gives rise to the Lyapunov exponent
\es{kap}{
\kappa(p)=(h-1)v\,,
}
where the momentum dependence of $h$ is defined via taking the $1\leq h\leq2$ branch of \eqref{eq:defh}. The relevant strong coupling limit of this result can be obtained by taking $v=1-\de v,\, p^2=\de v\, \tilde p^2$ and expanding \eqref{kap} in $\delta v$, which yields $\kappa(\tilde p)=1-\le(1+\ga \,\tilde p^2/3\ri)\de v+\dots$, whose analog for $q=4$ was derived in \cite{Gu:2016oyy}. Here we obtain $\kappa(p)$ for all values of the coupling and all momenta.

 The growing piece of the four point function in real lattice space $x$ is then represented by an integral of the form (here $t=i (Y-Y')=i(\tau_1+\tau_2-\tau_3-\tau_4)/2$)
\beq
\label{eq:OTOCpint}
\int dp \ \frac{1}{\cos \le(\frac{\pi \kappa(p)}{2}\ri)}e^{\kappa(p)t+i p x}\,,
\eeq
where we have used the ladder identity of \cite{Gu:2018jsv} relating the prefactor to the exponent. For large $t$, this integral is either dominated by a saddle point or a pole at $\kappa(p)=1$, depending on $x$. We associate the pole with the stress tensor contribution.\footnote{This identification is inspired by the analogy to what one gets in conformal Regge theory \cite{Mezei:2019dfv}. It would be interesting to establish this connection firmly. We will also encounter further poles for (pure imaginary) $p$, which never dominate the integral. It would be interesting to identify whether they are also stress tensor contributions or  correspond to other operators.  } This mechanism is quite generic \cite{Shenker:2014cwa,Gu:2016oyy,Xu:2018xfz,Gu:2018jsv,Lian:2019axs,Mezei:2019dfv}, and as explained in \cite{Mezei:2019dfv}, leads to the velocity dependent Lyapunov exponent (we use $u=x/t$ for velocity to avoid confusion with SYK coupling $v$)
\beq
\label{eq:VDLE}
\lambda(u)= \begin{cases}
\lambda_{\rm saddle}(u)\equiv \text{ext}_r (\kappa(ir)-r u) & \text{when } u<u_*\,, \\
1-\frac{u}{u^{(T)}_B} & \text{when } u>u_*\,,
\end{cases}
\eeq
where the stress tensor butterfly speed $u^{(T)}_B$ and critical velocity $u_*$ are defined via solving
\beq
\kappa\left( \frac{i}{u^{(T)}_B} \right)=1\,, \quad u_*=i\frac{d\kappa(p)}{d p}\bigg\vert_{p={i/u^{(T)}_B}}\,.
\eeq
Explicit expressions for these quantities read as
\beq
\label{eq:stresstensorVB}
u^{(T)}_B=\left(\text{arccosh} \frac{1+v+(\gamma -2)v^2}{\gamma v^2} \right)^{-1}\,, \quad u_*=\frac{\sqrt{(1+v-2v^2)(1+v+2(\gamma -1)v^2)}}{2+v}\,.
\eeq
Let us, for completeness, also write down the explicit formula for the Legendre transform part of the VDLE (first line of \eqref{eq:VDLE})
\beq
\small
\label{eq:VDLEsaddle}
\lambda_{\rm saddle}(u) &=
\frac{1}{2} v \left(\sqrt{-4 \gamma +\frac{4 \left(2 u^2+\sqrt{4 u^4+(9-4 \gamma ) u^2 v^2+\gamma ^2 v^4}\right)}{v^2}+9}-1\right)\\ &-u \arccosh\left(\frac{2 u^2+\sqrt{4 u^4+(9-4 \gamma ) u^2 v^2+\gamma ^2 v^4}}{\gamma  v^2}\right)\,.
\eeq

It is important that \eqref{eq:VDLE} is valid only for $u_*<u^{(T)}_B$, in which case the true butterfly speed (defined via $\lambda(u_B)=0$) is the stress tensor one $u_B=u^{(T)}_B$, and the velocity dependent exponent is maximal for $u>u_*$ \cite{Mezei:2019dfv}. On the other hand, when $u_*>u^{(T)}_B$, one only has the first branch of \eqref{eq:VDLE} and the exponent is nowhere maximal. In this case, the true butterfly speed is determined by solving $\lambda_{\rm saddle}(u_B)=0$. Note that due to concavity of $\lambda_{\rm saddle}(u)$ we have $u_B\leq u^{(T)}_B$ for all couplings. There is a critical line in coupling space marking the boundary between these two types of behaviors, given by equating the two velocities in  \eqref{eq:stresstensorVB}. This is shown on Fig.~\ref{fig:vdle}.

It is perhaps educational to give explicit formulas in the strong coupling limit \cite{Mezei:2019dfv}:
\es{strongCoup}{
\lam(u)&= \begin{cases}
1-{u_*\ov 2 u^{(T)}_B}\le[1+\le(u\ov u_*\ri)^2\ri] & \text{when } u<u_*\,, \\
1-\frac{u}{u^{(T)}_B} & \text{when } u>u_*\,,
\end{cases}\\
u_*&=\sqrt{\ga\, \de v}\,, \quad u^{(T)}_B=\sqrt{\ga\ov 6\, \de v}\,.
}
We see that $u_*$ goes to zero, while $u^{(T)}_B$ diverges. If take $u$ comparable to $u^{(T)}_B$, we see that the theory is maximally chaotic, while if we take very small $u$'s, we can detect the deviation from maximal chaos (that itself goes to zero as $\de v\to 0$).

We close this section by noting that \eqref{eq:OTOCpint} only determines the $t$ and $x$ dependence of the OTOC, where $t=i(\tau_1+\tau_2-\tau_3-\tau_4)/2$ is the separation between the ``center of mass" times, while $x$ is the spatial distance between the two operator pairs (see \eqref{eq:4ptfunc}). However, the OTOC also depends on the relative times $\xi=\frac{iv}{2}(\tau_1-\tau_2+\pi)$, $\xi'=\frac{iv}{2}(\tau_3-\tau_4+\pi)$. This dependence is encoded in the wave function factors in the retarded kernel eigenvalue equation \eqref{eq:retardedkernel} \cite{Maldacena:2016hyu,Kitaev:2017awl}. The full position dependence of the growing piece of the OTOC is therefore
\beq
\int dp \ \frac{1}{[\cosh\xi \cosh \xi' ]^{\kappa(p)/v}}\frac{1}{\cos \le(\frac{\pi \kappa(p)}{2}\ri)}e^{\kappa(p)t+i p x}\,.
\eeq
In the chaos limit, the relative positions $\xi$, $\xi'$ remain bounded, so this extra dependence on them does not affect the steepest descent contour and the discussion in this section.

\section{Four point function}
\label{sec:4point}

In this section we invert the kinetic kernel of \eqref{eq:Liouvilleaction}. Instead of starting from its spectral representation, we solve Green's equation with the appropriate delta sources on the right hand side.
We switch to Matsubara representation using $\delta(Y-Y')={1\ov 2\pi}\sum_n e^{i n (Y-Y')}$, and expand
\beq
\mathcal{G}_p(\theta,Y,\theta',Y')=\sum_n{1\ov  2\pi} e^{in(Y-Y')} \mathcal{G}_{p,n}(\theta,\theta')\,,
\eeq 
so that Green's equation for the kerenel \eqref{eq:kernel} reads as
\beq
\label{eq:greenseq}
{N\ov q^2}\left[{v^2\ov 4}\left( \pa_{\theta}^2- {h(h-1) \ov \sin^2({\theta})}\right)+{n^2 \ov 4 }\right] \mathcal{G}_{p,n}(\theta,\theta')={2v}\big[ \delta(\theta-\theta')+(-1)^n\delta(\theta+\theta'-\pi)\big]\,.
\eeq
This equation is supplemented with the boundary condition $\mathcal{G}_{p,n}(\theta_v ,\theta')=\mathcal{G}_{p,n}(\pi-\theta_v ,\theta')=0 $, where $\theta_v\equiv {(1-v)\pi\ov 2}$.

\begin{figure}[!h]
\begin{center}
\includegraphics[width=0.5\textwidth]{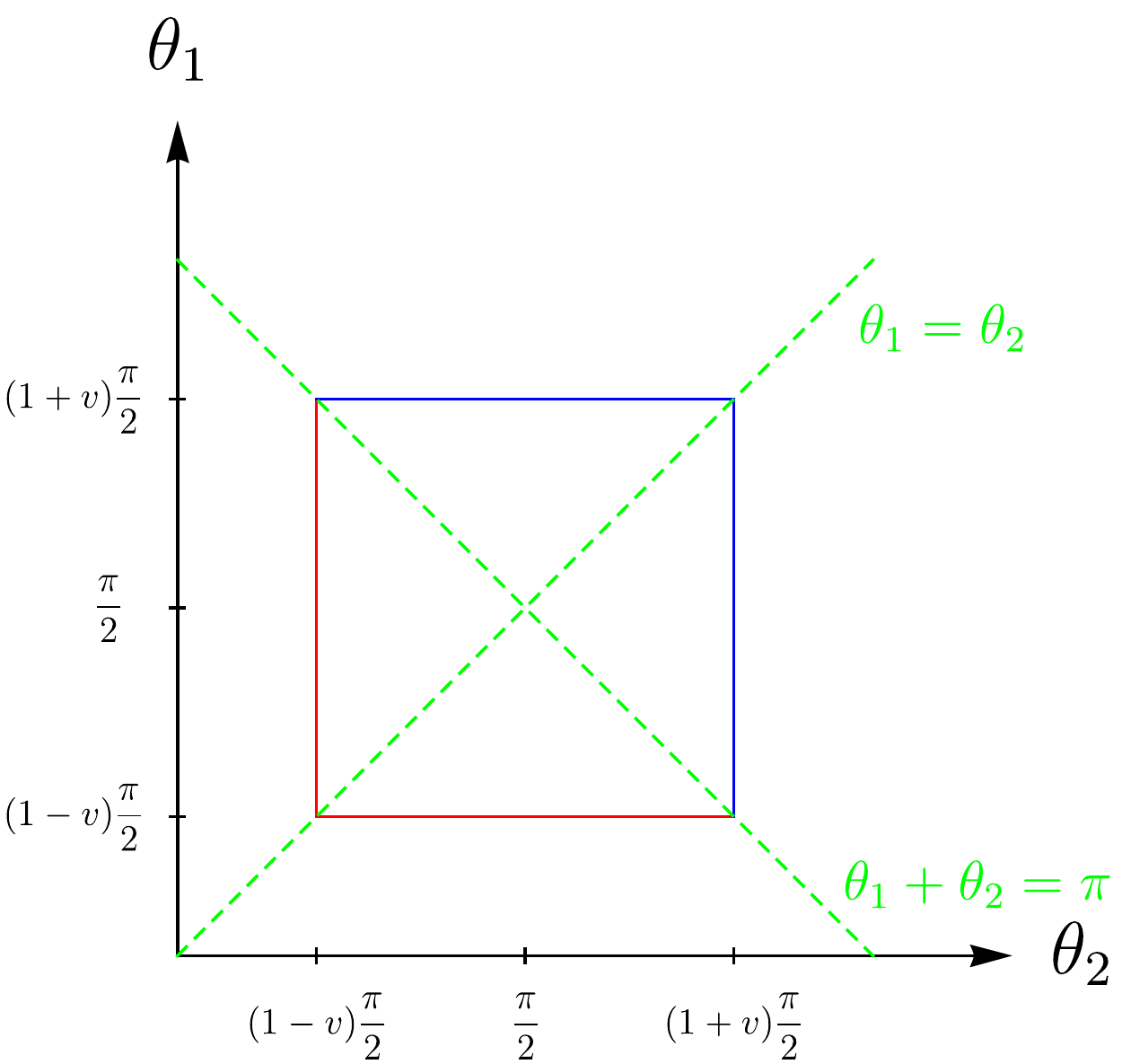}
\caption{We piece together the solution to \eqref{eq:greenseq} using the homogeneous solution $\Psi_n^-$ for regions with red boundaries and $\Psi_n^+$ for regions with blue boundaries. Along green dashed lines, there are jumps in the first derivatives giving rise to the delta functions in the RHS of \eqref{eq:greenseq}.}
\label{fig:regions}
\end{center}
\end{figure}

We can find $\mathcal{G}_n$ using the homogeneous solutions $\Psi_n^{\pm}$ that satisfy $\Psi_n^-(\theta_v,v)=\Psi_n^+(\pi-\theta_v,v)=0$. We explain how to do this on Fig.~\ref{fig:regions}. The resulting formula is
\beq
\label{eq:greensol1}
&{N\ov q^2}\mathcal{G}_{p,n}(\theta_1,\theta_2)={8\ov v \mathcal W}\big[ \Psi_n^-(\theta_1,v)\Psi_n^+(\theta_2,v) \Theta(-\theta_1+\theta_2) + \Psi_n^+(\theta_1,v)\Psi_n^-(\theta_2,v) \Theta(\theta_1-\theta_2)\big]
\\&+(-1)^n{8\ov v \mathcal W}\big[\Psi_n^-(\theta_1,v)\Psi_n^+(\pi- \theta_2,v) \Theta(\pi-\theta_1-\theta_2) + \Psi_n^+(\theta_1,v)\Psi_n^-(\pi-\theta_2,v) \Theta(-\pi+\theta_1+\theta_2)\big]\,.
\eeq
Here the Wronskian is given by $\mathcal{W}=\Psi_n^- \partial_\theta\Psi_n^+-\Psi_n^+ \partial_\theta \Psi_n^-$ and is independent of $\theta$ and is inserted to correctly normalize the delta functions on the RHS of \eqref{eq:greenseq}. We can easily construct homogeneous solutions with the required properties using the zero modes \eqref{eq:eigenfunctions}:
\beq
\label{eq:greensol2}
\Psi_n^-(\theta,v)&= \psi_n^e(\theta)\psi_n^o(\theta_v)-\psi_n^o(\theta)\psi_n^e(\theta_v)\,,
\\ \Psi_n^+(\theta,v)&=\psi_n^e(\theta)\psi_n^o(\theta_v)+\psi_n^o(\theta)\psi_n^e(\theta_v)\,.
\eeq
Note that $\Psi_n^-(\theta,v)=\Psi_n^+(\pi-\theta,v)$ due to $\psi_n^{e/o}(\theta)=\pm \psi_n^{e/o}(\pi-\theta)$. Since the Wronskian is independent of $\theta$, we may evaluate it at $\theta=\theta_v$ that results in 
\beq
\mathcal{W}&=-\Psi_n^+(\theta_v,v){\Psi_n^-}'(\theta_v,v)\\
&=-2 \psi_n^e(\theta_v) \psi_n^o(\theta_v) [{\psi_n^e}'(\theta_v)\psi_n^o(\theta_v)-{\psi_n^o}'(\theta_v)\psi_m^e(\theta_v)]\,,
\eeq
where prime denotes $\theta$ derivative. We recognize in the brackets the Wronskian between $\psi_n^e$ and $\psi_n^o$ which must be independent of $\theta_v$ since they solve the same second order homogeneous ODE without first order terms. We may therefore replace $\theta_v \rightarrow \pi/2$ in the square brackets, which (using \eqref{eq:eigenfunctions}) then evaluates to one. In summary, we  simply have
\beq
\label{eq:greensol3}
\mathcal{W}=-2 \psi_n^e(\theta_v) \psi_n^o(\theta_v) \,.
\eeq

It would be interesting to obtain a position space expression by summing \eqref{eq:greensol1}. We did not manage to do this in general, however, note that for $p=0$ ($h=2$) the result must agree with the four point function in the case of the SYK dot \cite{Streicher:2019wek,Choi:2019bmd}. We confirm that this is the case in Appendix~\ref{app:sykdot}.

\section{Energy density two point function}
\label{sec:2point}

\subsection{Prescription}

Here we would like to compute the Matsubara amplitudes of the connected energy density two point function (recall that we take the lattice spacing to be $1$ and set $\beta=2\pi$)
\beq \label{eq:gee}
G_{\varepsilon \varepsilon}(\tau,x)=\langle \mathcal T\varepsilon_{x+y}(\tau) \varepsilon_y(0)\rangle_{\rm conn} ={1\ov 2\pi}\sum_{n\in\mathbb Z} e^{i n \tau}\int_{-\pi}^{\pi} {dp\ov2\pi}~ e^{i p x} G^M_{\varepsilon \varepsilon}(n,p)\,,
\eeq
where we define the energy density as a local term in the Hamiltonian \eqref{eq:Hamiltonian}
\beq \label{eq:eden}
H&=\sum_{x=1}^M \varepsilon_x(0)\,,\\
\varepsilon_x(0)&=i^{q/2} \Big\lbrace \sum_{i_1<...<i_q}J_{i_1...i_q,x}\chi_{i_1,x}...\chi_{i_q,x}\\ &+ \frac{1}{2}\sum_{\substack{i_1<...<i_{q/2} \\ j_1<...<j_{q/2}}}\Big[J'_{i_1...i_{q/2}j_1...j_{q/2},x}\chi_{i_1,x}...\chi_{i_{q/2},x}\chi_{j_1,x+1}...\chi_{j_{q/2},x+1} \\
&+ J'_{i_1...i_{q/2}j_1...j_{q/2},x-1}\chi_{i_1,x-1}...\chi_{i_{q/2},x-1}\chi_{j_1,x}...\chi_{j_{q/2},x} \Big] \Big\rbrace\,.
\eeq 
The idea is to extract this correlator from an OPE limit of the time ordered four point function (TOC) . This method was used in \cite{Gu:2016oyy} in the conformal limit, with a prescription inspired by the factorization of the $p=0$ contribution to the TOC. We cannot rely on this here, away from the conformal limit. Instead we can find the right prescription using the observation of \cite{Sarosi:2017ykf}, i.e. that the Clifford anticommutation relations imply
\beq
\label{eq:cliffordidentity}
\sum_i \chi_{i,x}(0) [ \chi_{i,x}(0), H ] = q\, \varepsilon_x (0)\,.
\eeq
Note that the energy density is not a uniquely defined operator in a lattice model; our definition in \eqref{eq:eden} was chosen such that the above identity holds.

Next, we  represent the four point function \eqref{eq:4ptfunc} as a trace
\beq
\mathcal{F}_{xy}(\tau_1,\tau_2;\tau_3,\tau_4) \propto \sum_{i,j}\frac{1}{Z}\text{Tr}\big[e^{-\beta H} \chi_{i,x}(\tau_1)\chi_{i,x}(\tau_2)\chi_{j,y}(\tau_3)\chi_{j,y}(\tau_4)\big]- \text{disconnected}\,,
\eeq
and plug in the  formula  $\chi_{i,x}(t)=e^{i H t}\chi_{i,x}(0)e^{-iHt}$ for Heisenberg operators. We can derive the following identity:  
 \beq
\lim_{\substack{\tau_1 \rightarrow \tau_2 \\ \tau_3 \rightarrow \tau_4}}\partial_{\tau_1}\partial_{\tau_3}\frac{1}{N}\mathcal{F}_{xy}(\tau_1,\tau_2;\tau_3,\tau_4) = \frac{q^2}{N^2}G_{\varepsilon \varepsilon}(\tau_2-\tau_4,x-y)+\text{contact term}\,,
\eeq
where we exploit that time derivatives are implemented by commutators with $H$, which in the OPE limit $\tau_1 \rightarrow \tau_2$, $\tau_3 \rightarrow \tau_4$ can be combined to
 become two copies of the (Heisenberg evolved) energy density according to \eqref{eq:cliffordidentity}.  A contact term appears due to the non-triviality of taking the coincident limit $\tau_2-\tau_4\rightarrow 0$ first.
 
  We can write this formula in Fourier space in terms of \eqref{eq:greensol1} as
\beq
\label{eq:4pt to ee}
\lim_{\substack{\theta_1 \rightarrow \theta_v \\ \theta_2 \rightarrow \theta_v} } {N^2 v^2\ov 16q^4}\partial_{\theta_1}\partial_{\theta_2}\mathcal{G}_{p,n}(\theta_1,\theta_2) \sim G^{M}_{\varepsilon \varepsilon}(n,p)+\text{contact term} \,.
\eeq
 In Fourier space, the possible contact terms are polynomials in $n$ and $\cos( p)$. We will determine the right contact term in Sec.~\ref{sec:contact} after obtaining the analytically continued expression of the regular part of energy two point function.

\subsection{Calculation}

We proceed to evaluate \eqref{eq:4pt to ee}. Using \eqref{eq:greensol1},~\eqref{eq:greensol2},~\eqref{eq:greensol3} we find that up to a possible contact term
\beq
 G^{M}_{\varepsilon \varepsilon}(n,p)= {Nv\ov 4q^2 \psi_n^e(\theta_v)\psi^o_n(\theta_v)} \partial_\theta \Psi_n^-(\theta_v,v)\left(  \partial_\theta \Psi_n^-(\pi-\theta_v,v) -(-1)^n \partial_\theta \Psi_n^-(\theta_v,v)\right).
\eeq
Note that
\beq
\partial_\theta \Psi_n^-(\theta_v,v)&=\partial_\theta\psi_n^e(\theta_v)\psi_n^o(\theta_v)-\partial_\theta \psi_n^o(\theta_v)\psi_n^e(\theta_v)
\\&=\mathcal W(\psi_n^o(\theta_v),\psi_n^e(\theta_v))\\
&=1\,,
\eeq
where  we again used that the Wronskian of the two solutions is independent of $\theta$ and evaluated it at $\theta_v \rightarrow \pi/2$.
The other combination can be evaluated as
\beq
  \partial_\theta \Psi_n^-(\pi-\theta_v,v) -(-1)^n \partial_\theta \Psi_n^-(\theta_v,v)&=-(1+(-1)^n)\partial_\theta \psi_n^e(\theta_v) \psi_n^o(\theta_v)-(1-(-1)^n)\partial_\theta \psi_n^o(\theta_v) \psi_n^e(\theta_v)
  \\&=\begin{cases}-2\partial_\theta \psi_n^e(\theta_v) \psi_n^o(\theta_v) & n\in 2\mathbb Z\,,
  \\ -2\partial_\theta \psi_n^o(\theta_v) \psi_n^e(\theta_v)  & n\in 2\mathbb Z+1\,.  \end{cases}
\eeq
Therefore we find the Matsubara Green's function for the energy density
\beq
\label{eq:MatsubaraTT}
 {q^2 \ov N} G^{M}_{\varepsilon \varepsilon}(n,p)=\begin{dcases}-{v\partial_\theta \psi_n^e(\theta_v) \ov 2 \psi_n^e(\theta_v) }+\text{contact term}& n\in 2\mathbb Z\,,
  \\ -{v\partial_\theta \psi_n^o(\theta_v) \ov 2 \psi_n^o(\theta_v) }+\text{contact term}& n\in 2\mathbb Z +1 \,.\end{dcases}
\eeq
Next, we proceed to obtain the analytic continuation in $n$ after which we determine the contact term.

\subsection{Analytic continuation}

In order to obtain retarded correlators we need to determine the analytic continuation of \eqref{eq:MatsubaraTT} from integer $n$ to the complex $n$ plane. This is unique due to Carlson's theorem once we eliminate the exponential growth in the $n\rightarrow \pm i\infty$ directions. We can do this by defining a master function that unifies $\psi_n^{e/o}$ of \eqref{eq:eigenfunctions} for even/odd $n$ and does not grow exponentially in the limit $n\rightarrow \pm i\infty$  at $\theta=\theta_v$. It turns out that the right master function is\footnote{One may arrive at this by taking the $v\rightarrow 1$ limit in which case $\psi^{e/o}$ in \eqref{eq:eigenfunctions} are simple functions of $\theta$, but involves certain gamma function prefactors. We obtain the answer for general $v$ by replacing $n\rightarrow n/v$ inside the resulting gamma functions. This leaves us with an undetermined phase: in the 
$\sin\left({\pi h \ov 2}+{\pi n \ov 2v} \right)$ term in $c_o$ in \eqref{eq:cs}, the phase shift ${\pi h \ov 2}$ is not fixed by this argument (but it must agree with the phase shift in the $\cos\left({\pi h \ov 2}+{\pi n \ov 2v} \right)$  term in $c_e$). We guessed the right value based on matching with (rather high order) perturbation theory in $\de v\equiv 1-v$, where the analytic continuation is straightforward.} 
\beq
\label{eq:master}
\psi_n(\theta)&=c_o \psi_n^o(\theta)+c_e \psi_n^e(\theta)\,,
\eeq
where
\beq
\label{eq:cs}
c_o={\Gamma\left(1-{h\ov 2}-{n\ov 2v}\right) \sin\left({\pi h \ov 2}+{\pi n \ov 2v} \right) \sin\left({n\pi\ov 2}\right)\ov  \Gamma\left({1\ov 2}-{h\ov 2}+{n\ov 2v}\right)}\,, \quad c_e= {\Gamma\left({1\ov 2}-{h\ov 2}-{n\ov 2v}\right)\cos \left({\pi h \ov 2}+{\pi n \ov 2v} \right) \cos\left({n\pi\ov 2}\right)\ov 2\Gamma\left(1-{h\ov 2}+{n\ov 2v}\right)}\,.
\eeq
For integer $n$, $\psi_n(\theta)\propto \psi_n^{e/o}(\theta)$ depending on the parity of $n$, up to a $\theta$ independent prefactor that cancels in the ratio \eqref{eq:MatsubaraTT}. We choose to introduce this prefactor so that $c_{e/o}$ can be entire functions in $n$; this will soon be useful.
With the use of this master function, we may write the retarded correlator as
\beq
\label{eq:retTT}
{q^2 \ov N}G_{\varepsilon \varepsilon }^R(\omega)&=-{\frac{v}{2}\frac{\partial_\theta \psi_n(\theta_v)}{\psi_n(\theta_v)}\vline}_{\, n\rightarrow -i\omega+\epsilon}+\text{contact term}\\
&=-{{v\ov 2}{c_o \pa_\theta \psi_n^o(\theta_v)+c_e \pa_\theta \psi_n^e(\theta_v) \ov c_o  \psi_n^o(\theta_v)+c_e  \psi_n^e(\theta_v)}\vline}_{\, n\rightarrow -i\omega+\epsilon}+\text{contact term}\,.
\eeq
One may confirm that this is analytic in $\omega$ in the upper half plane for real momentum ($1\leq h \leq 2$), as it should be.

\subsection{Contact term} \label{sec:contact}

Now we determine the contact term in the energy density two point function. One necessary condition on the contact term comes from the Ward identity corresponding to energy conservation which gives \cite{Policastro:2002tn}:
\beq \label{eq:contact}
~&\lim_{p\rightarrow 0} G_{\varepsilon\varepsilon}^R(\omega \neq 0,p)=0\,.
\eeq
This condition however is not sufficient to fix the contact term, since it leaves the freedom of adding a $p$ dependent contact term which vanishes at $p=0$. We can fix this freedom by matching to the expected short time behavior of the energy correlator. Since we are analyzing a quantum mechanical model of Majorana fermions, which in the UV become asymptotically free (at leading order in $N$), and the operator $\varepsilon_x(\tau)$ in \eqref{eq:eden} is a polynomial of fermions, but not their time derivatives, we expect that
\es{finiteness}{
G_{\varepsilon\varepsilon}^R(t=0^+,p)=\text{finite}\,.
}
This is only possible, if the correlator decays fast enough in frequency space. For convenience we take the UV limit in the Euclidean theory, which corresponds to taking $\omega \rightarrow i\infty$ with $0<v<1$ and the momentum (or equivalently $h$) fixed in \eqref{eq:retTT}. We get:
\beq \label{eq:asym}
{q^2 \ov N}G_{\varepsilon \varepsilon }^R(\omega)\vert_{\omega \rightarrow i\infty}=\left[{v\ov 4}\tan\le({\pi v\ov 2}\ri)h\left(h-1\right)+O\le(1\ov \omega\ri)\right]+\text{contact term}\,.
\eeq 
The Fourier integral producing \eqref{finiteness} will only converge, if the contact term cancels the constant in the above equation. This can indeed be done by a contact term, since $h(h-1)$ is a (first order) polynomial in $\cos (p)$, see \eqref{eq:defh}. Fourier transforming to real lattice space, this expression produces a combination of $\delta(\tau_1-\tau_2)\delta_{x,x'}$ and $\delta(\tau_1-\tau_2)\delta_{x,x'\pm 1}$, which are contact terms in both time and space. Getting rid of these terms, we get the complete expression of the retarded energy density two point function as announced in \eqref{2ptCorr}:
\beq \label{eq:2ptfinal}
G_{\varepsilon \varepsilon }^R(\omega)=-{Nv\ov2q^2}\le(\p_\theta \log  \psi_n(\theta_v) +\tan\le({\pi v\ov 2}\ri){h(h-1)\ov2}\ri)\Big\vert_{\, n\rightarrow -i\omega+\epsilon}\,.
\eeq

\section{Pole skipping}\label{sec:poleskip}

\subsection{Pole skipping points}

We are interested in points where zero and pole lines of \eqref{eq:2ptfinal} meet, which are called pole skipping points. The necessary condition of pole skipping is that the denominator and the numerator of the first term in the parenthesis of \eqref{eq:2ptfinal} goes to zero simultaneously
\beq
\begin{pmatrix} \psi_n^o(\theta_v) &\psi_n^e(\theta_v)\\ \pa_\theta\psi_n^o(\theta_v)&\pa_\theta \psi_n^e(\theta_v)\end{pmatrix} \bpm c_o \\c_e \epm =\bpm 0\\ 0\epm .
\eeq
Let us denote this equation as $A\,\vec c=\vec 0$. Since $\det A=\mathcal{W}(\psi_n^o,\psi_n^e)(\theta)=1$, pole skipping can only happen when $c_e=c_o=0$. Inspecting \eqref{eq:cs} we find that there are six classes of solutions
\beq
\label{eq:poleskippingpoints}
 \text{(i) }&\cos({n\pi \ov 2})=0 ~\& ~ 1/\Gamma\left({1\ov 2}-{h\ov 2}+{n\ov 2v}\right)=0  
 \\  &\quad\Longrightarrow
n=2\mathbb Z+1 ~\&~ h=1+{n\ov v} +2\mathbb N_0
 \\
 \text{(ii) }& \cos({n\pi \ov 2})=0 ~\& ~  \Gamma\left(1-{h\ov 2}-{n\ov 2v}\right)\sin\left({\pi h \ov 2}+{\pi n \ov 2v} \right)=0 
 \\&\quad \Longrightarrow 
n=2\mathbb Z+1 ~\&~ h=-{n\ov v} -2\mathbb N_0
 \\
\text{(iii) }&\sin({n\pi \ov 2})=0~ \& ~ 1/\Gamma\left(1-{h\ov 2}+{n\ov 2v}\right)=0
\\ &\quad\Longrightarrow  n=2\mathbb Z  ~\& ~ h={2+{n\ov v}}+2\mathbb N_0
\\
\text{(iv) }&  \sin({n\pi \ov 2})=0~ \& ~\Gamma\left({1\ov 2}-{h\ov 2}-{n\ov 2v}\right)\cos \left({\pi h \ov 2}+{\pi n \ov 2v} \right)=0
\\ &\quad\Longrightarrow  n=2\mathbb Z  ~\& ~ h={-1-{n\ov v}}-2\mathbb N_0
\\ \text{(v) }&  1/\Gamma\left({1\ov 2}-{h\ov 2}+{n\ov 2v}\right)=0  ~\&~ ~\Gamma\left({1\ov 2}-{h\ov 2}-{n\ov 2v}\right)\cos \left({\pi h \ov 2}+{\pi n \ov 2v} \right)=0
\\&\quad\Longrightarrow n=-v(a+b+1) ~\&~h=-a+b,\quad a,b\in \mathbb N_0
\\  \text{(vi) }&1/\Gamma\left(1-{h\ov 2}+{n\ov 2v}\right)=0 ~\&~\Gamma\left(1-{h\ov 2}-{n\ov 2v}\right)\sin\left({\pi h \ov 2}+{\pi n \ov 2v} \right)=0
\\&\quad\Longrightarrow n=-v(a+b+1) ~\&~h=-a+b+1,\quad a,b\in \mathbb N_0
\eeq
THus we see that pole skipping points are only possible for real-valued $h$ and so it is natural to divide into the following three cases $\{p\in i \mathbb R$, $p\in \mathbb R$, $p\in (2\mathbb Z+1)\pi+i\mathbb R\}$. We note that only the first case has pole skipping points on the upper half $\omega$ plane, while all three cases have additional pole skipping points on the lower half $\omega$ plane.

\medskip
\noindent
$\bullet$ {\it Purely imaginary momentum $p\in i \mathbb R$: $h\geq 2$}
\medskip

\begin{figure}[!h]
\begin{center}
\includegraphics[width=0.45\textwidth]{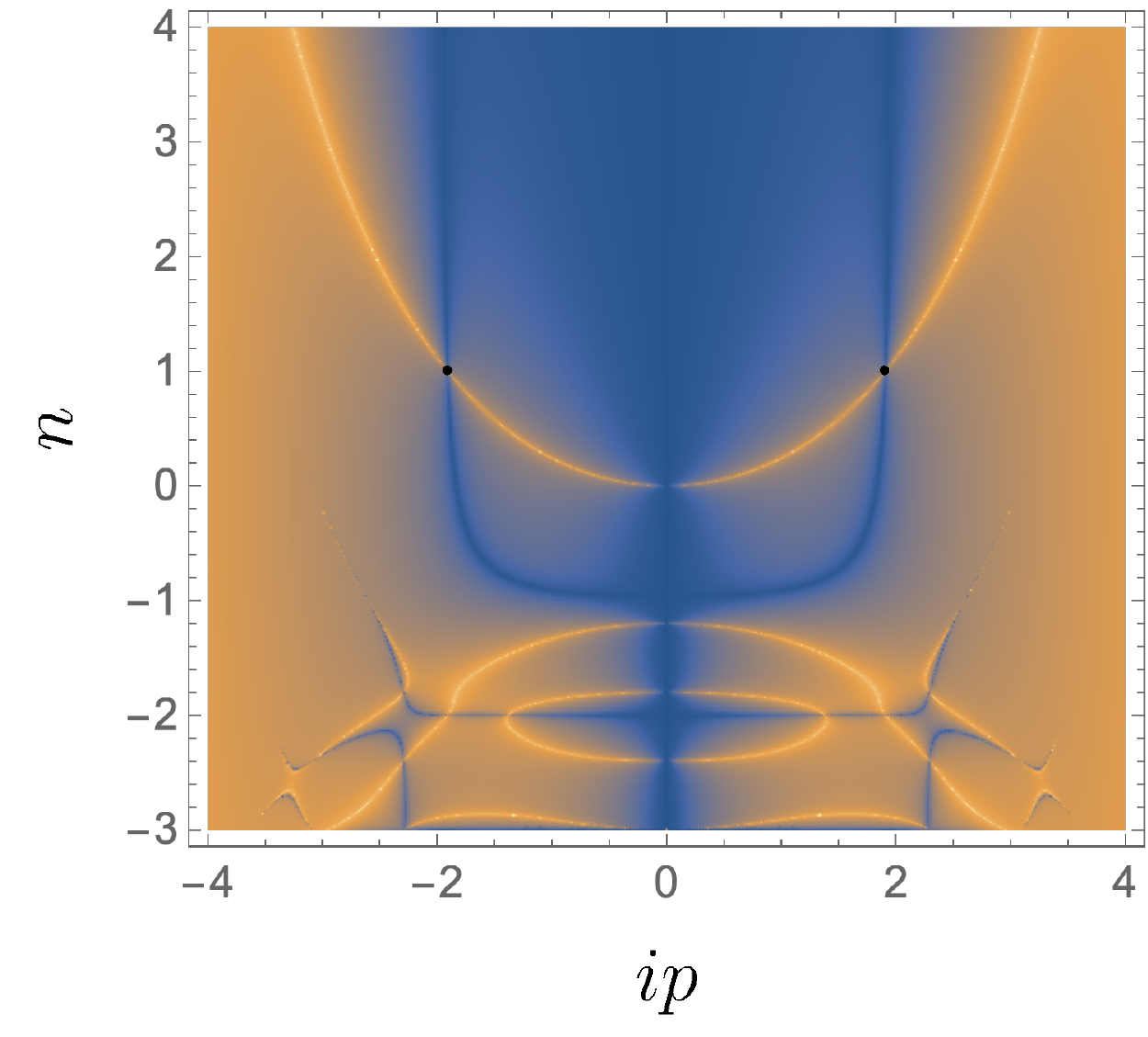}\hspace{0.5cm}
 \includegraphics[width=0.45\textwidth]{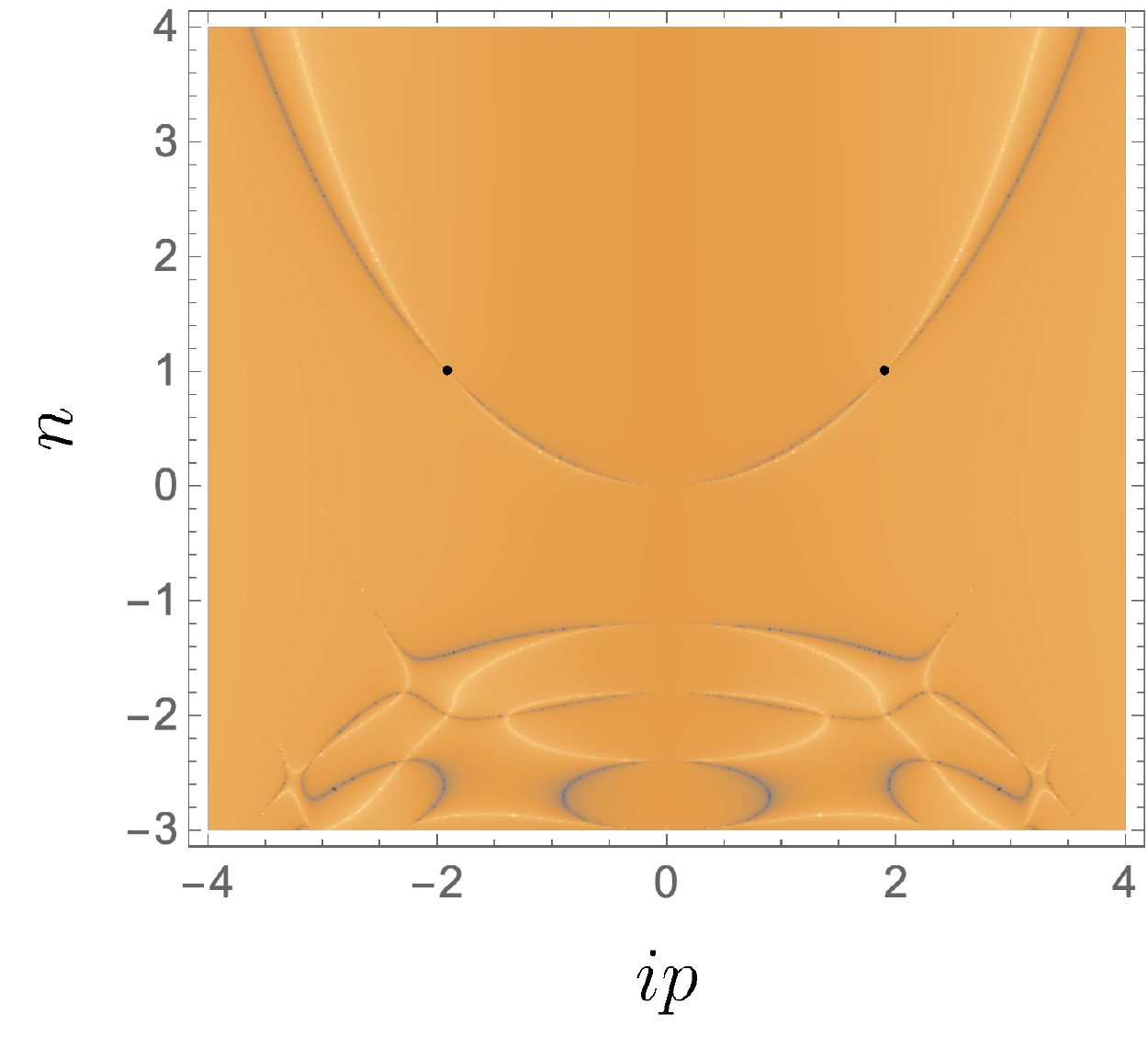} 
\caption{Density plot of the retarded energy-energy two point function for imaginary momentum and frequency for $v=0.6,\, \ga=1$. Hot lines (white) are pole lines, cold lines (blue) are zero lines. Black dots are the pole skipping points corresponding to chaos, whose location agrees with the proposal \eqref{PoleSkip2}. {\bf Left:} We plot the complete correlator.  {\bf Right:} We drop the momentum dependent contact term and plot only $\p_\theta \log  \psi_n(\theta_v)$, as we do in the rest of the density plots in this paper. The pole lines and the pole skipping points are unaffected by this, while the shape of the zero lines change. These shapes are therefore not physical. }
\label{fig:poleskip}
\end{center}
\end{figure}

Here pole skipping happens both on the upper and the lower half of the complex frequency plane. Let us focus on the pole skipping points on the upper half plane $n>0$ with $h\geq 2$. These points can only arise from (i), (iii) in \eqref{eq:poleskippingpoints} and can be further simplified to
\beq
\label{eq:poleskipupper}
\le\{(h,n)=\le({3+(-1)^n\ov 2} +{n\ov v}+2 k ,n\ri) ~\vert~  n,k\in \mathbb N\ri\}\,.
\eeq
The pole skipping point connected to the diffusion pole is the one at $n=1$ and $h=1+{1\ov v}$, we show this on Fig.~\ref{fig:poleskip}. 
Our modified pole skipping conjecture \eqref{PoleSkip2} is that pole skipping on this pole line happens at $n=1$ and $p=i/u^{(T)}_B$, which indeed translates to $h=1+1/v$ based on our discussion of the OTOC in Sec.~\ref{sec:Lyapunov}. Therefore, the conjecture \eqref{PoleSkip2} indeed holds exactly in the SYK chain at any coupling.

\begin{figure}[!h]
\begin{center}
\includegraphics[width=0.6\textwidth]{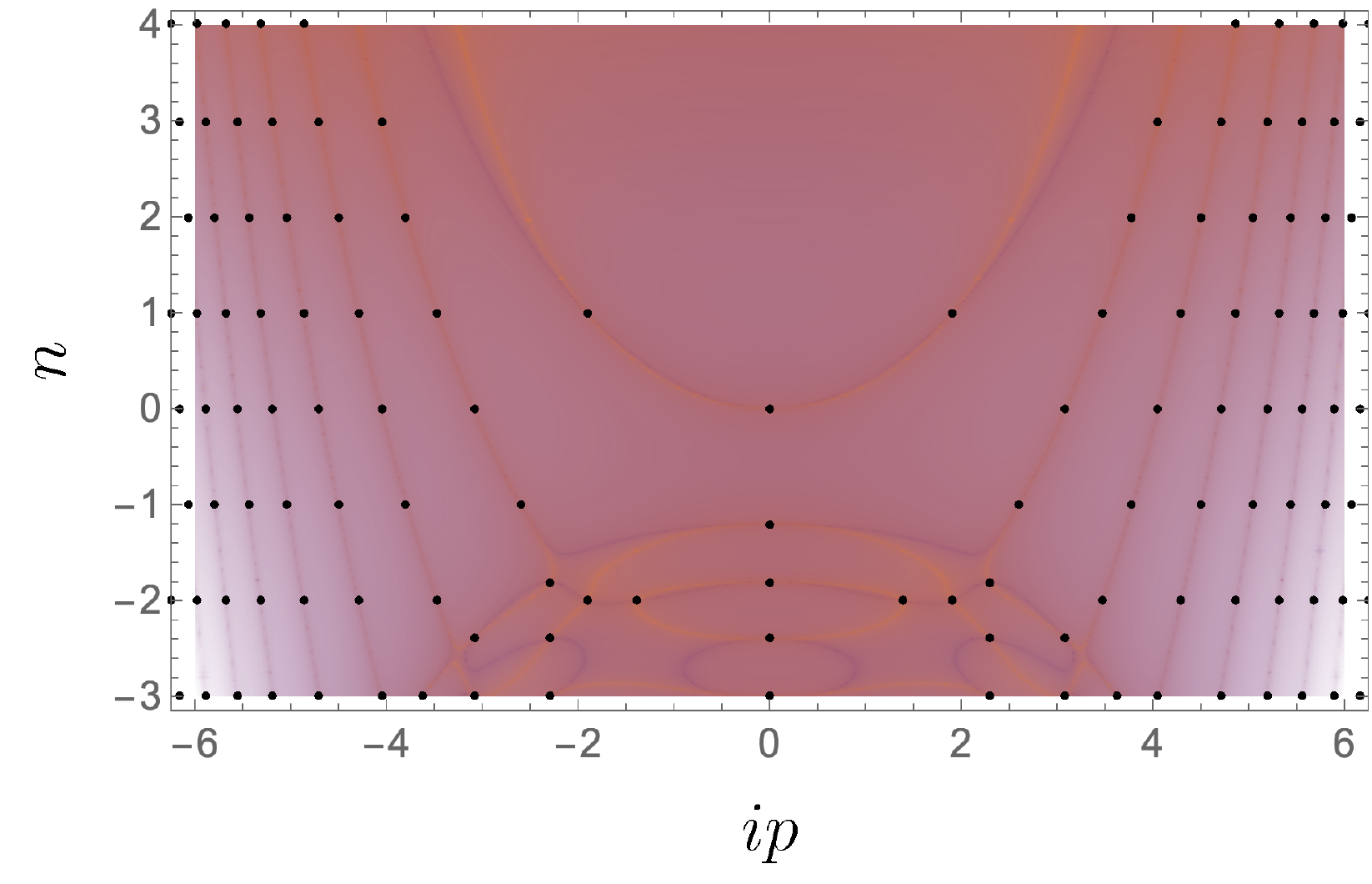}
\caption{ Density plot of the numerator and denominator of \eqref{eq:retTT} overlaid, with pole skipping points marked with black dot. There are pole/zero lines starting from negative $n$ that make it to positive $n$ that are not visible on the overall density plot of Fig.~\ref{fig:poleskip}. Each of the additional pole zero line pairs cross an odd number of times. The diffusion pair crosses once, the next one three times, the next one five times and so on. The pole skipping points that are not on the diffusion pole line do not contribute to chaos. 
}
\label{fig:imaginaryp}
\end{center}
\end{figure}

Furthermore, we note that there are additional pole skipping points on the lower half plane $n<0$ corresponding to (ii), (iv), (v), (vi) and the remaining solutions of (i), (iii) in \eqref{eq:poleskippingpoints} satisfying $h\geq 2$. The complete collection of pole skipping points for imaginary momentum can be observed in Fig.~\ref{fig:imaginaryp}, where we see that pole skipping points on the lower half plane exist for both integer and non-integer values of $n$, whereas on the upper half plane, all pole skipping points have integer $n$. 

\medskip
\noindent
$\bullet$ {\it Real momentum $p\in \mathbb R$: ${1+\sqrt{9-8\gamma}\ov 2}\leq h\leq 2$}
\medskip

\begin{figure}[!h]
\begin{center}
\includegraphics[width=0.45\textwidth]{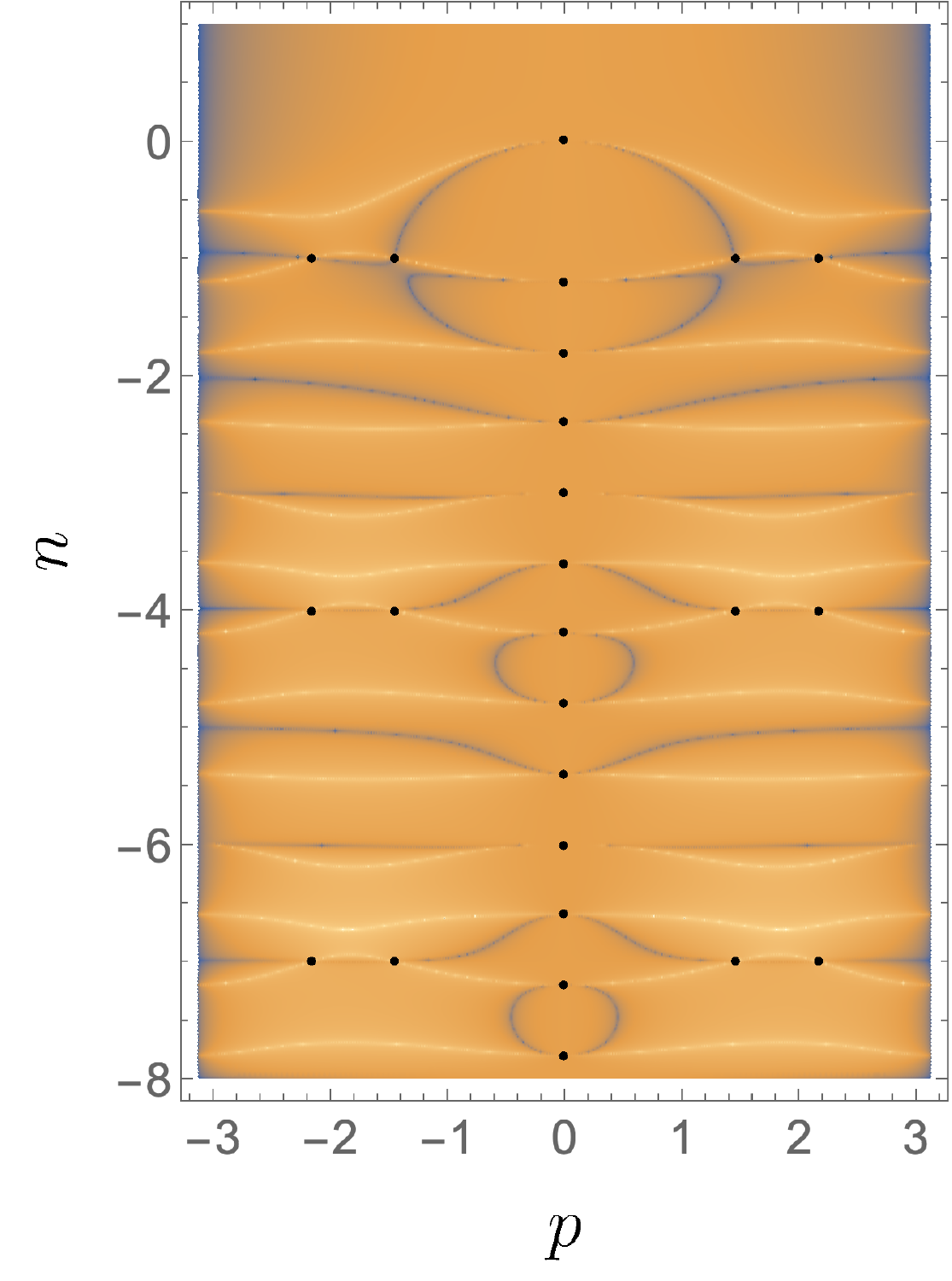} \hspace{0.5cm}
\includegraphics[width=0.45\textwidth]{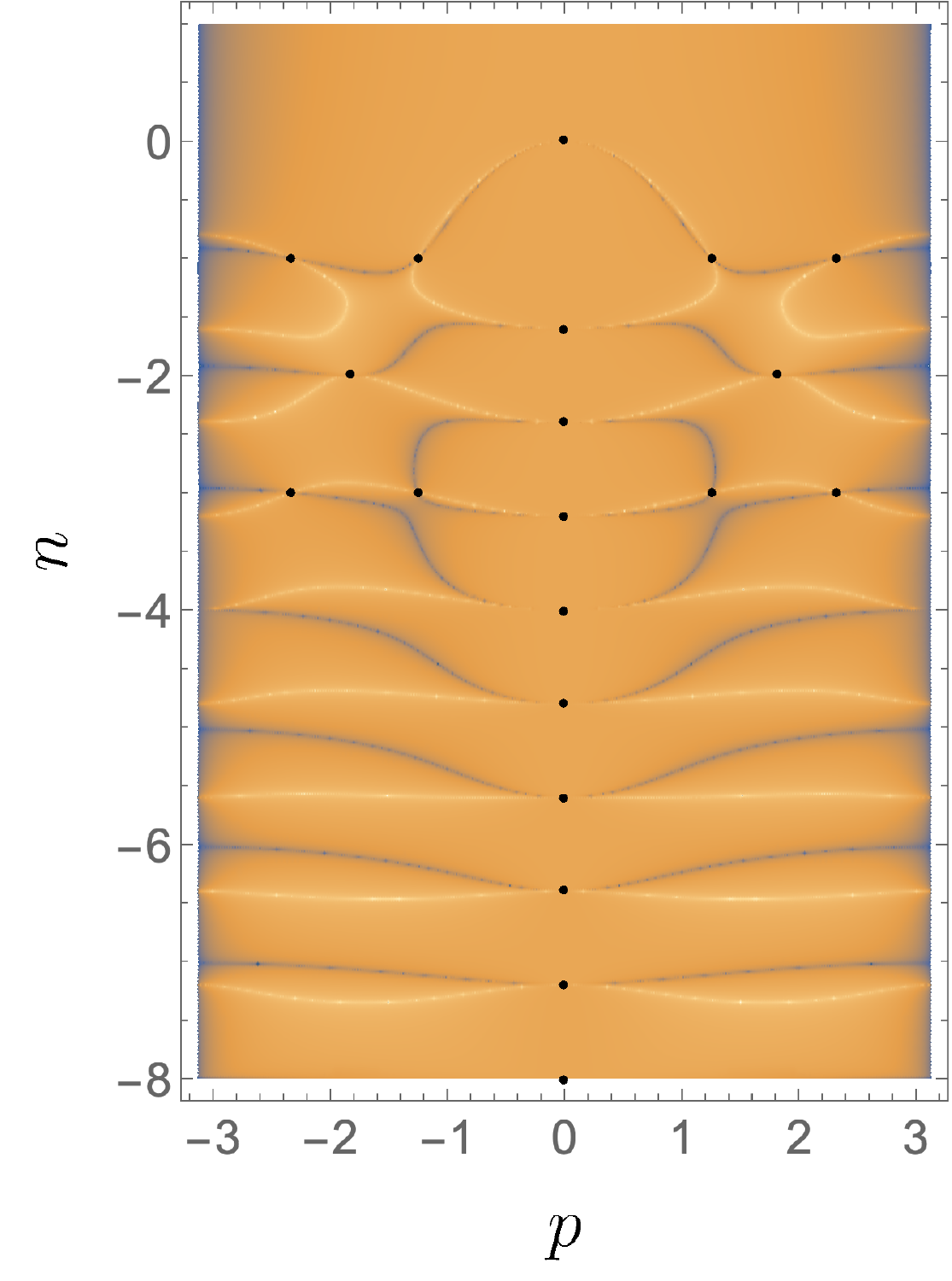}
\caption{Two point function for imaginary frequency and real momentum. The pole lines (hot) are confined to negative $n$ as they should be. There are various pole skipping points in this case too, marked by black dots. The left plot is for $v=0.6$, the right one is for $v=0.8$ with $\ga=1$.}
\label{fig:realp}
\end{center}
\end{figure}

For real $p$ the retarded Green's function can only have poles (and hence pole skipping points) in the lower half plane. In this case, the list of pole skipping points coming from (i), (ii), (iii), (iv) of \eqref{eq:poleskippingpoints} simplifies into the following expression which corresponds to integer $n$
\beq
\left\{ h=2+{n\ov v}-\lceil{n\ov v}\rceil , ~1-{n\ov v}+\lceil{n\ov v}\rceil \right\} \cap\left \{ n\in -\mathbb N_0 ~\vert ~n- \lceil{n\ov v}\rceil =0  ~\text{mod }2\right \}.
\eeq
Furthermore (v) and (vi) of \eqref{eq:poleskippingpoints} generate pole skipping points at $h=2$ (equivalently $p=0$) and generically non-integer $n$ given by
\beq\label{eq:nonint1}
\left\{ (h,n)=(2,-v(1+r))~ \vert~ r\in \mathbb N_+ \right\}\,.
\eeq
We plot the pole skipping point for the real momentum case on Fig.~\ref{fig:realp}.

\medskip
\noindent
$\bullet$ {\it$p\in (2\mathbb Z+1) \pi +i\mathbb R$: ${1\ov 2} \leq h\leq {1+\sqrt{9-8\gamma}\ov 2}$}\,.
\medskip

Finally, there is one more class of pole skipping points which happens on the line of complex momentum whose real part lies on the edge of the Brillouin zone $\text{Re}(p)=(2\mathbb Z+1)\pi$. Similarly to the real momentum case, pole skipping only happens on the lower half of the complex frequency plane and (i), (ii), (iii), (iv) of \eqref{eq:poleskippingpoints} correspond to the following unified expression
\beq
~&\left\{\left\{ h=1+{n\ov v}-\lceil{n\ov v}\rceil , ~-{n\ov v}+\lceil{n\ov v}\rceil ~\vert~{1\ov 2}\leq h\leq 1 \right\} \cap\left \{ n\in -\mathbb N_0 ~\vert ~n- \lceil{n\ov v}\rceil =1  ~\text{mod }2\right \}  \right\}\cup
\\ &\left\{\left\{ h=2+{n\ov v}-\lceil{n\ov v}\rceil , ~1-{n\ov v}+\lceil{n\ov v}\rceil ~\vert~1\leq h\leq {1+\sqrt{9-8\gamma}\ov 2}\right\} \cap\left \{ n\in -\mathbb N_0 ~\vert ~n- \lceil{n\ov v}\rceil =0  ~\text{mod }2\right \}\right\}\,.
\eeq
And finally, (v), (vi) of \eqref{eq:poleskippingpoints} generate pole skipping points at $h=1$ with non-integer $n$ 
\beq \label{eq:nonint2}
\left\{ (h,n)=(1,-vr)~ \vert~ r\in \mathbb N_+ \right\}\,.
\eeq

\subsection{Chaos and higher pole skipping on the upper half plane}

\begin{figure}[!h]
\begin{center}
\includegraphics[width=0.6\textwidth]{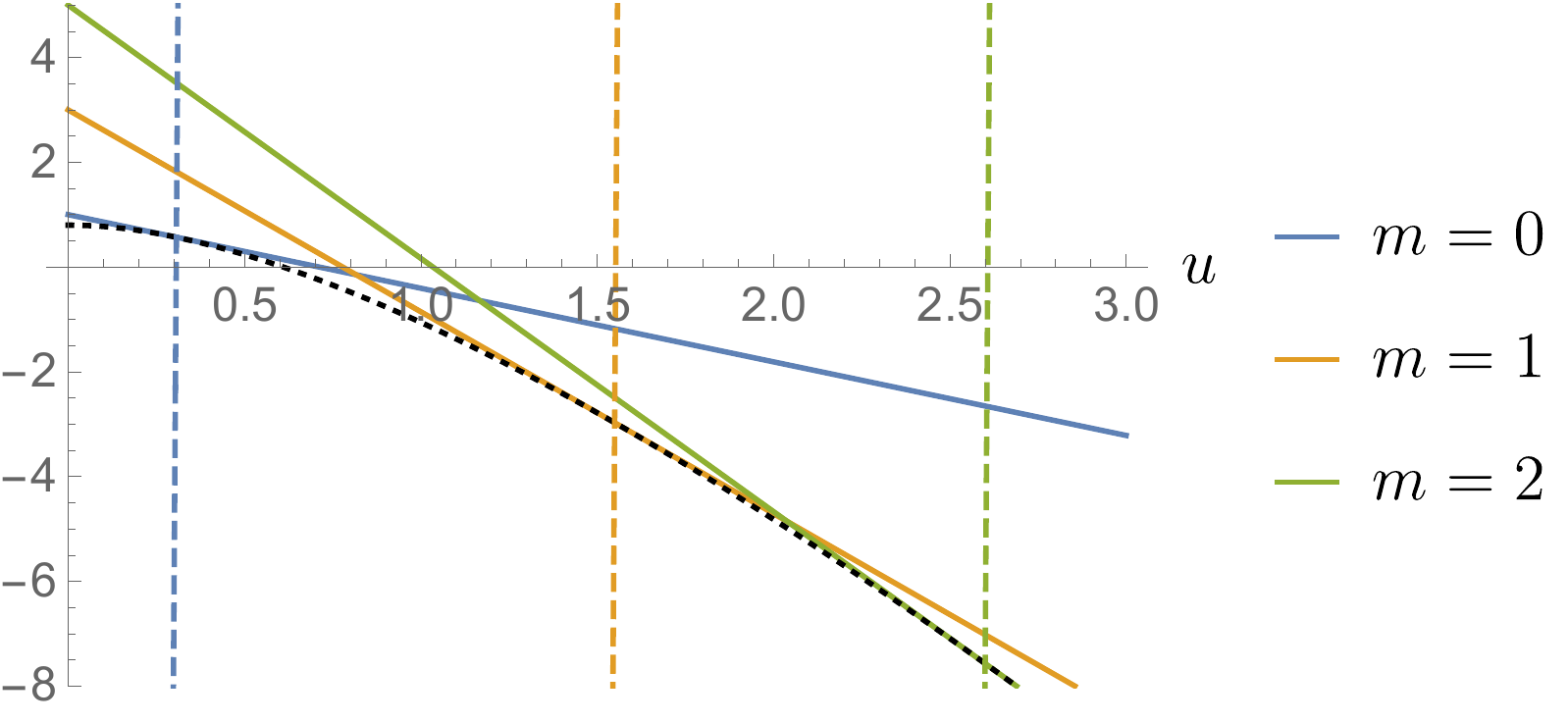}
\caption{The solid colored lines are contribution of the higher poles $h=1+\frac{2m+1}{v}$ to the VDLE, while the dashed gridlines are their respective critical velocities where they get activated. The only pole that gets activated where it gives a positive exponent, and hence can contribute to the growth of the OTOC is the one with $m=0$. (Even when they get 
activated, they do not dominate over the $m=0$ contribution.) The dotted black line is the saddle contribution to the VDLE that touches the pole contributions at critical velocities marked with dashed gridlines. The plot is for $v=0.8$, $\gamma=0.6$.  }
\label{fig:higherpoles}
\end{center}
\end{figure}

One may wonder how the sequence of pole skipping points \eqref{eq:poleskipupper} on the upper half plane is related to Lyapunov growth. Examining \eqref{eq:OTOCpint} the poles that can be picked up in the OTOC are at $(h-1)v=2m+1$, $m\in \mathbb{Z}$, or equivalently, $h=1+\frac{2m+1}{v}$. This is only a subset of the pole skipping points in \eqref{eq:poleskipupper}. We may ask if these poles ever dominate the OTOC. It turns out that their contribution is already negative at their respective critical velocities where they are activated (where the steepest descent contour in \eqref{eq:OTOCpint} crosses them), or in other words, their critical velocities are higher than $u_B^{(T)}$. Therefore they do not contribute to the growth of the OTOC, except for $m=0$. We show this in Fig.~\ref{fig:higherpoles}.

\subsection{Pole skipping on the lower half plane}

In the discussion above, we have found that the large $q$ SYK chain has pole skipping points on the lower half plane. These points can be divided into two classes: one class involves negative integer Matsubara frequencies, while the others are at non-integer values of $n$ frequencies. Note that both classes occur both at real and complex momenta $p$. While pole skipping points on the lower half plane cannot contribute to an exponential growth of the OTOC, we remark that the existence of pole skipping at non-integer multiples of the unit Matsubara frequency is interesting in its own right.

Pole skipping in the thermal energy two point function at negative integer multiples of the unit Matsubara frequency was discovered in \cite{Blake:2019otz} in the context of holography. The authors found from the near horizon expansion that at special values of $(\omega,p)$, the quasinormal modes of linearized Einstein gravity are not unique, implying that at these frequencies, the holographically dual retarded Green's function \cite{Son:2002sd} is indefinite, which is another indication of pole skipping. In the discussion of \cite{Blake:2019otz}, the universal structure of the black hole metric and the near horizon expansion always seems to yield such locations at negative Matsubara frequencies $\omega=-i{2\pi  n\ov \beta},~n\in \mathbb  N$ (this is also robust under higher derivative corrections \cite{Wu:2019esr}). 
It would be interesting to see whether non-integral valued pole skipping points like \eqref{eq:nonint1}, \eqref{eq:nonint2} van be understood in the context of holography by finding singular quasinormal modes at non-integral frequencies.

\section{Hydrodynamics and analytic structure of the  two point function} \label{sec:hydro}

The energy density retarded two point function determines the linear response behavior of the system. If we take $\abs{\om},\,\abs{p}\ll T$, we are in the hydrodynamic regime, and from the pole closest to the origin, we can determine the hydrodynamic transport coefficients. We find that for all values of the coupling the transport of energy is diffusive, and is controlled by a diffusion pole. 

We find that the two point function is meromorphic: it only has poles, but no branch cuts. We investigate the motion of  the hydrodynamic and non-hydrodynamic poles on the complex $\omega$ plane as we change $p$. We already analyzed this problem partially: it is the continuation of the diffusion pole line to $p\sim T$ that participates in pole skipping \eqref{PoleSkip2}. Here we ask, whether poles collide as we change $p$; the collision between the diffusion pole and a non-hydrodynamic pole delineates the applicability of the (all order) hydrodynamic expansion. These phenomena have been thoroughly analyzed in the planar four-dimensional ${\cal N}=4$ super Yang-Mills  (SYM) theory both at weak and at strong coupling \cite{Son:2002sd,Starinets:2002br,Kovtun:2005ev,Hartnoll:2005ju,Grozdanov:2016vgg,Grozdanov:2019kge,Grozdanov:2019uhi}. Our system allows us to perform the analysis at all values of the coupling, and we comment on similarities and differences between the SYK model and  SYM theory.

\subsection{Diffusion}\label{sec:Diff}

We can extract the diffusion constant by examining the $\omega=in \ll 1$, $p\ll 1 \rightarrow h\approx 2$ limit of \eqref{eq:retTT}. In this limit one has 
\beq
{q^2 \ov N}G_{\varepsilon\varepsilon}^R &\approx \frac{\sec ^2\left(\frac{\pi  v}{2}\right) \left(\sin (\pi  v) ((h-2) v-n)+\pi  (h-2) v^2\right)}{(2-h)[\pi   v^2 \tan \left(\frac{\pi  v}{2}\right)+2 v]+2 n}\,,
\eeq
which  by using the relation between $h$ and $p$ in \eqref{eq:defh} leads to the diffusion constant:
\beq
D=\frac{1}{12}\gamma v \left(\pi  v \tan \left(\frac{\pi  v}{2}\right)+2\right)\,.
\eeq
The strong coupling limit of this result is $D={\ga \ov 6\, \de v}$ with $\de v\equiv 1- v$, whose analog for $q=4$ was derived already in \cite{Gu:2016oyy}. (See also the discussion around \eqref{kap}.) Note that $D$ is an increasing function of the coupling $v$, unlike in the more familiar cases of field theories, where the diffusion constant diverges at weak coupling. 

Note that one has $D\leq (u_B^{(T)})^2$ (or $D\leq \frac{\beta}{2\pi}  (u_B^{(T)})^2$ when we reinstate the temperature) for all values of the couplings $v$ and $\gamma$, with saturation at strong coupling $v\rightarrow 1$. This is consistent with (in fact stronger than) the diffusivity bound of \cite{Hartman:2017hhp}.  One may also examine the bound in terms if the true butterfly speed when it is not given by $u_B^{(T)}$. In these cases, $u_B$ can be determined numerically by equating \eqref{eq:VDLEsaddle} to zero. It turns out that $D\leq u_B^2$ can be violated at weak coupling, but we get a correct bound by dividing with the Lyapunov exponent, that is, $D\leq u_B^2/v$ is always true, again consistently with \cite{Hartman:2017hhp}.
One may confirm this analytically in the weak coupling limit, where we can solve for the true butterfly speed
\beq
u_B=\frac{1}{2}\sqrt{\frac{\gamma}{2}}\,e v + O(v^2)\,, \quad \implies \quad \frac{u_B^2}{v}=\frac{1}{8}e \gamma v + O(v^2)\,,
\eeq
where $e$ is the natural number, while $D=\frac{1}{6}\gamma v + O(v^2)$ in this limit. 

Note that the pole line giving rise to the diffusion pole is the same as the pole line participating in pole skipping at \eqref{PoleSkip2}.

\subsection{Movement of poles}

Let us concentrate on the movement of poles for real $p$ first. To  explore the full range of $h\in [1,2]$ as we move around in the Brillouin zone $p\in[-\pi,\pi]$, we set $\ga=1$. All the plots in this section can be easily converted to any value of $\ga$ using the relation $p_\ga=\arccos\le[\cos(p_\text{here})-1+\ga\ov \ga\ri]$ that follows from \eqref{eq:retTT}. Relatedly, for $\ga<1$ we only cover part of the possible $h$ range.

\begin{figure}[!h]
\begin{center}
\includegraphics[width=0.3\textwidth]{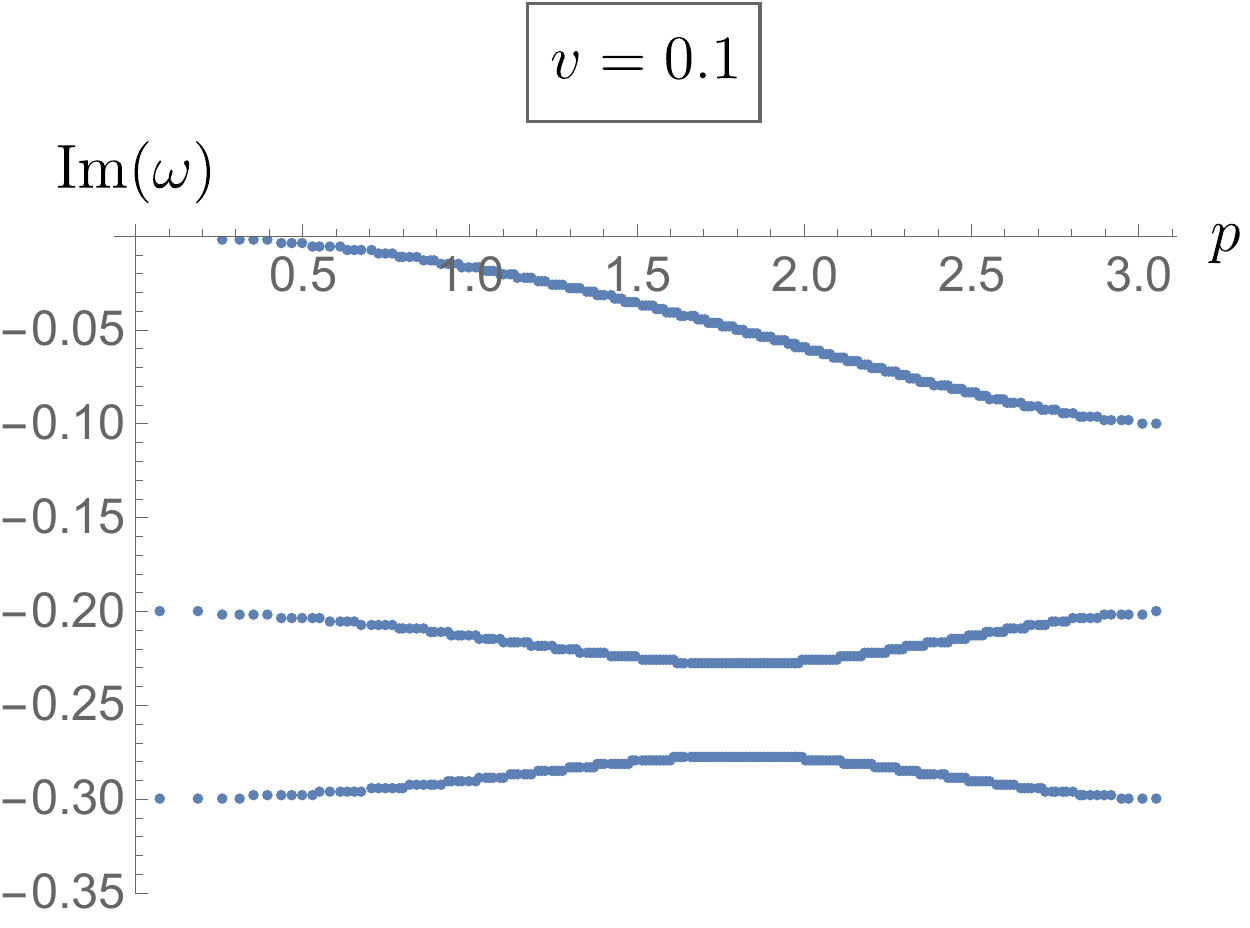}\hspace{0.5cm}
\includegraphics[width=0.3\textwidth]{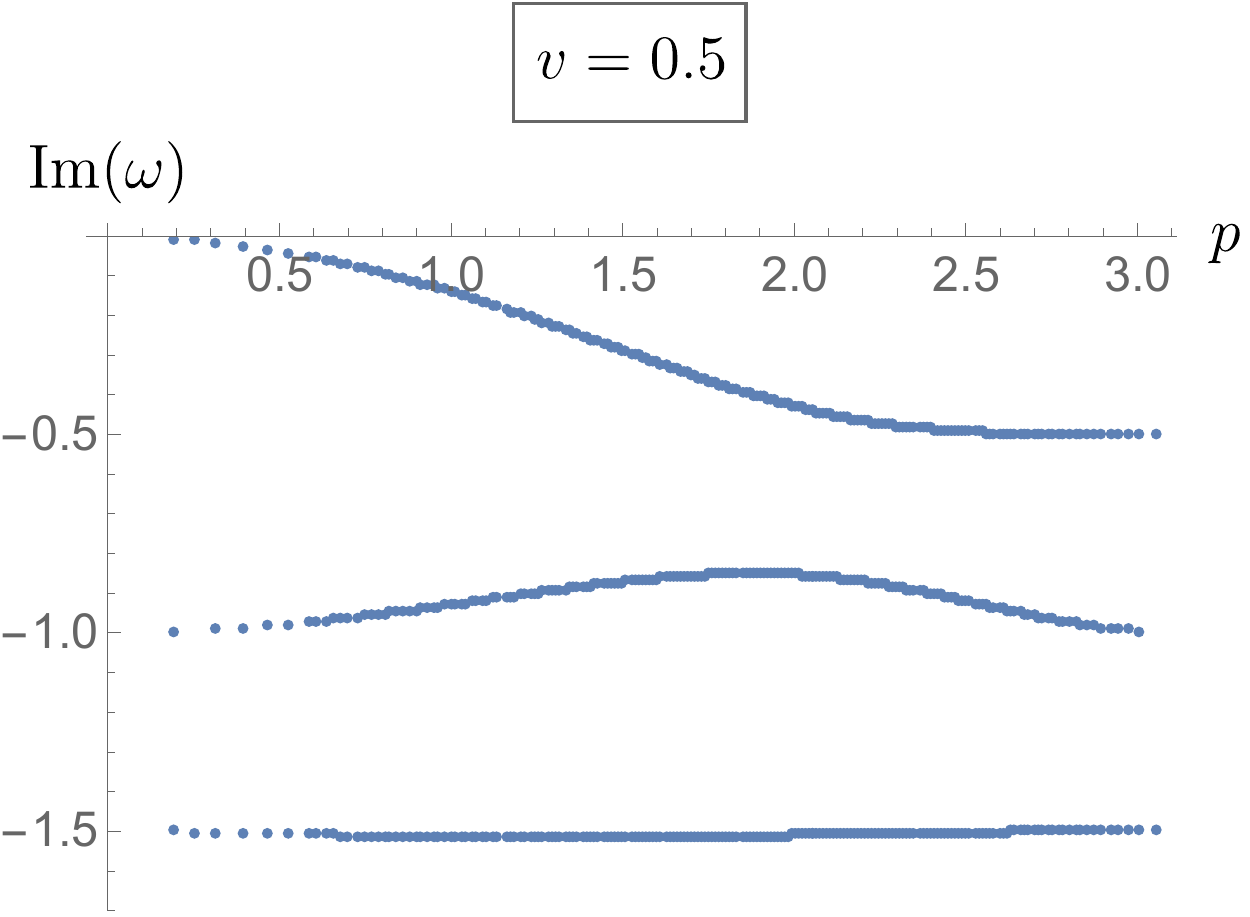}\hspace{0.5cm}
\includegraphics[width=0.3\textwidth]{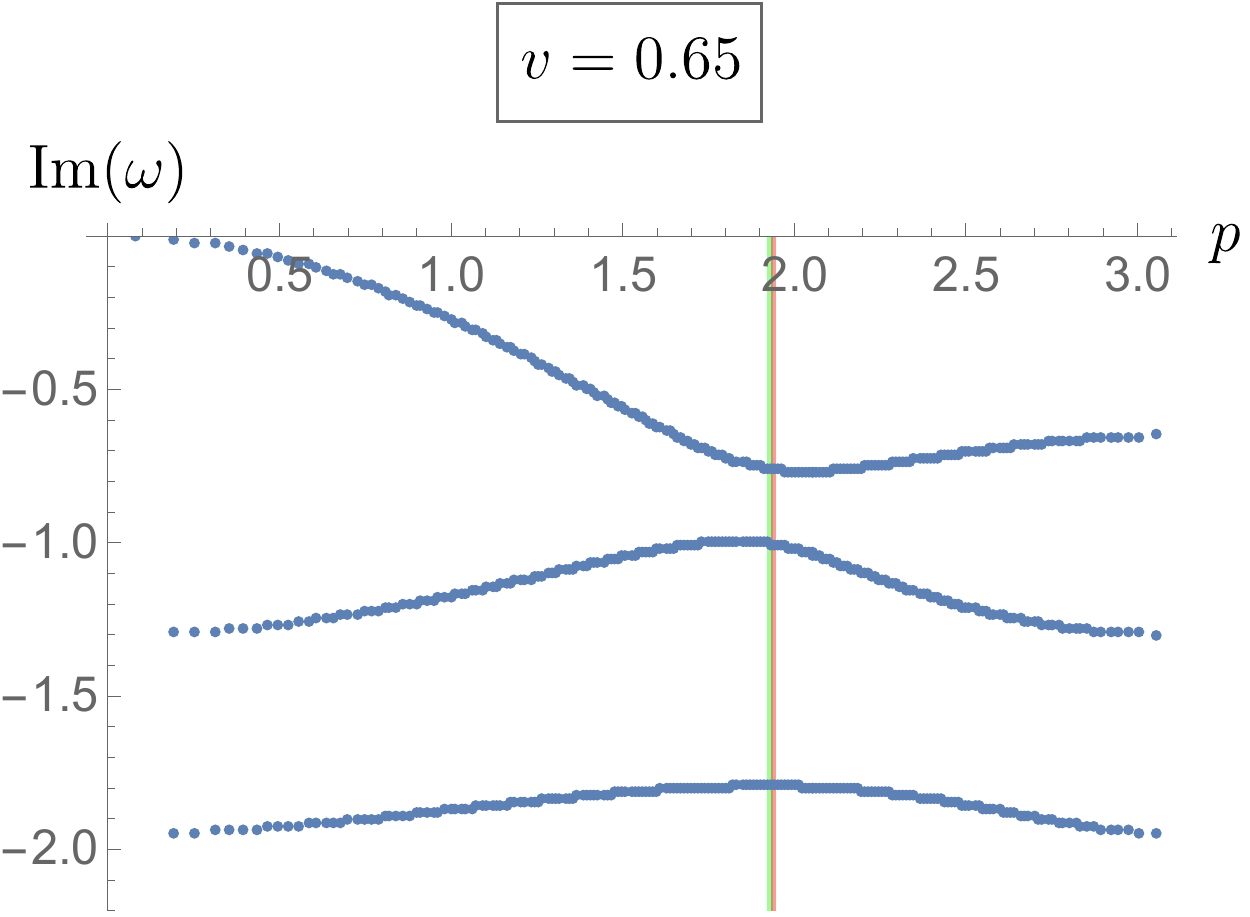}\vspace{0.5cm}
\includegraphics[width=0.3\textwidth]{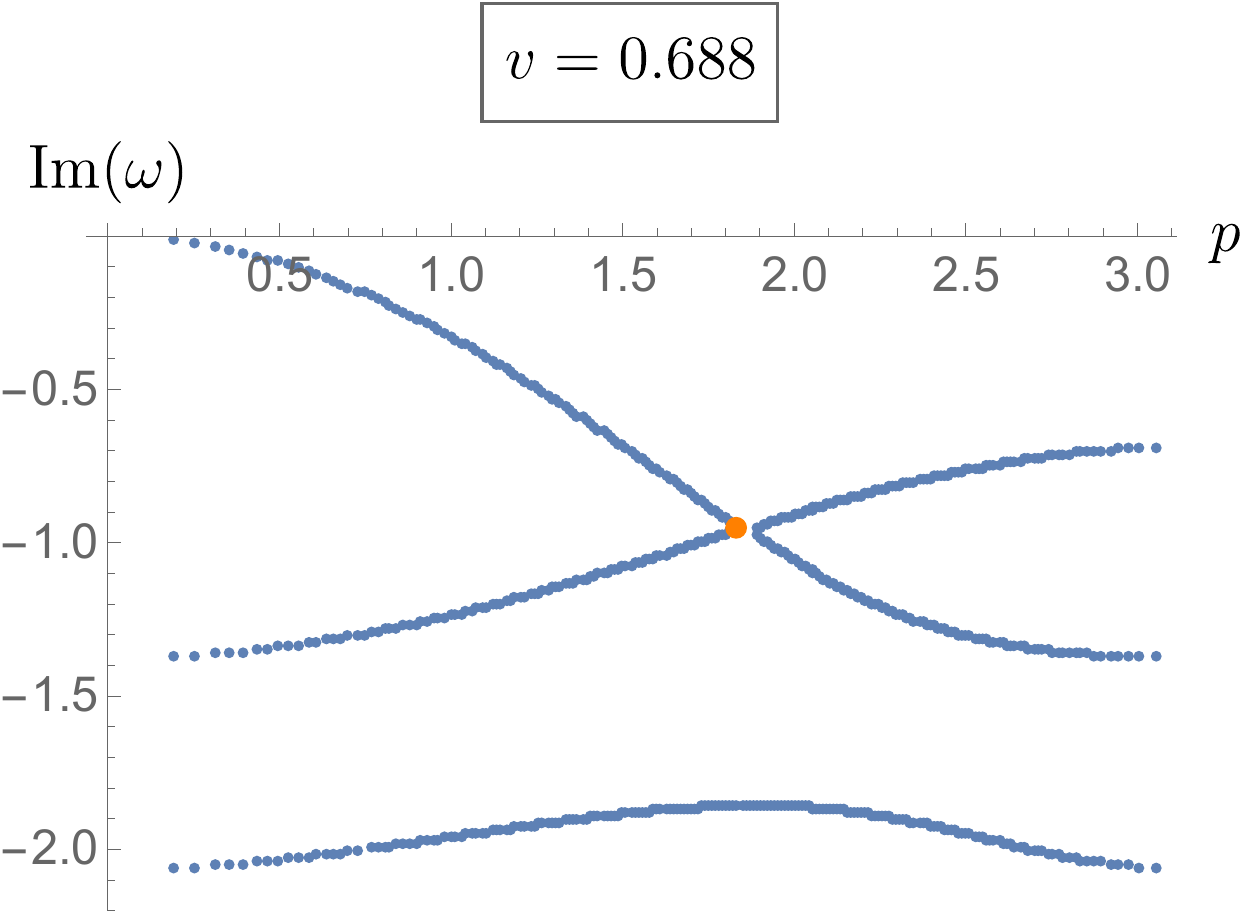}\hspace{0.5cm}
\includegraphics[width=0.3\textwidth]{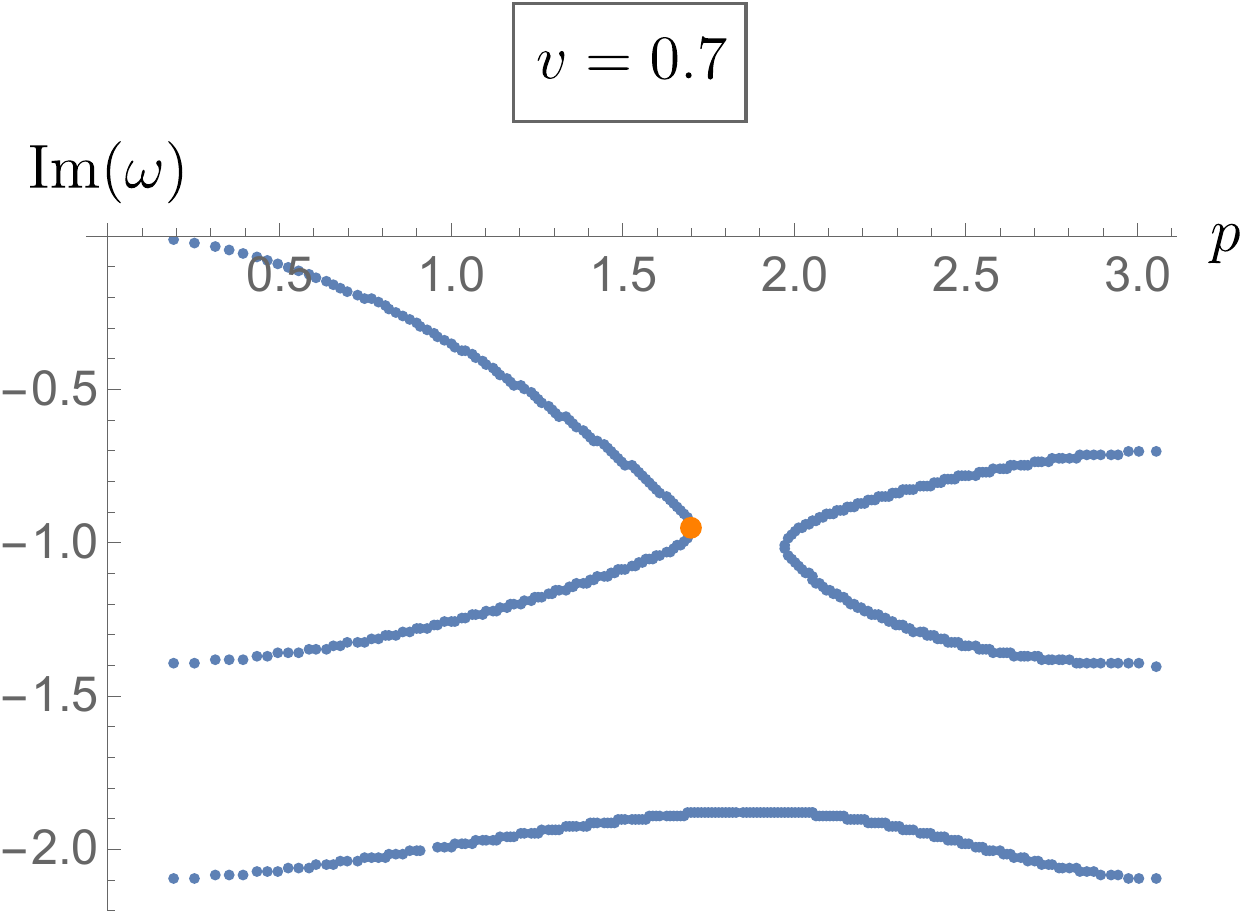}\hspace{0.5cm}
\includegraphics[width=0.3\textwidth]{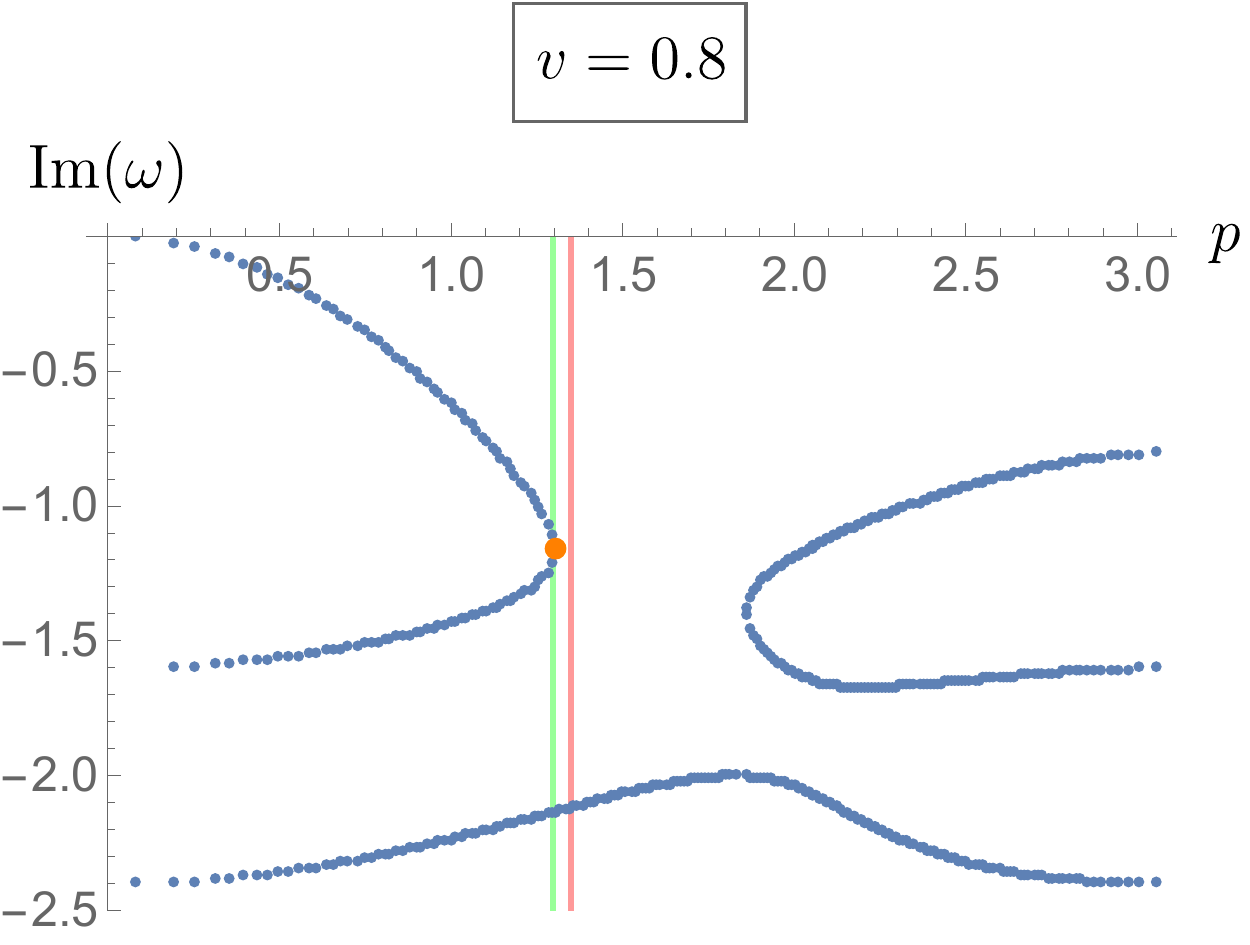}\vspace{0.5cm}
\includegraphics[width=0.3\textwidth]{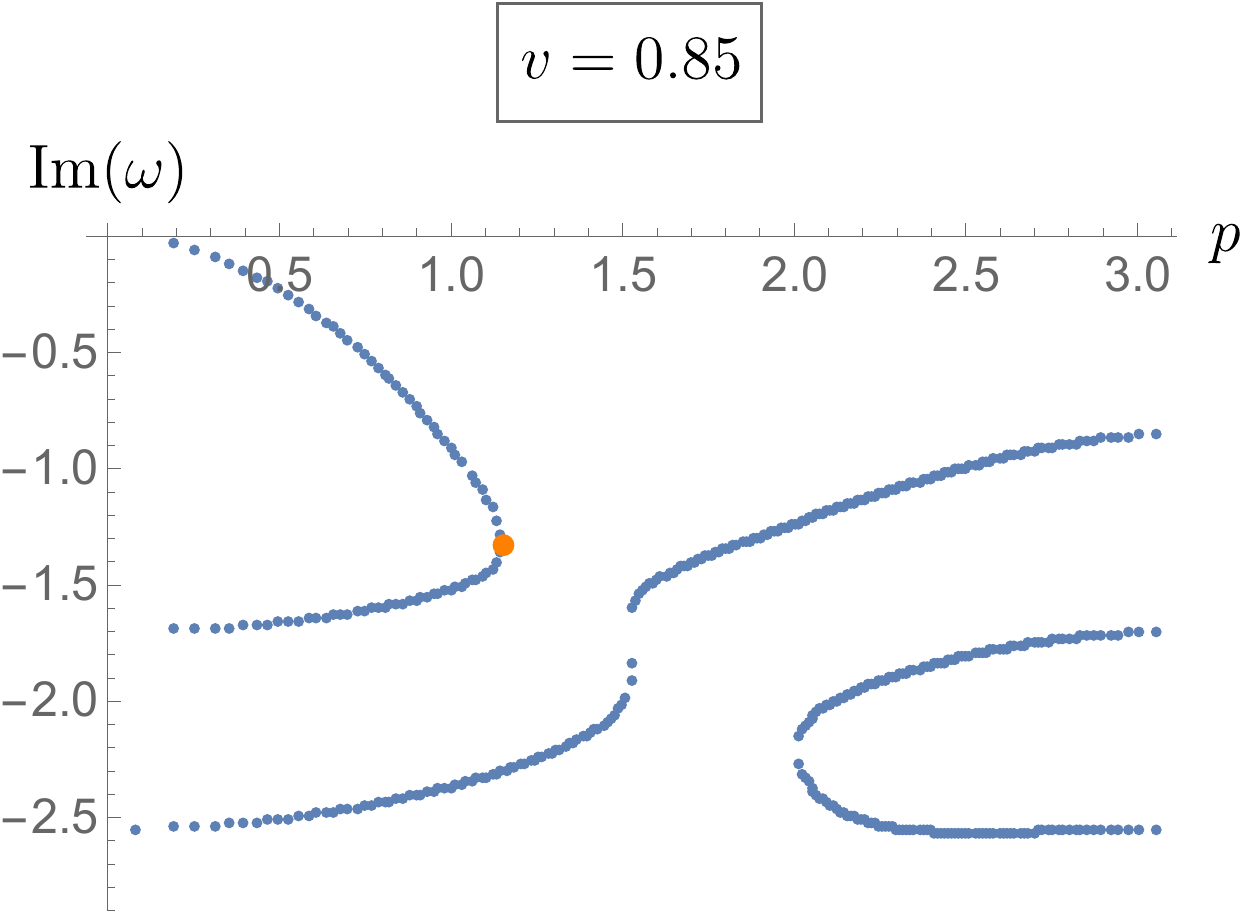}\hspace{0.5cm}
\includegraphics[width=0.3\textwidth]{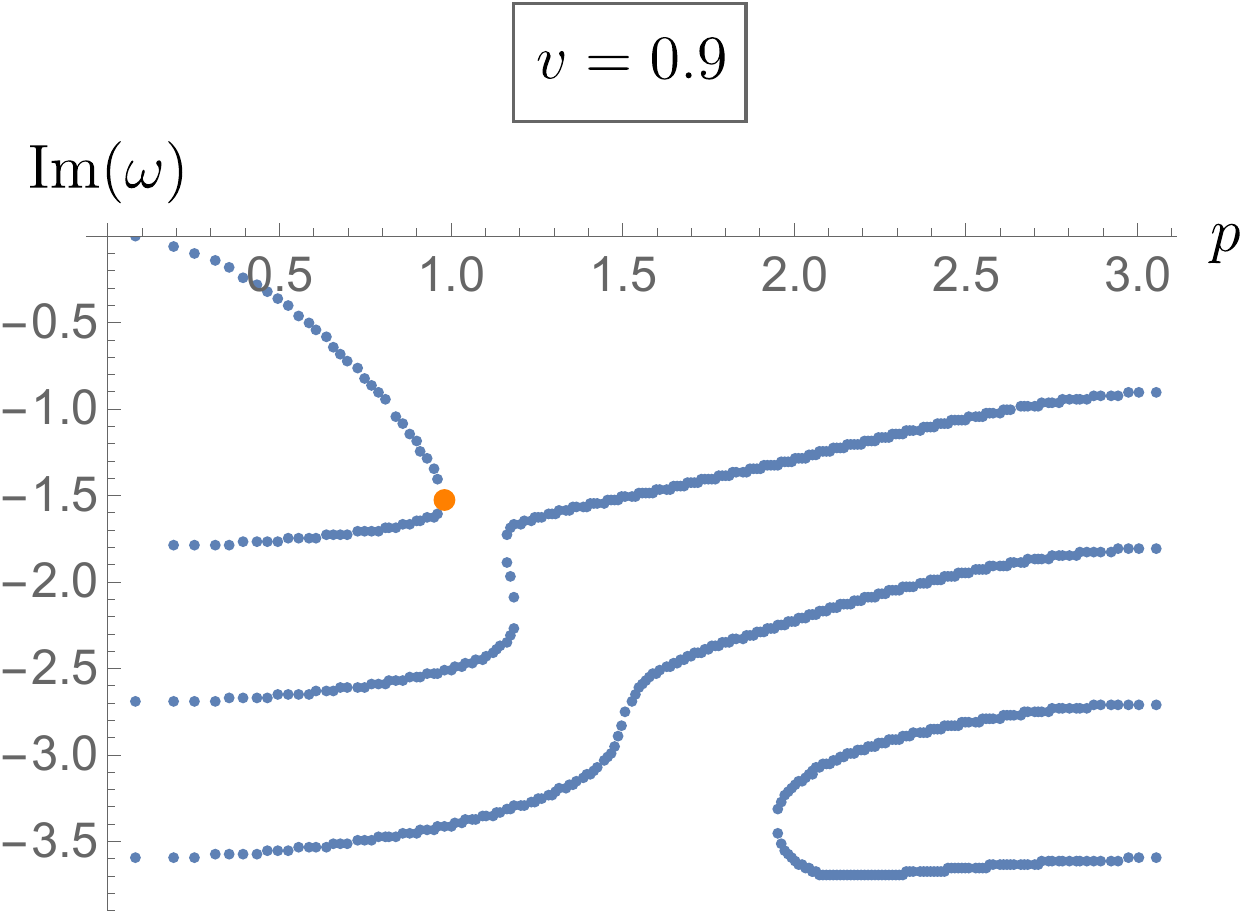}\hspace{0.5cm}
\includegraphics[width=0.3\textwidth]{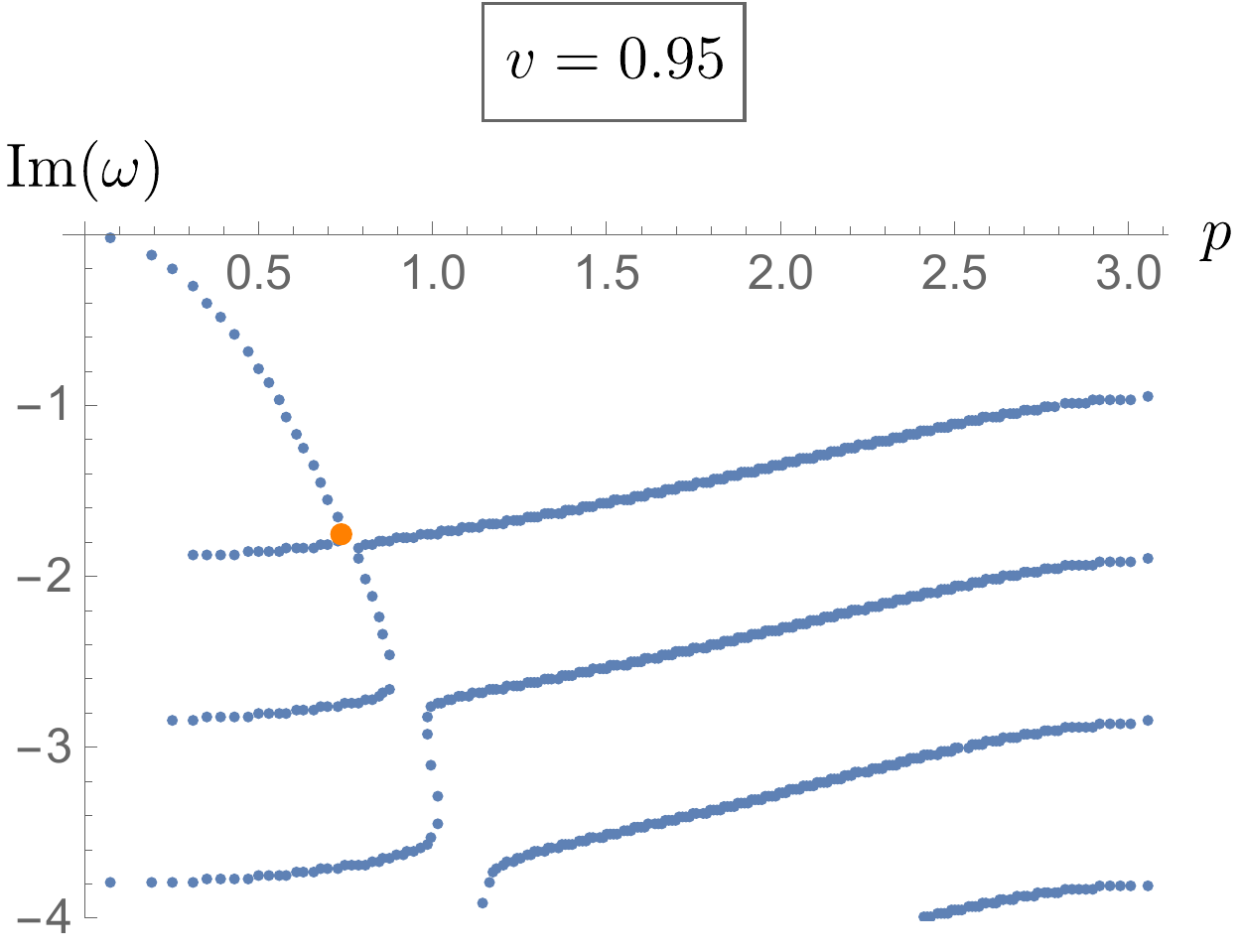}
\caption{The dispersion relation of the first few poles from weak to strong coupling. Orange dots mark the collision of two pole lines and determine the radius of convergence of all order hydrodynamics. For $v=0.65$ and $v=0.8$ we included two gridlines that mark the values of $p$ that we analyze further in Figs.~\ref{fig:reconnection2} and \ref{fig:polemotion2},~\ref{fig:reconnection} respectively. }
\label{fig:polemotion}
\end{center}
\end{figure}

From Fig.~\ref{fig:realp}, we can already start building intuition about the movement of pole trajectories (hot lines). In Fig.~\ref{fig:polemotion} we plot the dispersion relations of the first few poles for representative values of $v$ as we go from weak to strong coupling.  We see that at weak coupling the first few poles stay at  pure imaginary $\om$. At stronger coupling, we encounter collisions of poles, which is followed by a gap in momentum, after which another pair of poles appears. Next we seek to understand the details of  pole collisions.

\begin{figure}[!h]
\begin{center}
\includegraphics[width=0.45\textwidth]{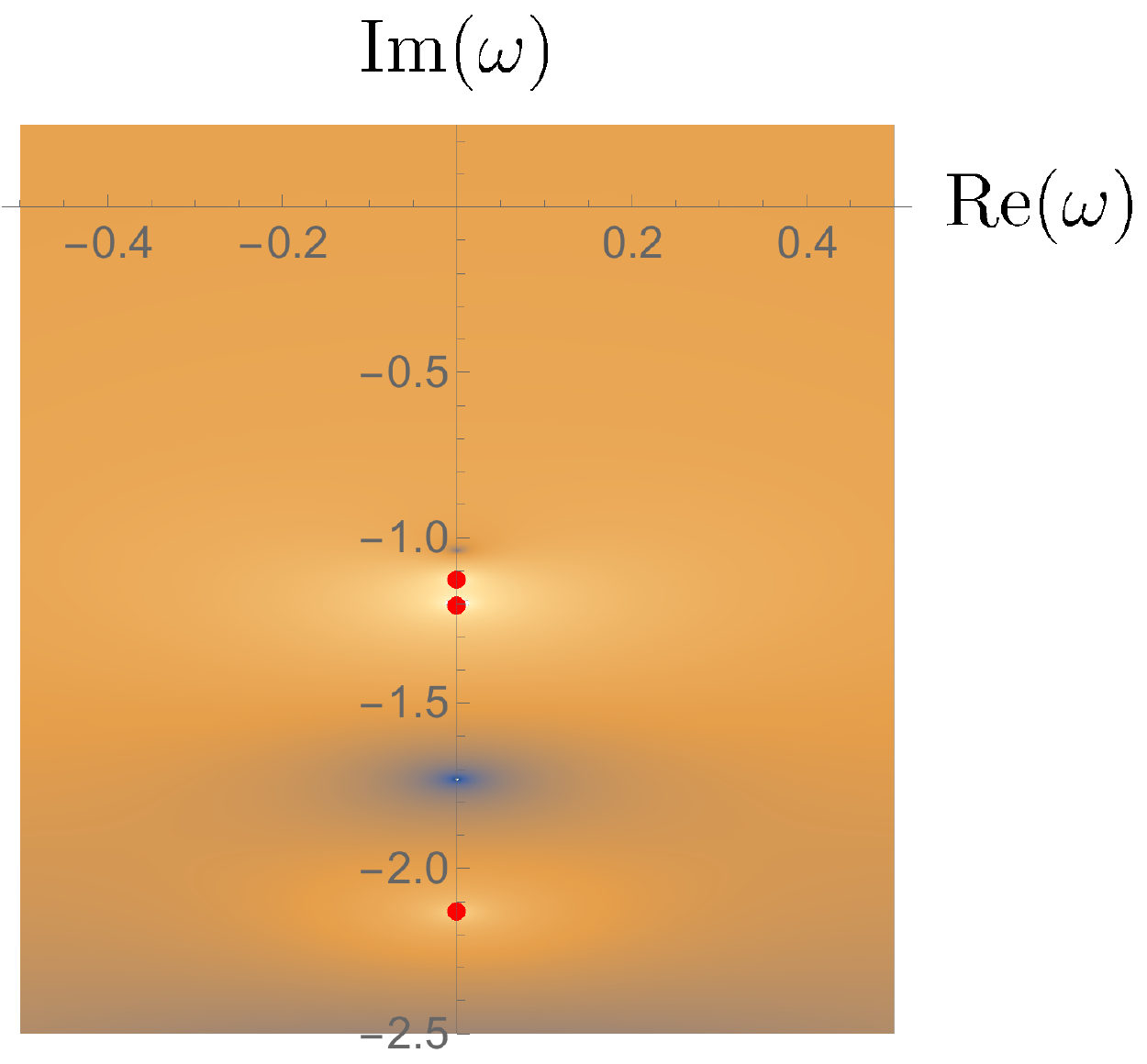}\hspace{0.5cm}
\includegraphics[width=0.45\textwidth]{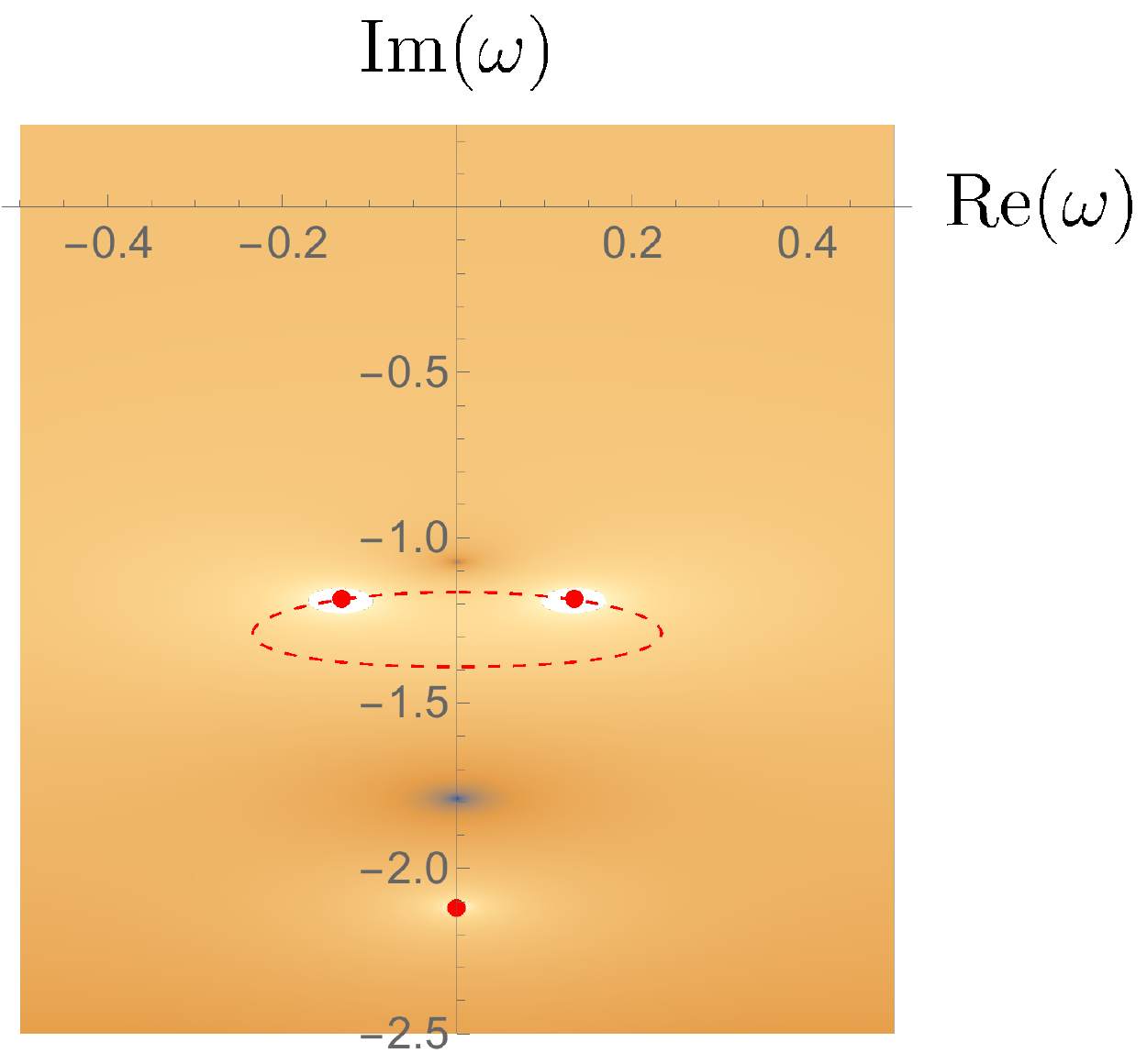}
\caption{Movement of poles in the complex plane for $v=0.8$. The hot locations are poles (also marked by red dots), and the cold ones are zeros of the two point function. {\bf Left:} For $p=1.297$ the poles are on the negative imaginary axis, as we have learned from Fig.~\ref{fig:polemotion}. {\bf Right:} As we increase the momentum, the poles collide and depart the imaginary axis. They follow the trajectory marked by the red dashed line, and eventually return to the imaginary axis, as can also be seen clearly from  Fig.~\ref{fig:polemotion}. Here we show the density plot for $p=1.35$. }
\label{fig:polemotion2}
\end{center}
\end{figure}

It turns out that after the collision the poles move off to the complex plane. We illustrate this on Fig.~\ref{fig:polemotion2}, by examining the location of poles on the complex $\om$ plane for the two values of momenta marked by gridlines on the $v=0.8$ figure of Fig.~\ref{fig:polemotion}. As the dashed lines show on Fig.~\ref{fig:polemotion2}, as we increase $p$ the pair of poles return to the imaginary axis and continue to live there until we reach the boundary of the Brillouin zone. 

\begin{figure}[!h]
\begin{center}
\includegraphics[width=0.45\textwidth]{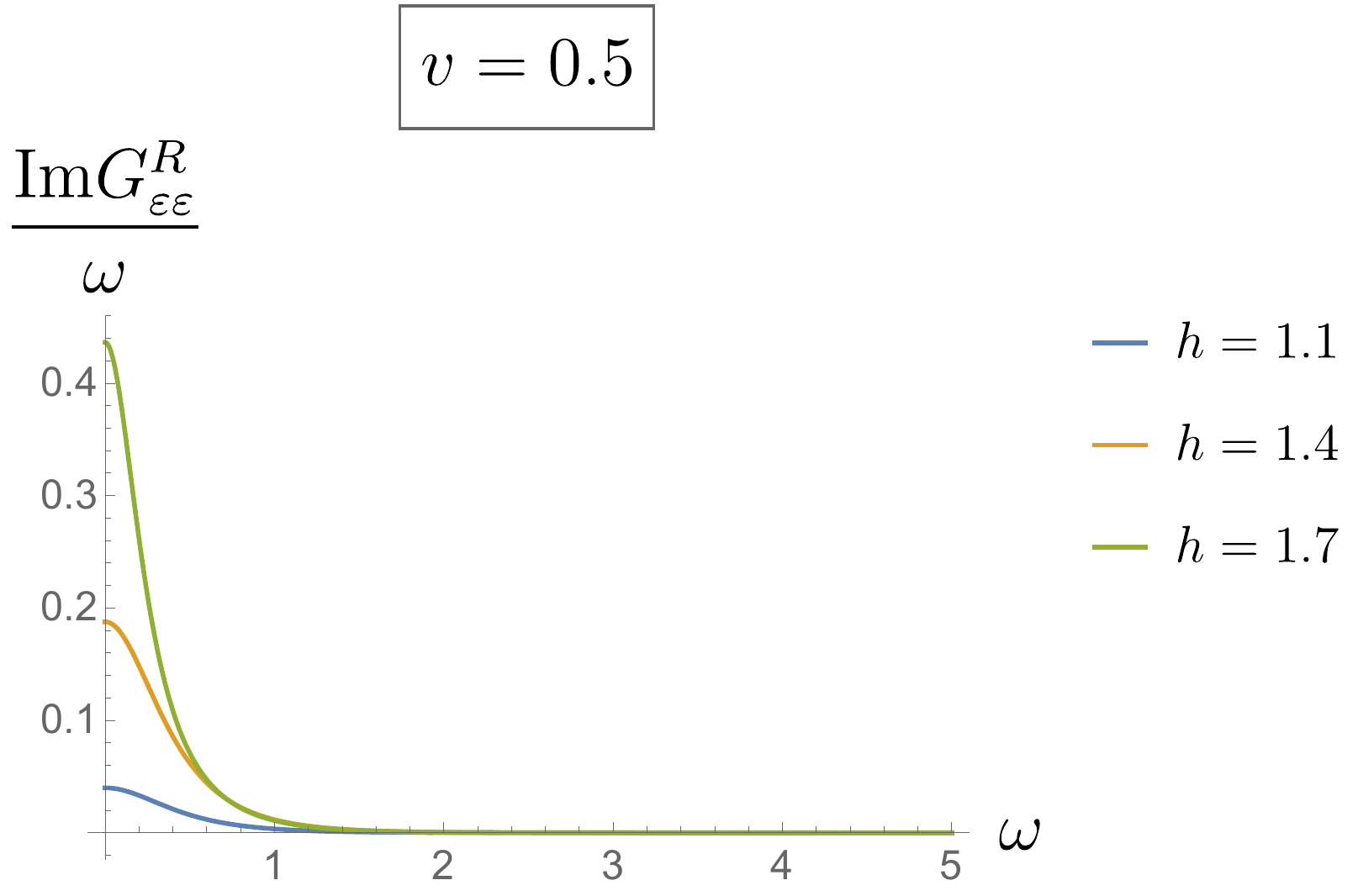}\hspace{0.5cm}\includegraphics[width=0.45\textwidth]{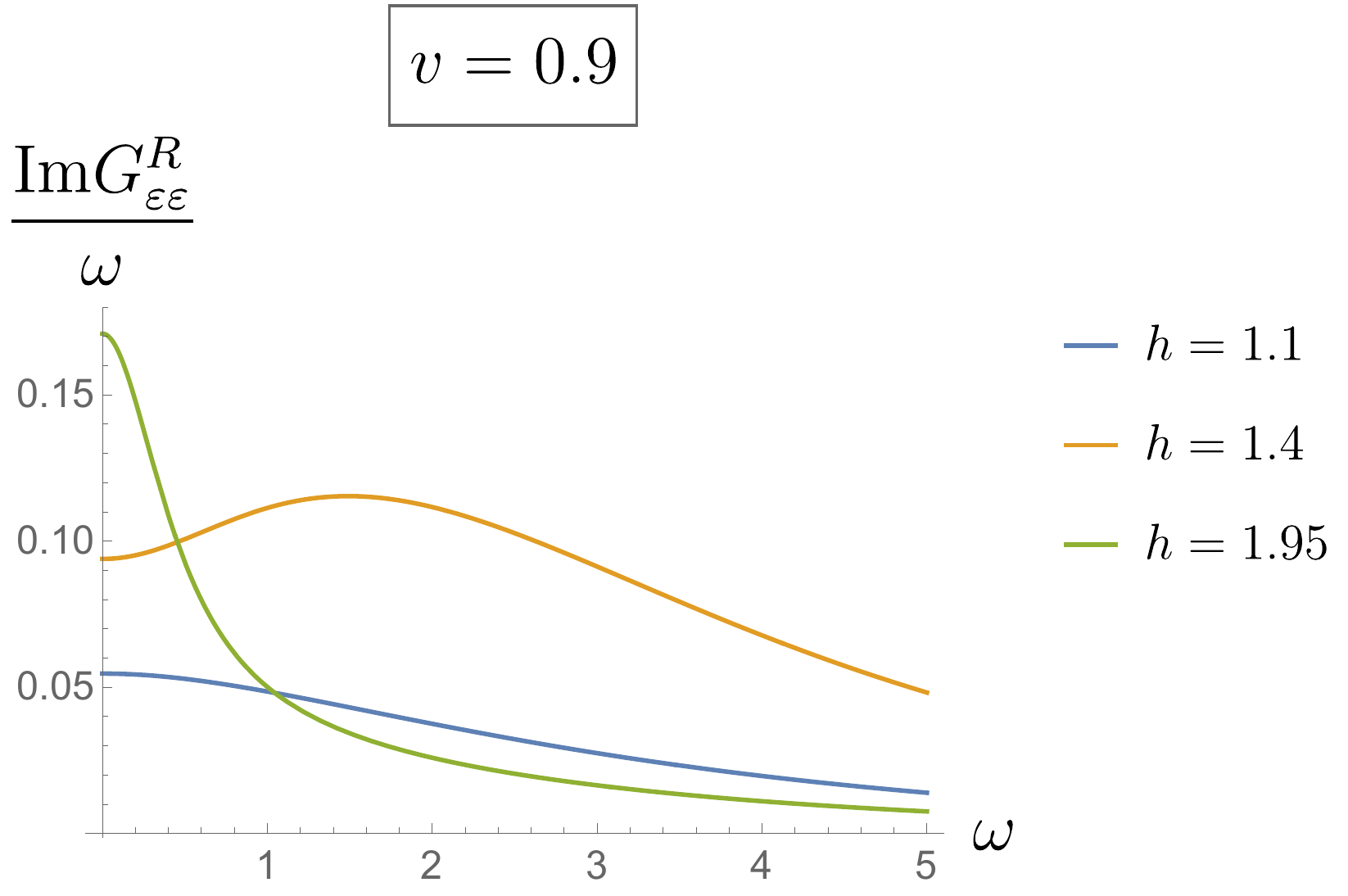}
\caption{The spectral function for various momenta $h$ and couplings $v$.}
\label{fig:spectral}
\end{center}
\end{figure}

The fact that poles never wander too far from the imaginary axis implies that the spectral function $\rho(\omega,p)=2\text{Im}G^R_{\varepsilon\varepsilon}(\omega, p)$ is rather featureless: there are no quasiparticle peaks emerging even at weak coupling, see Fig.~\ref{fig:spectral}. Besides the usual sharp Drude peak for small fixed momentum (at small frequency), which is the signature of the hydrodynamic diffusion pole close to the real $\om$ axis, a noteworthy feature is that for larger momentum and sufficiently strong coupling, the peak can shift away from zero frequency. This can be observed on the right panel of Fig.~\ref{fig:spectral}. This happens when the retarded correlator has a zero on the negative imaginary $\om$ axis that is closer to the real axis than the first pole. Note that this is very different behavior  than what one finds in SYM theory slightly away from infinite coupling \cite{Solana:2018pbk}.

\begin{figure}[!h]
\begin{center}
\includegraphics[width=0.6\textwidth]{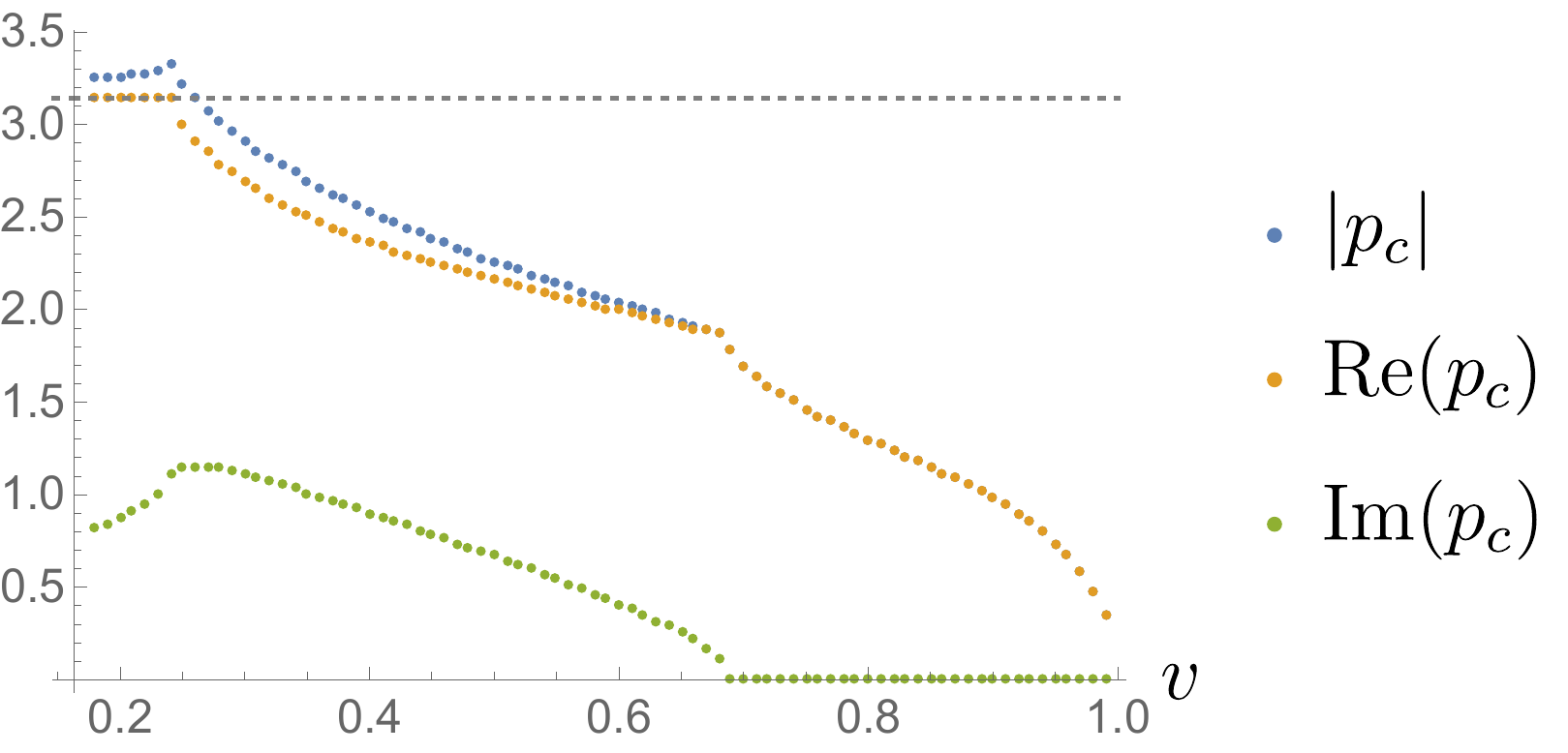} 
\caption{The numerical solutions to \eqref{eq:implicit} for various values of $v$. The absolute value $|p_c|$ is the radius of convergence in momentum space of the hydrodynamic expansion. For $v>v_c \approx 0.69$, the resulting $p_c$ is real, while for $v<v_c$ it develops an imaginary part. The real part of $p_c$ reaches $\pi$ shown in dashed (i.e. the edge of the Brillouin zone) at $v_c'=0.241\pm 0.005$ and then it stays there.} 
\label{fig:convrad}
\end{center}
\end{figure}

Next, following \cite{Grozdanov:2019kge,Grozdanov:2019uhi}, we ask about the convergence radius of all order hydrodynamics. As explained in \cite{Grozdanov:2019kge}, we can use the analytic implicit function theorem to find the radius of convergence. This amounts to finding a point on the diffusion pole line $n(p)$ such that 
\beq
\label{eq:implicit}
\partial_n \psi_n(\theta_v)|_{n(p_c),p_c}=0\,.
\eeq
We plot the solution $p_c$ to this equation on Fig.~\ref{fig:convrad}. We find that the radius of convergence $|p_c|$ increases as we \textit{decrease} the coupling.\footnote{The case of the limit $v\to 1$ is special. See the discussion in Sec.~\ref{sec:strong}.}${}^{,}$\footnote{An analogous result for real fluids recently appeared in \cite{Baggioli:2020loj}.}
For strong coupling $v_c<v<1$, where $v_c=0.6875\pm 0.0005$, the critical momentum $p_c$ is real, while for $v<v_c$ it develops an imaginary part. This corresponds to the fact that we have a dispersion relation for the diffusion mode that has finite $p$ support for $v>v_c$, while it reaches the edge of the Brillouin zone for $v<v_c$, see Figs.~\ref{fig:polemotion} and \ref{fig:realp}. There is another distinguished value of the coupling which we denote by $v_c'=0.241\pm 0.005$. For $v<v_c'$ we have $\text{Re}\,p_c=\pi$, therefore the region of convergence of all order hydrodynamics contains the entire Brillouin zone. 

Note that the convergence radius shrinks to zero size as we go to maximal coupling $v\to 1$. To gain more intuition about how this happens, we sketch the analog  of Fig.~\ref{fig:polemotion} for very strong coupling in Fig.~\ref{fig:strong}. We see that the diffusive hydrodynamic pole with $D$ diverging as $v\to1$ is intersecting with a non-hydrodynamic pole that is sitting at $i\om=O(1)$ that has mild momentum dependence. This clearly leads to a decreasing $p_c$ as we increase the coupling.

\begin{figure}[!h]
\begin{center}
\includegraphics[width=0.45\textwidth]{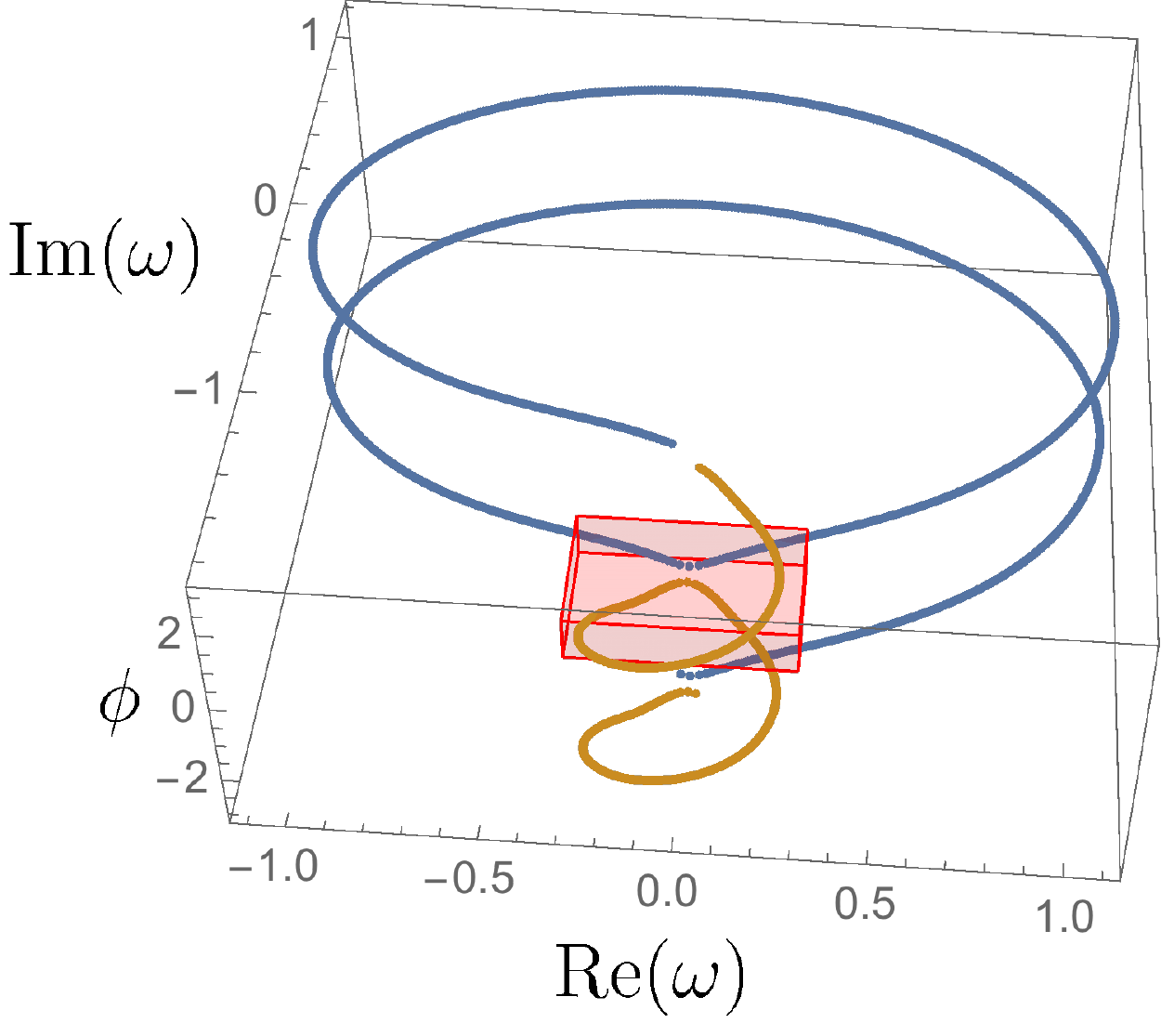}\hspace{0.5cm}
\includegraphics[width=0.45\textwidth]{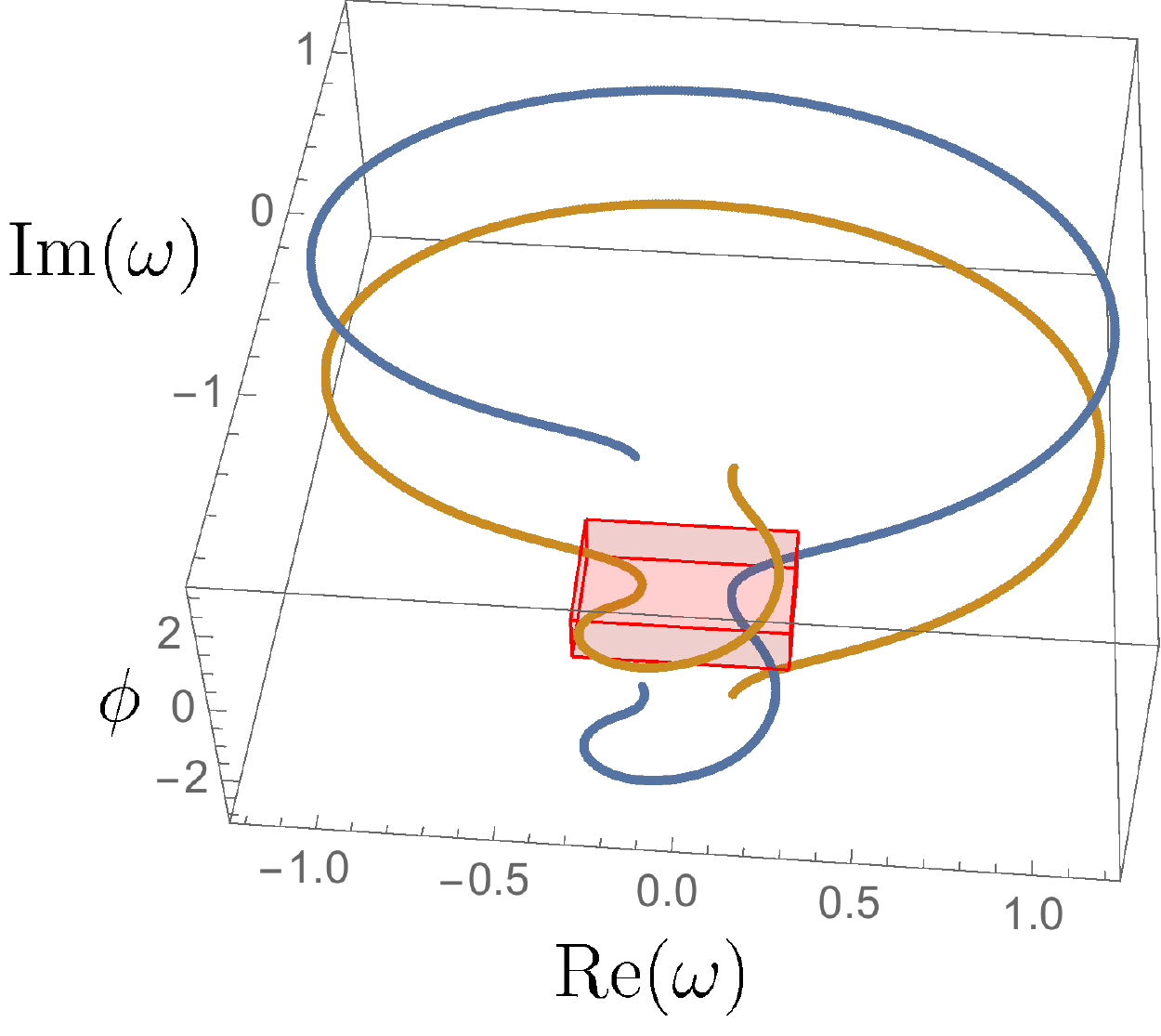}\vspace{0.5cm}
\includegraphics[width=0.45\textwidth]{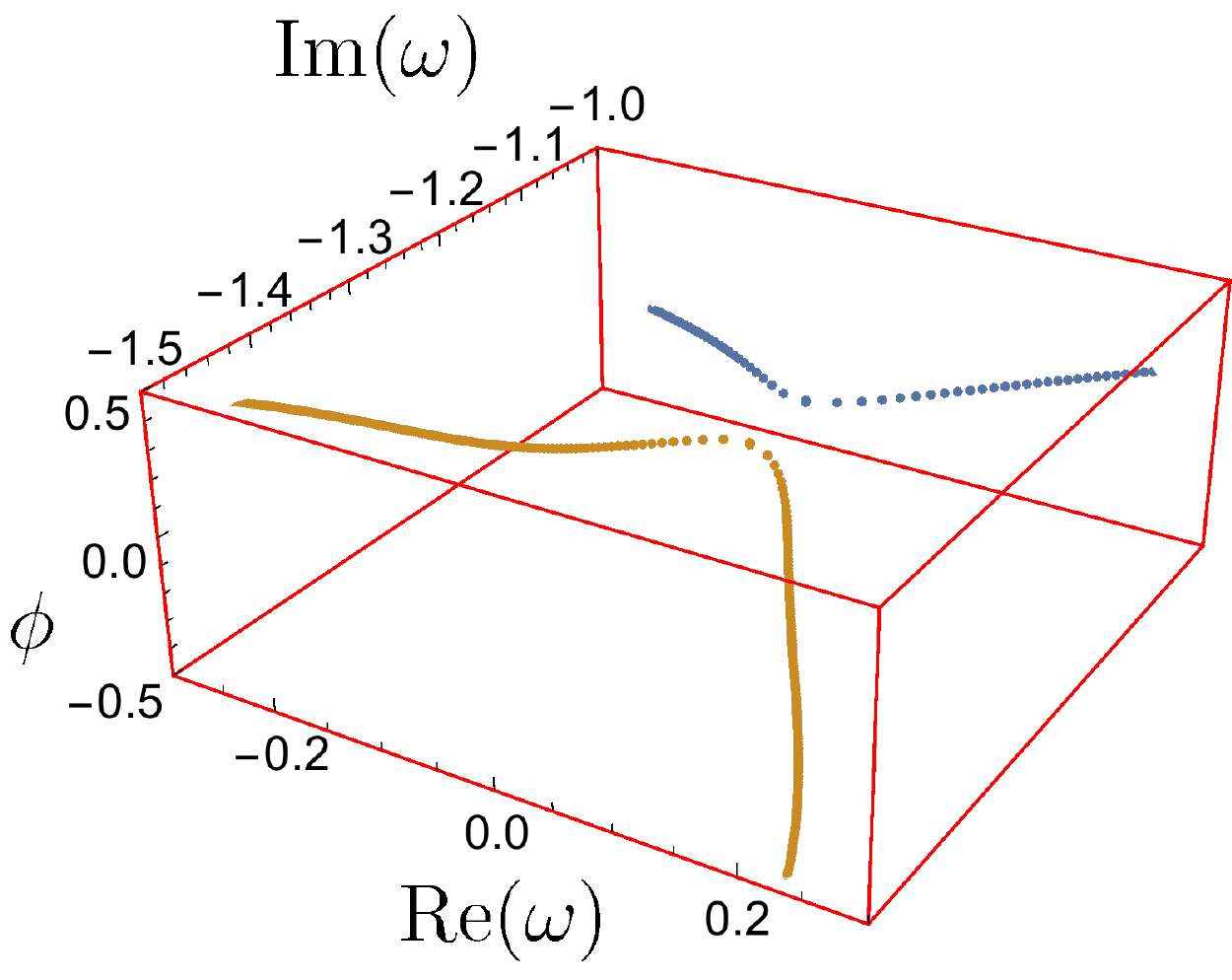}\hspace{0.5cm}
\includegraphics[width=0.45\textwidth]{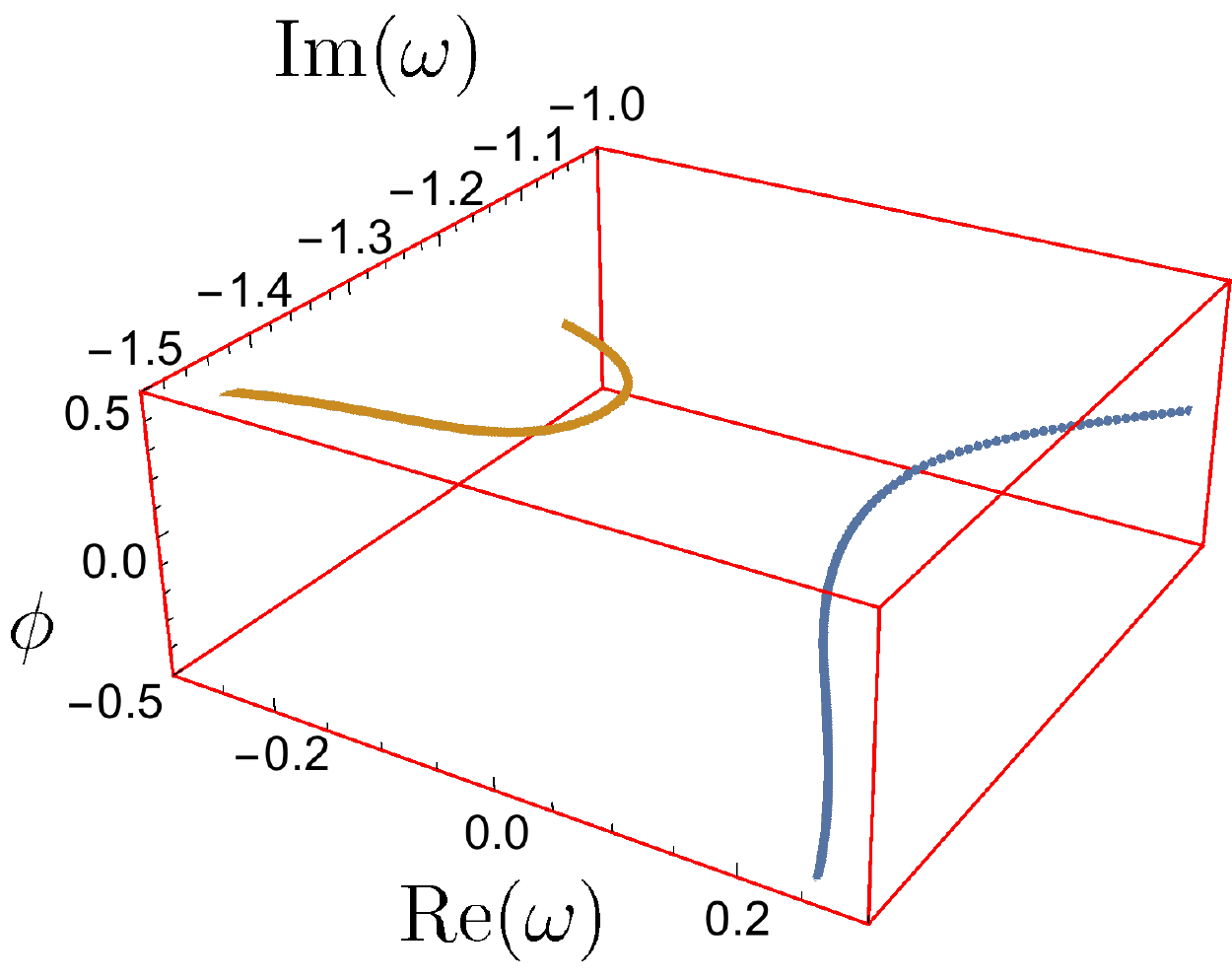}
\caption{Motion of the hydrodynamic and a non-hydrodynamic pole on the complex $\om$ plane for $v=0.8$ at fixed $p_m$ and varying phase of the momentum $\phi$. {\bf Left:} For $p_m=1.297$ the poles sit on the imaginary axis of the complex $\om$ plane (horizontal plane) close to each other for $\phi=0,\pi$. As we increase $\phi$ (vertical axis) from $0$ to $2\pi$ the projection of their trajectory onto the $\om$ plane goes around a circular curve twice, since the function only depends on $p^2$. Note that the vertical axis represents a periodic direction, so the upper and lower end planes are identified.
 {\bf Right:} For $p=1.35$ the two curves reconnect, and we wind around the projection of the curve to the $\om$ plane once as we change the phase from $0$ to $2\pi$. There is also another curve whose plot coincides with this one, but has phase shifted by $\pi$. {\bf Insets:} The insets zooms onto the point where the reconnection takes place, and are marked with red cuboids on the large figures.
 }
\label{fig:reconnection}
\end{center}
\end{figure}

It is the collision of the hydrodynamic pole with a non-hydrodynamic pole that delineates the region in which hydrodynamics converges.  
 To understand this, we have to explore complex  $p$'s. We fix the modulus and vary the phase of $p=p_m\, e^{i\phi}$. We obtain the reconnection of pole trajectories as shown in Fig.~\ref{fig:reconnection} for the two selected values of $p_m$ we used in Fig.~\ref{fig:polemotion2}. At strong coupling, the reconnection  always happens at real $p$, and we can read off the convergence radius of all order hydrodynamics from Fig.~\ref{fig:polemotion}.

\begin{figure}[!h]
\begin{center}
\includegraphics[width=0.45\textwidth]{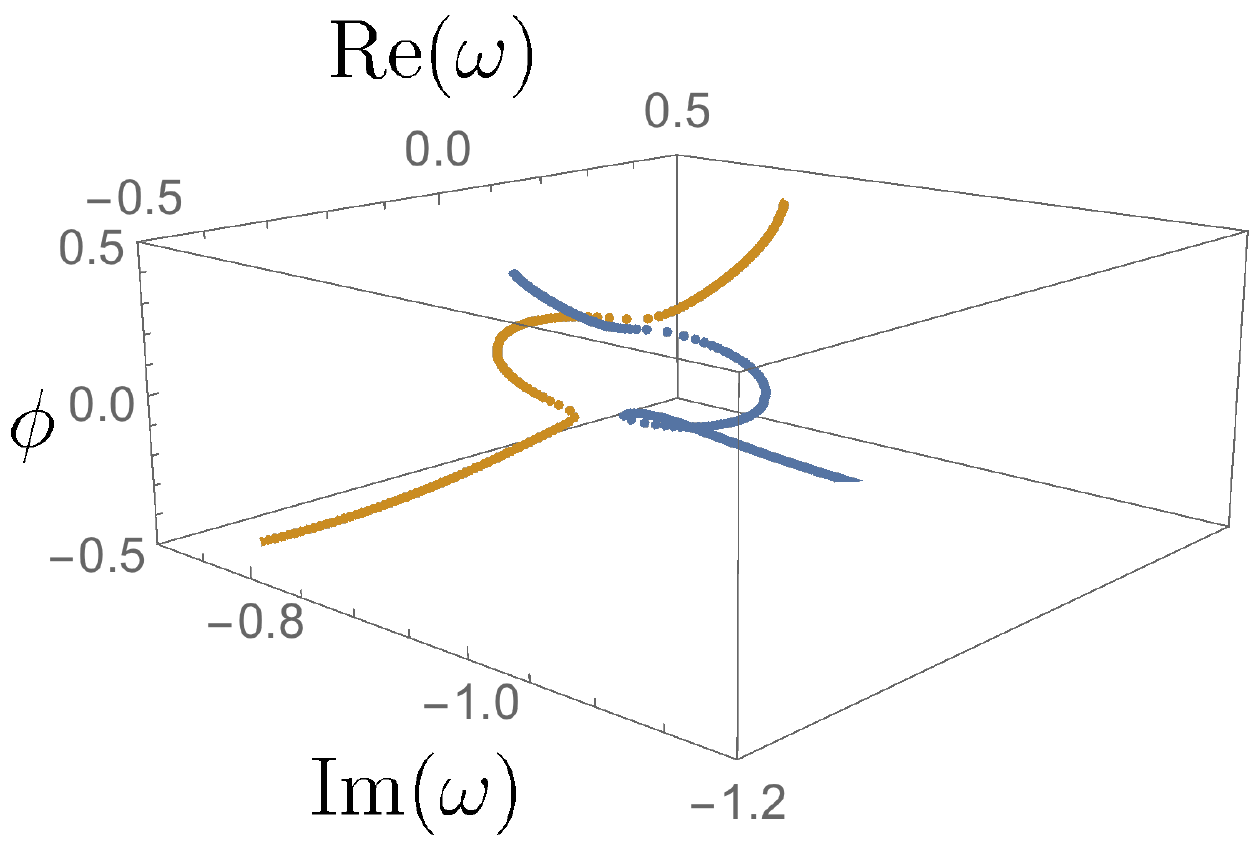}\hspace{0.5cm}
\includegraphics[width=0.45\textwidth]{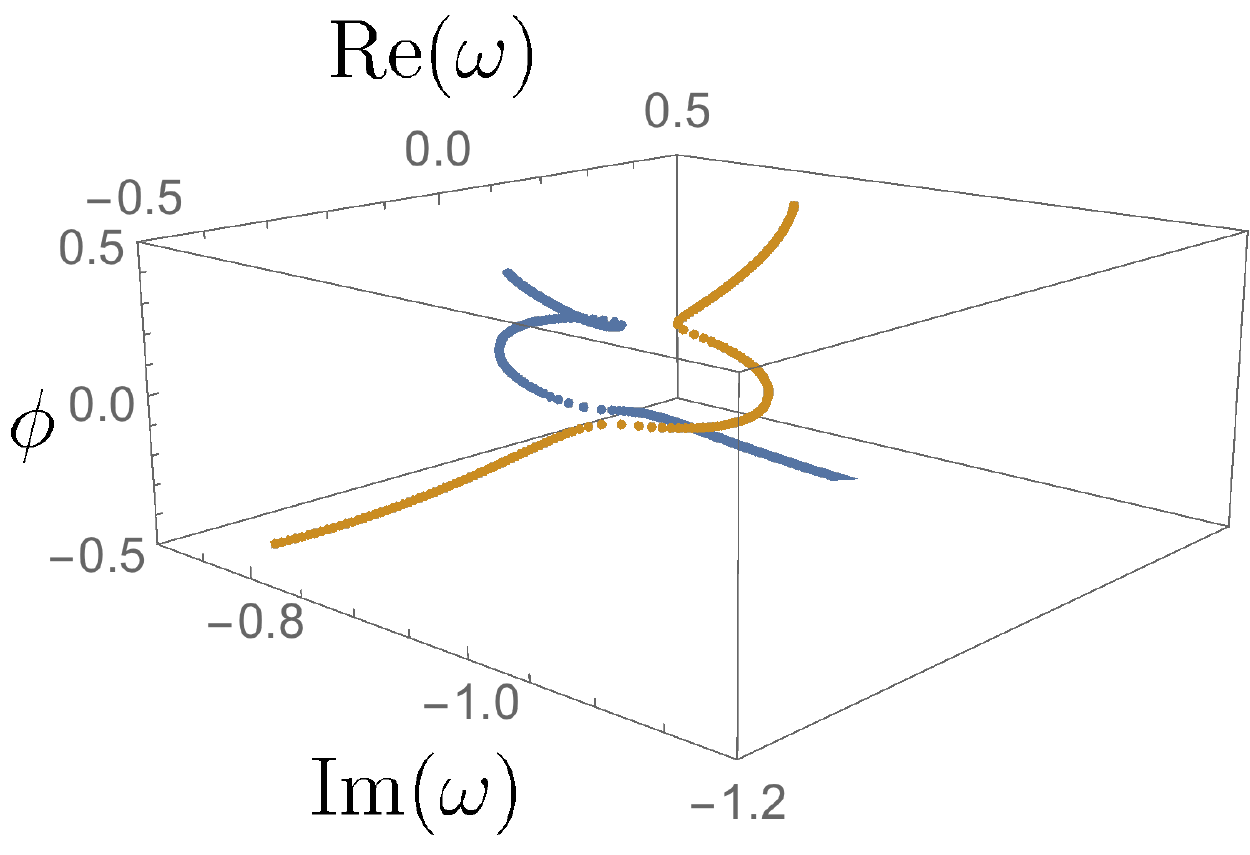}
\caption{The analog of Fig.~\ref{fig:reconnection} for $v=0.65$. Here we show only the region near $p_c$, the whole curve is too complicated to comprehend.  {\bf Left:} For $p_m=1.93<\abs{p_c}$ the pole trajectories are unlinked. {\bf Right:} They become linked for $p_m=1.94>\abs{p_c}$. Both the blue and the orange curve only change significantly near the crossing point, and remain virtually unchanged far away from it.
 }
\label{fig:reconnection2}
\end{center}
\end{figure}

At weaker coupling, we find a pole collision of a different flavor. Instead of the reconnection of pole trajectories, they are unlinked at small $p$, and become linked at a complex $p_c$ plotted on Fig.~\ref {fig:convrad}, without the reorganization of each individual curve. We show this on Fig.~\ref{fig:reconnection2}.

\subsection{The strong coupling limit}\label{sec:strong}

\begin{figure}[!h]
\begin{center}
\includegraphics[width=0.5\textwidth]{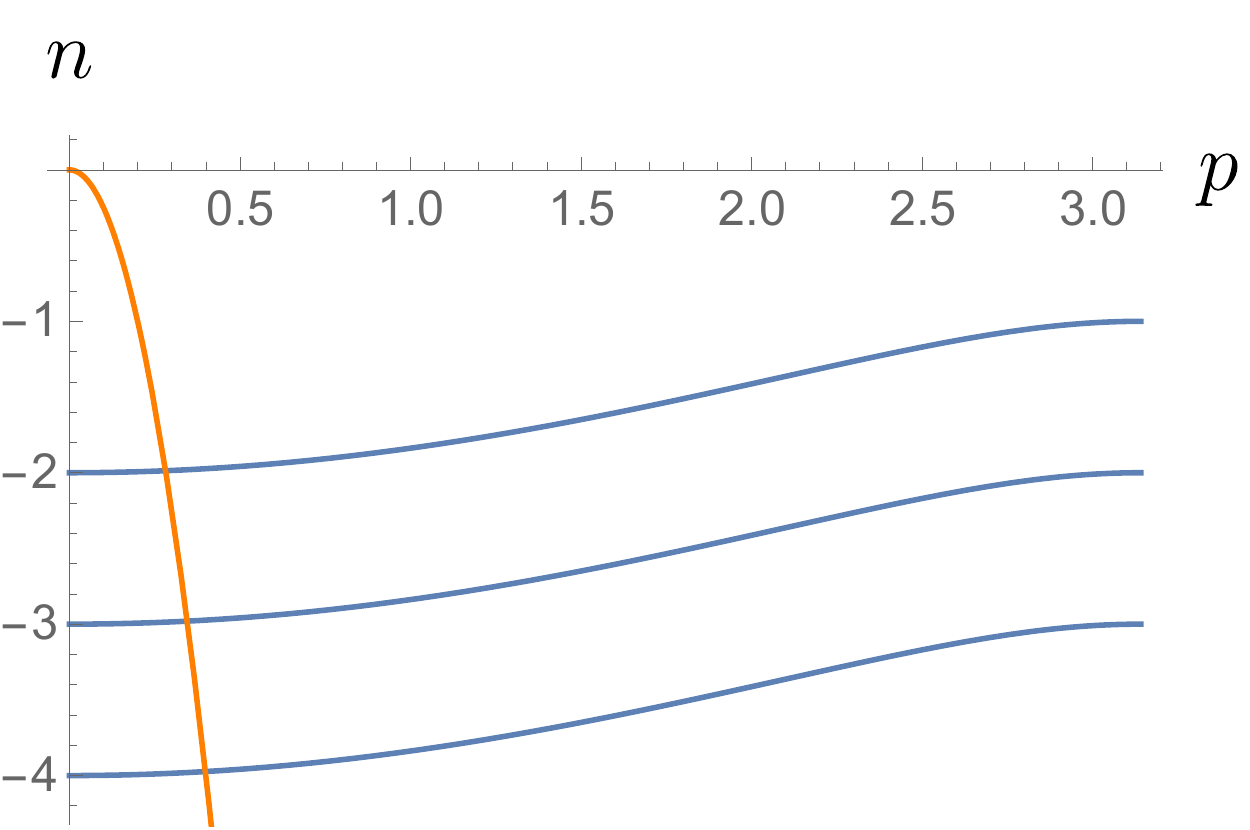}
\caption{The dispersion relation of the poles of \eqref{StrongCouplingLimit} is drawn with blue lines for real momenta. We also include a sketch of the diffusion pole with an orange line, which is not captured by the formula (since the pole degenerates to the vertical axis). Compare with the last plot of Fig.~\ref{fig:polemotion}.
 }
\label{fig:strong}
\end{center}
\end{figure}

For fixed $h,\, n$ we can take the strong coupling limit corresponding to small $\de v\equiv 1-v$, and obtain the following simple formula valid for $2>{\rm Re}(h)> 1/2$:
\es{StrongCouplingLimit}{
{q^2 \ov N}G_{\varepsilon \varepsilon }^R(\omega)=&{(2-h)(h-1)\ov 2\pi \de v}-{(2-h)(h-1)\ov 2\pi }+(\pi \de v)\le({h(h-1)(2h-1)+6\om^2\ov 48(h-3/2)}\ri)\\
&+{\le(\pi \de v\ov 4\ri)^{2h-2}}\, {\cot\le(\pi h\ov 2\ri)\,\Gamma\le(\frac32-h\ri)\Gamma\le(h-i \om\ri)\ov 2\,\Gamma\le(-\frac12+h\ri)\Gamma\le(1-h-i \om\ri)}+O\le(\de v^2,\, \de v^{2h-1}\ri)\,.
}
Note that the terms in the first line are analytic in $\om$; they play the important role of cancelling the singular contribution of the second line at $h=3/2$, where the powers $\de v$ and $\de v^{2h-2}$ coincide. 
We can simply verify pole skipping on this function, and we plot the motion of the poles for real momentum on Fig.~\ref{fig:strong}. This plots is helpful in that it explains where some features we saw on Fig.~\ref{fig:polemotion} originate from.

 Note however that this formula misses the important diffusion pole, as that happens for $2-h=O(\de v)$. If we take another limit, we can recover the results of \cite{Gu:2016oyy} including the diffusion pole, but we miss the higher poles discussed above. To analyze the hydrodynamic limit ($p,\omega\rightarrow 0$) of the retarded correlator at strong coupling, we rescale the momentum as $p^2=\tilde p^2\delta v$ and expand \eqref{eq:retTT} in terms of $\delta v$ (see also the discussion around \eqref{kap}). Together with a first order approximation $h=2-{\gamma \ov 6}p^2$ and the diffusion constant $D= {\gamma \ov 6\,\de v}$ we  get 
 \es{StrongCouplinghydrolimit}{
{q^2 \ov N}G_{\varepsilon \varepsilon }^R(\omega)&=\de v\, {\pi Dp^2 (1+\omega^2)\ov 4(-i \omega+Dp^2)}+\text{(contact terms)}+O(\delta v^2)\,,\\
\text{(contact terms)}&={D p^2\ov 2\pi}-\de v\,{D p^2+(D p^2)^2\ov 2\pi}+O(\delta v^2)\,.
}
This matches the result obtained in \cite{Gu:2016oyy} up to the contact terms that they did not keep track of.
We note that we do not see the poles that were plotted in Fig.~\ref{fig:strong}, as they have subleading residues compared to the diffusion pole. This leads to the curious conclusion that the convergence radius of hydrodynamics in this scaling limit is infinite, as opposed to the result of Fig.~\ref{fig:convrad} that took into account the collision with the aforementioned subleading poles. Another consequence is that the dispersion relation of the pole, $\om(p)=-iD p^2$ is an entire function of $p$, and according to  the result of \cite{Grozdanov:2020koi} this implies the relation $D=(u_B^{(T)})^2$ that we found in the strong coupling limit Sec.~\ref{sec:Diff}.\footnote{As shown in \cite{Grozdanov:2020koi} various bounds can be derived for $D$ using the analyticity properties of the dispersion relation $\om(p)$ and pole skipping, and it would be interesting to explore them for our system.}

\subsection{Analytic structure at weak coupling}

Here we wish to analyze \eqref{2ptCorr} in the weak coupling limit $v\rightarrow 0$. In analogy with the strong coupling limit, we will consider two different scalings of the parameters of the Green function.

Let us first consider fixed real momentum, $h\in[1,2]$. Already from Fig.~\ref{fig:polemotion} it is visible that the poles on the negative imaginary axis become denser as we decrease $v$, and in the limit $v\to 0$ they proliferate. We did not find a nice limiting function, in particular the accumulation of poles does not seem to form a branch cut.

Since the density of poles on the imaginary axis diverges in the weak coupling limit, it is reasonable to consider another limit, in which we define the rescaled imaginary frequency $\tilde n=n/v$ and keep this fixed as $v\rightarrow 0$. If we do this, the hypergeometric functions ${}_2F_1(a,b,c;z)$ in \eqref{masterfuncDef} can be approximated by $1$, since their $z$ argument goes to $0$, while their $a,b,c$ arguments remains finite. We get the simple approximation:
\beq
\frac{q^2}{N} G^R_{\varepsilon \varepsilon}(in)\approx \frac{1}{4} \pi  v^2 \left(\frac{2 \tilde{n} \Gamma \left(\frac{1}{2} \left(-h+\tilde{n}+2\right)\right) \Gamma \left(\frac{1}{2} \left(h+\tilde{n}+1\right)\right)}{\Gamma \left(\frac{1}{2} \left(-h+\tilde{n}+1\right)\right) \Gamma \left(\frac{1}{2} \left(h+\tilde{n}\right)\right)}-\tilde{n}^2+{h(h-1)\ov2}\right)\,.
\eeq
Similarly to \eqref{StrongCouplingLimit}, this expression is useful, as the movement of poles and pole skipping can be easily analyzed.

\subsection{Assorted comments on the SYK chain}\label{sec:comments}

At the level of the fermion two point function, the SYK chain is ultralocal in space, i.e. $\<\chi_{i,x}(\tau)\chi_{i,x'}(0)\>\propto \de_{x,x'}$. It is locally critical \cite{Gu:2016oyy}, since in the vacuum the correlator decays as a power law, and at small temperature it takes the form dictated by finite temperature conformal quantum mechanics \eqref{Saddle}.\footnote{These properties make it analogous to extremal black holes in holography, whose dual is a semi-local quantum liquid \cite{Iqbal:2011in}. A related phenomenon is the slow spreading of quantum information (measured by R\'enyi entropies) in the system \cite{Gu:2017njx}.} If we move on to the study of the energy density, we saw that our state of matter exhibits diffusive energy transport.\footnote{The SYK chain built from charged fermions also transports charge diffusively \cite{Davison:2016ngz}. }

Next, we ask, if our system has a continuum limit. We have been working with dimensionless positions and momenta, $x$ and $p$. If we worked with dimensionful versions, $y\equiv x a,\, k\equiv p/a$, we would find that the diffusion constant is
\es{DiffPhys}{
D_\text{phys}=\le(2\pi  a^2 \ov \beta \ri)\,\frac{1}{12}\gamma v \left(\pi  v \tan \left(\frac{\pi  v}{2}\right)+2\right) =\begin{cases}
{\pi {\cal J} \ga\, a^2\ov 6}\qquad \text{(for $\beta {\cal J}\to\infty$)}\,,\vspace{0.3cm}\\
{{\cal J} \ga\, a^2\ov 3}\qquad \text{(for $\beta {\cal J}\to0$)}\,.
\end{cases}
}
The only meaningful way to take the continuum limit of the system is to implement the double scaling limit
\es{ContLim}{
{\beta\ov a}\to \infty\,, \quad \beta {\cal J}\to \infty\,, \quad {D_\text{phys}\ov \beta}={\pi\ga\ov 6}\,(\beta {\cal J})\, \le({a\ov \beta}\ri)^2=\text{fixed}\,.
}
We find that for all other $O(1)$ values of the dimensionless coupling constant $\beta {\cal J}$, i.e. all values of $v$ not infinitesimally close to $1$, we get a lattice scale diffusion constant. 

Let us now examine the propagation of chaos in the model. The physical butterfly velocity corresponding to the spatial coordinate $y$ is $u_{B,\text{phys}} = {2\pi\, a\ov \beta}\, u_B $, which in the $\beta {\cal J}\to \infty$ ($v\to 1$) limit takes the value
\es{uBCont}{
u_{B,\text{phys}} = \sqrt{{\pi^2 \ga\ov 3 }\,(\beta {\cal J}) \le({a\ov \beta}\ri)^2}\,,
}
which is exactly the same combination as appearing in \eqref{ContLim}, and hence remains finite in the continuum limit. This is how it had to be since the remarkable relation $D_\text{phys}={\beta\ov 2\pi} u_{B,\text{phys}}^2$ has to be satisfied in the strong coupling limit as discussed in Sec.~\ref{sec:Diff}. 

In the continuum limit the theory is maximally chaotic, with $\lam(u)={2\pi\ov \beta}\le(1-{\abs{u}\ov u_B}\ri)$, see \eqref{strongCoup}. The energy density Green function given by \eqref{StrongCouplinghydrolimit} is extremely simple: it only has a pole with dispersion relation $\om=-i D_\text{phys} k^2$ and an infinite radius of convergence in $k$.

That the continuum limit is so simple, clearly demonstrates that momentum is not an approximately conserved quantity at any scale in the SYK chain. While the collision of the diffusion and a non-hydrodynamic pole on the imaginary $\om$ axis shown on Fig.~\ref{fig:polemotion2} is reminiscent of the scenario articulated in \cite{Grozdanov:2018fic}, whereby at large $k$ the collision of poles create two sound modes with $\om=\pm c_s\, k+\dots$, here the poles only depart the imaginary $\om$ axis for a short while, they always have comparable real and imaginary parts, and return to the  imaginary $\om$ axis for larger values of the momentum. At fixed momentum, tuning the coupling from weak to strong has a dramatic effect in SYM theory: the closely spaced branch cuts at zero 't Hooft coupling \cite{Hartnoll:2005ju} break up into families of poles that form multiple branches of a ``Christmas tree'' at strong coupling \cite{Grozdanov:2018gfx}, only for the top branch to remain at infinite coupling. The motion of poles in our model is less rich, but fully calculable and should complement the recent studies of the analytic structure of thermal correlators at small but finite coupling \cite{Romatschke:2015gic,Kurkela:2017xis,Moore:2018mma,Grozdanov:2018atb}: changing the coupling at fixed $p$ leads to occasional collisions of poles that move them out to the complex $\om$ plane in some window of $v$, and sometimes pole skipping happens, when a line of zeros intersects with a line of poles.

\section*{Acknowledgements}

We thank Felix Haehl, Hong Liu, Shiraz Minwalla, Douglas Stanford, Alexander Zamolodchikov, and especially Sa\v{s}o Grozdanov for useful discussions and comments on earlier versions of the draft. CC is supported in part by the Simons Foundation grant 488657 (Simons Collaboration on the
Non-Perturbative Bootstrap) and also by the KITP Graduate Fellowship. MM is supported by the Simons Center for Geometry and Physics.  

\pagebreak

\appendix
 
\section{Four point function for $p=0$}
\label{app:sykdot}

As mentioned in the main text, the four point function \eqref{eq:greensol1} must agree with that of the SYK dot for $p=0$ ($h=2$). Here we confirm this by doing the $n$ sum. For $h=2$ the zero modes \eqref{eq:eigenfunctions} reduce to
\beq
\psi_n^e(\theta) &= \frac{v}{n} \left[ \frac{n \cos \left(\frac{n (\pi -2 \theta )}{2 v}\right)}{v}+\cot (\theta ) \sin \left(\frac{n (\pi -2 \theta )}{2 v}\right) \right]\,,\\
\psi_n^o(\theta) &=\frac{1}{\frac{n^2}{v^2}-1}\left[  \frac{n \sin \left(\frac{n (\pi -2 \theta )}{2 v}\right)}{v}-\cot (\theta ) \cos \left(\frac{n (\pi -2 \theta )}{2 v}\right) \right]\,.
\eeq
These agree with the eigenfunctions considered in \cite{Choi:2019bmd}, but with a different normalization. Let us introduce a rescaled version of the solutions \eqref{eq:greensol2}
\beq
\Phi_n^-(\theta)&=\frac{n}{v}\left( \frac{n^2}{v^2}-1 \right)\Psi_n^-(\theta)\,,
\\ \Phi_n^+(\theta)&=\frac{n}{v}\left( \frac{n^2}{v^2}-1 \right)\Psi_n^-(\theta) \,.
\eeq
The advantage of scaling out the $n$ dependence from the denominator is that now these modes depend polynomially on $n$ apart from the Fourier modes in $\theta$. That is, one can write
\beq
\label{eq:polynomprop}
\Phi_n^\pm(\theta) = p^{\pm}_+(n,\theta)e^{i\frac{n}{2v}\theta} + p^{\pm}_-(n,\theta)e^{-i\frac{n}{2v}\theta}\,,
\eeq
where $p^{\pm}_{\pm}(n,\theta)$ are degree two polynomials in $n$. Now \eqref{eq:greensol1} is invariant under rescaling the modes, so we have the expression 
\beq
{N\ov q^2}\mathcal{G}_{0,n}&(\theta_1,\theta_2)=v^3 \cot \frac{\pi  v}{2} {8\ov n^2 \left(n^2-v^2\right) }\Big[ (-1)^n\Phi_n^-(\theta_1,v)\Phi_n^+(\theta_2,v) \Theta(-\theta_1+\theta_2) \\&+ (-1)^n\Phi_n^+(\theta_1,v)\Phi_n^-(\theta_2,v) \Theta(\theta_1-\theta_2)
+\Phi_n^-(\theta_1,v)\Phi_n^-(\theta_2,v) \Theta(\pi-\theta_1-\theta_2) \\&+ \Phi_n^+(\theta_1,v)\Phi_n^+(\theta_2,v) \Theta(-\pi+\theta_1+\theta_2)\Big]\,,
\eeq
where we used the Wrosnkian
\beq
\mathcal{W}(\Phi^-_n,\Phi^+_n)= \frac{(-1)^n n^2 \left(n^2-v^2\right) \tan \left(\frac{\pi  v}{2}\right)}{v^4}\,.
\eeq
Using the property \eqref{eq:polynomprop}, we may pull out the polynomial dependence on $n$ as derivatives and write the position space expression as
\beq
\label{eq:dotfourpt}
{N\ov q^2}\sum_n e^{i n Y} \mathcal{G}_{0,n}&(\theta_1,\theta_2) = 8 v^3 \cot \frac{\pi  v}{2} \Big[ \Theta(\pi-\theta_1-\theta_2) \\ &\times  \sum_{a,b=\pm}p^-_a(-i \partial_Y,\theta_1)p^-_b(-i \partial_Y,\theta_2)F(Y+a\frac{n}{2v}\theta_1+b\frac{n}{2v}\theta_2) \\ &+ 
\Theta(-\pi+\theta_1+\theta_2) \sum_{a,b=\pm}p^+_a(-i \partial_Y,\theta_1)p^+_b(-i \partial_Y,\theta_2)F(Y+a\frac{n}{2v}\theta_1+b\frac{n}{2v}\theta_2) 
\\ &+ 
\Theta(-\theta_1+\theta_2) \sum_{a,b=\pm}p^-_a(-i \partial_Y,\theta_1)p^+_b(-i \partial_Y,\theta_2)F(Y+\pi+a\frac{n}{2v}\theta_1+b\frac{n}{2v}\theta_2) 
\\ &+ 
\Theta(\theta_1-\theta_2)\sum_{a,b=\pm}p^+_a(-i \partial_Y,\theta_1)p^-_b(-i \partial_Y,\theta_2)F(Y+\pi+a\frac{n}{2v}\theta_1+b\frac{n}{2v}\theta_2)  \Big]\,,
\eeq
where
\beq
F(Y) &=\sum_{n\neq 0}\frac{1}{n^2(n^2-v^2)} e^{i n Y}\\
&= \frac{\left(-3 Y^2+6 \pi  |Y|-2 \pi ^2\right) v^2-6 \pi  v \csc (\pi  v) \cos ((|Y|-\pi ) v)+6}{6 v^4}\,,
\eeq
and the sum for $F(Y)$ was done by Sommerfeld-Watson resummation. Since all the $p^{\pm}_a$ in \eqref{eq:dotfourpt} are degree two polynomials in the derivatives, it is easy to now explicitly evaluate \eqref{eq:dotfourpt} and confirm that it agrees with the results in \cite{Streicher:2019wek,Choi:2019bmd}.

\bibliographystyle{JHEP}
\bibliography{refs.bib}

\providecommand{\href}[2]{#2}\begingroup\raggedright\begin{thebibliography}{10}

\bibitem{Grozdanov:2017ajz}
S.~Grozdanov, K.~Schalm, and V.~Scopelliti, {\it {Black hole scrambling from
  hydrodynamics}},  {\em Phys. Rev. Lett.} {\bf 120} (2018), no.~23 231601,
  \href{http://xxx.lanl.gov/abs/1710.00921}{{\tt 1710.00921}}.

\bibitem{Blake:2017ris}
M.~Blake, H.~Lee, and H.~Liu, {\it {A quantum hydrodynamical description for
  scrambling and many-body chaos}},  {\em JHEP} {\bf 10} (2018) 127,
  \href{http://xxx.lanl.gov/abs/1801.00010}{{\tt 1801.00010}}.

\bibitem{Blake:2018leo}
M.~Blake, R.~A. Davison, S.~Grozdanov, and H.~Liu, {\it {Many-body chaos and
  energy dynamics in holography}},  {\em JHEP} {\bf 10} (2018) 035,
  \href{http://xxx.lanl.gov/abs/1809.01169}{{\tt 1809.01169}}.

\bibitem{Haehl:2018izb}
F.~M. Haehl and M.~Rozali, {\it {Effective Field Theory for Chaotic CFTs}},
  {\em JHEP} {\bf 10} (2018) 118,
  \href{http://xxx.lanl.gov/abs/1808.02898}{{\tt 1808.02898}}.

\bibitem{Grozdanov:2018kkt}
S.~Grozdanov, {\it {On the connection between hydrodynamics and quantum chaos
  in holographic theories with stringy corrections}},  {\em JHEP} {\bf 01}
  (2019) 048, \href{http://xxx.lanl.gov/abs/1811.09641}{{\tt 1811.09641}}.

\bibitem{Haehl:2019eae}
F.~M. Haehl, W.~Reeves, and M.~Rozali, {\it {Reparametrization modes, shadow
  operators, and quantum chaos in higher-dimensional CFTs}},  {\em JHEP} {\bf
  11} (2019) 102, \href{http://xxx.lanl.gov/abs/1909.05847}{{\tt 1909.05847}}.

\bibitem{Maldacena:2015waa}
J.~Maldacena, S.~H. Shenker, and D.~Stanford, {\it {A bound on chaos}},  {\em
  JHEP} {\bf 08} (2016) 106, \href{http://xxx.lanl.gov/abs/1503.01409}{{\tt
  1503.01409}}.

\bibitem{Gu:2016oyy}
Y.~Gu, X.-L. Qi, and D.~Stanford, {\it {Local criticality, diffusion and chaos
  in generalized Sachdev-Ye-Kitaev models}},  {\em JHEP} {\bf 05} (2017) 125,
  \href{http://xxx.lanl.gov/abs/1609.07832}{{\tt 1609.07832}}.

\bibitem{Maldacena:2016hyu}
J.~Maldacena and D.~Stanford, {\it {Remarks on the Sachdev-Ye-Kitaev model}},
  {\em Phys. Rev.} {\bf D94} (2016), no.~10 106002,
  \href{http://xxx.lanl.gov/abs/1604.07818}{{\tt 1604.07818}}.

\bibitem{Ramirez:2020qer}
D.~M. Ramirez, {\it {Chaos and pole skipping in CFT$_2$}},
  \href{http://xxx.lanl.gov/abs/2009.00500}{{\tt 2009.00500}}.

\bibitem{Mezei:2019dfv}
M.~Mezei and G.~S\'arosi, {\it {Chaos in the butterfly cone}},  {\em JHEP} {\bf
  01} (2020) 186, \href{http://xxx.lanl.gov/abs/1908.03574}{{\tt 1908.03574}}.

\bibitem{Xu:2018xfz}
S.~Xu and B.~Swingle, {\it {Accessing scrambling using matrix product
  operators}},  \href{http://xxx.lanl.gov/abs/1802.00801}{{\tt 1802.00801}}.

\bibitem{Khemani:2018sdn}
V.~Khemani, D.~A. Huse, and A.~Nahum, {\it {Velocity-dependent Lyapunov
  exponents in many-body quantum, semiclassical, and classical chaos}},  {\em
  Phys. Rev.} {\bf B98} (2018), no.~14 144304,
  \href{http://xxx.lanl.gov/abs/1803.05902}{{\tt 1803.05902}}.

\bibitem{Perlmutter:2016pkf}
E.~Perlmutter, {\it {Bounding the Space of Holographic CFTs with Chaos}},  {\em
  JHEP} {\bf 10} (2016) 069, \href{http://xxx.lanl.gov/abs/1602.08272}{{\tt
  1602.08272}}.

\bibitem{Roberts:2014isa}
D.~A. Roberts, D.~Stanford, and L.~Susskind, {\it {Localized shocks}},  {\em
  JHEP} {\bf 03} (2015) 051, \href{http://xxx.lanl.gov/abs/1409.8180}{{\tt
  1409.8180}}.

\bibitem{Mezei:2016wfz}
M.~Mezei and D.~Stanford, {\it {On entanglement spreading in chaotic systems}},
   {\em JHEP} {\bf 05} (2017) 065,
  \href{http://xxx.lanl.gov/abs/1608.05101}{{\tt 1608.05101}}.

\bibitem{Alishahiha:2016cjk}
M.~Alishahiha, A.~Davody, A.~Naseh, and S.~F. Taghavi, {\it {On Butterfly
  effect in Higher Derivative Gravities}},  {\em JHEP} {\bf 11} (2016) 032,
  \href{http://xxx.lanl.gov/abs/1610.02890}{{\tt 1610.02890}}.

\bibitem{Chowdhury:2019kaq}
S.~D. Chowdhury, A.~Gadde, T.~Gopalka, I.~Halder, L.~Janagal, and S.~Minwalla,
  {\it {Classifying and constraining local four photon and four graviton
  S-matrices}},  {\em JHEP} {\bf 02} (2020) 114,
  \href{http://xxx.lanl.gov/abs/1910.14392}{{\tt 1910.14392}}.

\bibitem{Shenker:2014cwa}
S.~H. Shenker and D.~Stanford, {\it {Stringy effects in scrambling}},  {\em
  JHEP} {\bf 05} (2015) 132, \href{http://xxx.lanl.gov/abs/1412.6087}{{\tt
  1412.6087}}.

\bibitem{Kitaev:2017awl}
A.~Kitaev and S.~J. Suh, {\it {The soft mode in the Sachdev-Ye-Kitaev model and
  its gravity dual}},  {\em JHEP} {\bf 05} (2018) 183,
  \href{http://xxx.lanl.gov/abs/1711.08467}{{\tt 1711.08467}}.

\bibitem{Hartnoll:2005ju}
S.~A. Hartnoll and S.~Kumar, {\it {AdS black holes and thermal Yang-Mills
  correlators}},  {\em JHEP} {\bf 12} (2005) 036,
  \href{http://xxx.lanl.gov/abs/hep-th/0508092}{{\tt hep-th/0508092}}.

\bibitem{Romatschke:2015gic}
P.~Romatschke, {\it {Retarded correlators in kinetic theory: branch cuts, poles
  and hydrodynamic onset transitions}},  {\em Eur. Phys. J. C} {\bf 76} (2016),
  no.~6 352, \href{http://xxx.lanl.gov/abs/1512.02641}{{\tt 1512.02641}}.

\bibitem{Kurkela:2017xis}
A.~Kurkela and U.~A. Wiedemann, {\it {Analytic structure of nonhydrodynamic
  modes in kinetic theory}},  {\em Eur. Phys. J. C} {\bf 79} (2019), no.~9 776,
  \href{http://xxx.lanl.gov/abs/1712.04376}{{\tt 1712.04376}}.

\bibitem{Moore:2018mma}
G.~D. Moore, {\it {Stress-stress correlator in $\phi^{4}$ theory: poles or a
  cut?}},  {\em JHEP} {\bf 05} (2018) 084,
  \href{http://xxx.lanl.gov/abs/1803.00736}{{\tt 1803.00736}}.

\bibitem{Grozdanov:2018atb}
S.~Grozdanov, K.~Schalm, and V.~Scopelliti, {\it {Kinetic theory for classical
  and quantum many-body chaos}},  {\em Phys. Rev. E} {\bf 99} (2019), no.~1
  012206, \href{http://xxx.lanl.gov/abs/1804.09182}{{\tt 1804.09182}}.

\bibitem{Son:2002sd}
D.~T. Son and A.~O. Starinets, {\it {Minkowski space correlators in AdS / CFT
  correspondence: Recipe and applications}},  {\em JHEP} {\bf 09} (2002) 042,
  \href{http://xxx.lanl.gov/abs/hep-th/0205051}{{\tt hep-th/0205051}}.

\bibitem{Starinets:2002br}
A.~O. Starinets, {\it {Quasinormal modes of near extremal black branes}},  {\em
  Phys. Rev. D} {\bf 66} (2002) 124013,
  \href{http://xxx.lanl.gov/abs/hep-th/0207133}{{\tt hep-th/0207133}}.

\bibitem{Kovtun:2005ev}
P.~K. Kovtun and A.~O. Starinets, {\it {Quasinormal modes and holography}},
  {\em Phys. Rev. D} {\bf 72} (2005) 086009,
  \href{http://xxx.lanl.gov/abs/hep-th/0506184}{{\tt hep-th/0506184}}.

\bibitem{Alday:2020eua}
L.~F. Alday, M.~Kologlu, and A.~Zhiboedov, {\it {Holographic Correlators at
  Finite Temperature}},  \href{http://xxx.lanl.gov/abs/2009.10062}{{\tt
  2009.10062}}.

\bibitem{Grozdanov:2018fic}
S.~Grozdanov, A.~Lucas, and N.~Poovuttikul, {\it {Holography and hydrodynamics
  with weakly broken symmetries}},  {\em Phys. Rev. D} {\bf 99} (2019), no.~8
  086012, \href{http://xxx.lanl.gov/abs/1810.10016}{{\tt 1810.10016}}.

\bibitem{Grozdanov:2019kge}
S.~Grozdanov, P.~K. Kovtun, A.~O. Starinets, and P.~Tadi\'c, {\it {Convergence
  of the Gradient Expansion in Hydrodynamics}},  {\em Phys. Rev. Lett.} {\bf
  122} (2019), no.~25 251601, \href{http://xxx.lanl.gov/abs/1904.01018}{{\tt
  1904.01018}}.

\bibitem{Grozdanov:2019uhi}
S.~Grozdanov, P.~K. Kovtun, A.~O. Starinets, and P.~Tadi\'c, {\it {The complex
  life of hydrodynamic modes}},  {\em JHEP} {\bf 11} (2019) 097,
  \href{http://xxx.lanl.gov/abs/1904.12862}{{\tt 1904.12862}}.

\bibitem{Streicher:2019wek}
A.~Streicher, {\it {SYK Correlators for All Energies}},  {\em JHEP} {\bf 02}
  (2020) 048, \href{http://xxx.lanl.gov/abs/1911.10171}{{\tt 1911.10171}}.

\bibitem{Choi:2019bmd}
C.~Choi, M.~Mezei, and G.~S\'arosi, {\it {Exact four point function for large
  $q$ SYK from Regge theory}},  \href{http://xxx.lanl.gov/abs/1912.00004}{{\tt
  1912.00004}}.

\bibitem{sachdev1993gapless}
S.~Sachdev and J.~Ye, {\it Gapless spin-fluid ground state in a random quantum
  heisenberg magnet},  {\em Physical review letters} {\bf 70} (1993), no.~21
  3339.

\bibitem{kitaev2014hidden}
A.~Kitaev, {\it Hidden correlations in the hawking radiation and thermal
  noise},  in {\em Talk given at the Fundamental Physics Prize Symposium},
  vol.~10, 2014.

\bibitem{Polchinski:2016xgd}
J.~Polchinski and V.~Rosenhaus, {\it {The Spectrum in the Sachdev-Ye-Kitaev
  Model}},  {\em JHEP} {\bf 04} (2016) 001,
  \href{http://xxx.lanl.gov/abs/1601.06768}{{\tt 1601.06768}}.

\bibitem{Michel:2016kwn}
B.~Michel, J.~Polchinski, V.~Rosenhaus, and S.~Suh, {\it {Four-point function
  in the IOP matrix model}},  {\em JHEP} {\bf 05} (2016) 048,
  \href{http://xxx.lanl.gov/abs/1602.06422}{{\tt 1602.06422}}.

\bibitem{Berkooz:2018qkz}
M.~Berkooz, P.~Narayan, and J.~Simon, {\it {Chord diagrams, exact correlators
  in spin glasses and black hole bulk reconstruction}},  {\em JHEP} {\bf 08}
  (2018) 192, \href{http://xxx.lanl.gov/abs/1806.04380}{{\tt 1806.04380}}.

\bibitem{Cotler:2016fpe}
J.~S. Cotler, G.~Gur-Ari, M.~Hanada, J.~Polchinski, P.~Saad, S.~H. Shenker,
  D.~Stanford, A.~Streicher, and M.~Tezuka, {\it {Black Holes and Random
  Matrices}},  {\em JHEP} {\bf 05} (2017) 118,
  \href{http://xxx.lanl.gov/abs/1611.04650}{{\tt 1611.04650}}, [Erratum:
  JHEP09,002(2018)].

\bibitem{Gu:2018jsv}
Y.~Gu and A.~Kitaev, {\it {On the relation between the magnitude and exponent
  of OTOCs}},  {\em JHEP} {\bf 02} (2019) 075,
  \href{http://xxx.lanl.gov/abs/1812.00120}{{\tt 1812.00120}}.

\bibitem{Lian:2019axs}
B.~Lian, S.~Sondhi, and Z.~Yang, {\it {The chiral SYK model}},  {\em JHEP} {\bf
  09} (2019) 067, \href{http://xxx.lanl.gov/abs/1906.03308}{{\tt 1906.03308}}.

\bibitem{Sarosi:2017ykf}
G.~S\'arosi, {\it {AdS$_{2}$ holography and the SYK model}},  {\em PoS} {\bf
  Modave2017} (2018) 001, \href{http://xxx.lanl.gov/abs/1711.08482}{{\tt
  1711.08482}}.

\bibitem{Policastro:2002tn}
G.~Policastro, D.~T. Son, and A.~O. Starinets, {\it {From AdS / CFT
  correspondence to hydrodynamics. 2. Sound waves}},  {\em JHEP} {\bf 12}
  (2002) 054, \href{http://xxx.lanl.gov/abs/hep-th/0210220}{{\tt
  hep-th/0210220}}.

\bibitem{Blake:2019otz}
M.~Blake, R.~A. Davison, and D.~Vegh, {\it {Horizon constraints on holographic
  Green\textquoteright{}s functions}},  {\em JHEP} {\bf 01} (2020) 077,
  \href{http://xxx.lanl.gov/abs/1904.12883}{{\tt 1904.12883}}.

\bibitem{Wu:2019esr}
X.~Wu, {\it {Higher curvature corrections to pole-skipping}},  {\em JHEP} {\bf
  12} (2019) 140, \href{http://xxx.lanl.gov/abs/1909.10223}{{\tt 1909.10223}}.

\bibitem{Grozdanov:2016vgg}
S.~Grozdanov, N.~Kaplis, and A.~O. Starinets, {\it {From strong to weak
  coupling in holographic models of thermalization}},  {\em JHEP} {\bf 07}
  (2016) 151, \href{http://xxx.lanl.gov/abs/1605.02173}{{\tt 1605.02173}}.

\bibitem{Hartman:2017hhp}
T.~Hartman, S.~A. Hartnoll, and R.~Mahajan, {\it {Upper Bound on Diffusivity}},
   {\em Phys. Rev. Lett.} {\bf 119} (2017), no.~14 141601,
  \href{http://xxx.lanl.gov/abs/1706.00019}{{\tt 1706.00019}}.

\bibitem{Solana:2018pbk}
J.~Casalderrey-Solana, S.~Grozdanov, and A.~O. Starinets, {\it {Transport Peak
  in the Thermal Spectral Function of $\mathcal N=4$ Supersymmetric Yang-Mills
  Plasma at Intermediate Coupling}},  {\em Phys. Rev. Lett.} {\bf 121} (2018),
  no.~19 191603, \href{http://xxx.lanl.gov/abs/1806.10997}{{\tt 1806.10997}}.

\bibitem{Baggioli:2020loj}
M.~Baggioli, {\it {How small hydrodynamics can go}},
  \href{http://xxx.lanl.gov/abs/2010.05916}{{\tt 2010.05916}}.

\bibitem{Grozdanov:2020koi}
S.~Grozdanov, {\it {Bounds on transport from univalence and pole-skipping}},
  \href{http://xxx.lanl.gov/abs/2008.00888}{{\tt 2008.00888}}.

\bibitem{Iqbal:2011in}
N.~Iqbal, H.~Liu, and M.~Mezei, {\it {Semi-local quantum liquids}},  {\em JHEP}
  {\bf 04} (2012) 086, \href{http://xxx.lanl.gov/abs/1105.4621}{{\tt
  1105.4621}}.

\bibitem{Gu:2017njx}
Y.~Gu, A.~Lucas, and X.-L. Qi, {\it {Spread of entanglement in a
  Sachdev-Ye-Kitaev chain}},  {\em JHEP} {\bf 09} (2017) 120,
  \href{http://xxx.lanl.gov/abs/1708.00871}{{\tt 1708.00871}}.

\bibitem{Davison:2016ngz}
R.~A. Davison, W.~Fu, A.~Georges, Y.~Gu, K.~Jensen, and S.~Sachdev, {\it
  {Thermoelectric transport in disordered metals without quasiparticles: The
  Sachdev-Ye-Kitaev models and holography}},  {\em Phys. Rev. B} {\bf 95}
  (2017), no.~15 155131, \href{http://xxx.lanl.gov/abs/1612.00849}{{\tt
  1612.00849}}.

\bibitem{Grozdanov:2018gfx}
S.~Grozdanov and A.~O. Starinets, {\it {Adding new branches to the
  \textquotedblleft{}Christmas tree\textquotedblright{} of the quasinormal
  spectrum of black branes}},  {\em JHEP} {\bf 04} (2019) 080,
  \href{http://xxx.lanl.gov/abs/1812.09288}{{\tt 1812.09288}}.

\end{thebibliography}\endgroup

\end{document}